\documentclass[onecolumn, longauth]{aa}
\pdfoutput=1
\usepackage{euclid}
\usepackage[T1]{fontenc}
\usepackage[utf8]{inputenc}
\usepackage{geometry}   
\usepackage{float}
\usepackage[section]{placeins}
\usepackage{amsmath}
\usepackage{mathtools}
\usepackage[switch, mathlines]{lineno}
\usepackage[usenames,dvipsnames]{xcolor} 
\usepackage[normalem]{ulem}
\usepackage{enumitem}
\usepackage{csquotes}
\usepackage{xspace}

\usepackage{listings}
\usepackage{color}
\definecolor{dkgreen}{rgb}{0,0.6,0}
\definecolor{gray}{rgb}{0.5,0.5,0.5}
\definecolor{mauve}{rgb}{0.58,0,0.82}
\definecolor{burntorange}{rgb}{0.8, 0.33, 0.0}

\newcommand{\cosmicfish}{\texttt{CosmicFish}}
\newcommand{\CF}{\texttt{CF}}
\newcommand{\montepython}{\texttt{MontePython}}
\newcommand{\MP}{\texttt{MP}}
\newcommand{\class}{\texttt{CLASS}}
\newcommand{\camb}{\texttt{CAMB}}
\newcommand{\AP}{Alcock-Paczy\'nski}

\newcommand{\CFextCAMB}{\texttt{CF/ext/CAMB}}
\newcommand{\CFintCAMB}{\texttt{CF/int/CAMB}}
\newcommand{\CFextCLASS}{\texttt{CF/ext/CLASS}}
\newcommand{\CFintCLASS}{\texttt{CF/int/CLASS}}
\newcommand{\MPFisher}{\texttt{MP/Fisher}}
\newcommand{\MPMCMC}{\texttt{MP/MCMC}}

\newcommand{\de}{\mathrm{d}}

\defcitealias{Blanchard:2019oqi}{EP:VII}

\newcommand{\istfisher}{\citetalias{Blanchard:2019oqi}}

\usepackage[colorlinks=true, allcolors=blue]{hyperref}
\usepackage[capitalise]{cleveref}
\usepackage{orcidlink}

\crefname{chapter}{Chap.}{Chaps.}
\crefname{section}{Sect.}{Sects.}
\crefname{figure}{Fig.}{Figs.}
\Crefname{chapter}{Chapter}{Chapters}
\Crefname{section}{Section}{Sections}
\Crefname{figure}{Figure}{Figures}

\begin{document}

\title{\Euclid: Validation of the MontePython forecasting tools\thanks{This paper is published on behalf of the Euclid Consortium.}}


\newcommand{\orcid}[1]{\orcidlink{#1}} 
\author{S.~Casas\orcid{0000-0002-4751-5138}$^{1}$\thanks{\email{s.casas@protonmail.com}}, J.~Lesgourgues\orcid{0000-0001-7627-353X}$^{1}$, N.~Sch\"{o}neberg\orcid{0000-0002-7873-0404}$^{2}$, Sabarish~V.~M.\orcid{0000-0001-5677-0838}$^{1,3}$, L.~Rathmann\orcid{0000-0003-1035-1923}$^{1}$, M.~Doerenkamp$^{1,4}$, M.~Archidiacono$^{5}$, E.~Bellini$^{6,7,8,9}$, S.~Clesse\orcid{0000-0001-5079-1785}$^{10}$, N.~Frusciante$^{11}$
% ...
}

\institute{$^{1}$ Institute for Theoretical Particle Physics and Cosmology (TTK), RWTH Aachen University, 52056 Aachen, Germany\\
$^{2}$ Institut de Ci\`{e}ncies del Cosmos (ICCUB), Universitat de Barcelona (IEEC-UB), Mart\'{i} i Franqu\`{e}s 1, 08028 Barcelona, Spain\\
$^{3}$ Hamburger Sternwarte, University of Hamburg, Gojenbergsweg 112, 21029 Hamburg, Germany\\
$^{4}$ Physikalisches Institut, Ruprecht-Karls-Universit\"at Heidelberg, Im Neuenheimer Feld 226, 69120 Heidelberg, Germany\\
$^{5}$ Dipartimento di Fisica "Aldo Pontremoli", Universit\'a degli Studi di Milano, Via Celoria 16, 20133 Milano, Italy\\
$^{6}$ IFPU, Institute for Fundamental Physics of the Universe, via Beirut 2, 34151 Trieste, Italy\\
$^{7}$ SISSA, International School for Advanced Studies, Via Bonomea 265, 34136 Trieste TS, Italy\\
$^{8}$ INAF-Osservatorio Astronomico di Trieste, Via G. B. Tiepolo 11, 34143 Trieste, Italy\\
$^{9}$ INFN, Sezione di Trieste, Via Valerio 2, 34127 Trieste TS, Italy\\
$^{10}$ Universit\'e Libre de Bruxelles (ULB), Service de Physique Th\'eorique CP225, Boulevard du Triophe, 1050 Bruxelles, Belgium\\
$^{11}$ Department of Physics "E. Pancini", University Federico II, Via Cinthia 6, 80126, Napoli, Italy\\
$^{12}$ INAF-Osservatorio Astronomico di Roma, Via Frascati 33, 00078 Monteporzio Catone, Italy\\
$^{13}$ INFN-Sezione di Roma, Piazzale Aldo Moro, 2 - c/o Dipartimento di Fisica, Edificio G. Marconi, 00185 Roma, Italy\\
$^{14}$ Dipartimento di Fisica, Universit\'a degli Studi di Torino, Via P. Giuria 1, 10125 Torino, Italy\\
$^{15}$ INFN-Sezione di Torino, Via P. Giuria 1, 10125 Torino, Italy\\
$^{16}$ INAF-Osservatorio Astrofisico di Torino, Via Osservatorio 20, 10025 Pino Torinese (TO), Italy\\
$^{17}$ Departamento de F\'isica, FCFM, Universidad de Chile, Blanco Encalada 2008, Santiago, Chile\\
$^{18}$ Universit\'e St Joseph; Faculty of Sciences, Beirut, Lebanon\\
$^{19}$ Institut f\"ur Theoretische Physik, University of Heidelberg, Philosophenweg 16, 69120 Heidelberg, Germany\\
$^{20}$ Institut de Recherche en Astrophysique et Plan\'etologie (IRAP), Universit\'e de Toulouse, CNRS, UPS, CNES, 14 Av. Edouard Belin, 31400 Toulouse, France\\
$^{21}$ Dipartimento di Fisica e Scienze della Terra, Universit\'a degli Studi di Ferrara, Via Giuseppe Saragat 1, 44122 Ferrara, Italy\\
$^{22}$ Istituto Nazionale di Fisica Nucleare, Sezione di Ferrara, Via Giuseppe Saragat 1, 44122 Ferrara, Italy\\
$^{23}$ INAF-IASF Milano, Via Alfonso Corti 12, 20133 Milano, Italy\\
$^{24}$ Universit\'{e} Paris-Saclay, CNRS/IN2P3, IJCLab, 91405 Orsay, France\\
$^{25}$ Centre National d'Etudes Spatiales -- Centre spatial de Toulouse, 18 avenue Edouard Belin, 31401 Toulouse Cedex 9, France\\
$^{26}$ Jet Propulsion Laboratory, California Institute of Technology, 4800 Oak Grove Drive, Pasadena, CA, 91109, USA\\
$^{27}$ Universit\'e Paris-Saclay, Universit\'e Paris Cit\'e, CEA, CNRS, Astrophysique, Instrumentation et Mod\'elisation Paris-Saclay, 91191 Gif-sur-Yvette, France\\
$^{28}$ Universit\'e Paris-Saclay, CNRS, Institut d'astrophysique spatiale, 91405, Orsay, France\\
$^{29}$ Institute of Cosmology and Gravitation, University of Portsmouth, Portsmouth PO1 3FX, UK\\
$^{30}$ INAF-Osservatorio di Astrofisica e Scienza dello Spazio di Bologna, Via Piero Gobetti 93/3, 40129 Bologna, Italy\\
$^{31}$ Dipartimento di Fisica e Astronomia "Augusto Righi" - Alma Mater Studiorum Universit\`{a} di Bologna, via Piero Gobetti 93/2, 40129 Bologna, Italy\\
$^{32}$ INFN-Sezione di Bologna, Viale Berti Pichat 6/2, 40127 Bologna, Italy\\
$^{33}$ Dipartimento di Fisica, Universit\`{a} di Genova, Via Dodecaneso 33, 16146, Genova, Italy\\
$^{34}$ INFN-Sezione di Genova, Via Dodecaneso 33, 16146, Genova, Italy\\
$^{35}$ INAF-Osservatorio Astronomico di Capodimonte, Via Moiariello 16, 80131 Napoli, Italy\\
$^{36}$ Instituto de Astrof\'isica e Ci\^encias do Espa\c{c}o, Universidade do Porto, CAUP, Rua das Estrelas, PT4150-762 Porto, Portugal\\
$^{37}$ Institut de F\'{i}sica d'Altes Energies (IFAE), The Barcelona Institute of Science and Technology, Campus UAB, 08193 Bellaterra (Barcelona), Spain\\
$^{38}$ Port d'Informaci\'{o} Cient\'{i}fica, Campus UAB, C. Albareda s/n, 08193 Bellaterra (Barcelona), Spain\\
$^{39}$ INFN section of Naples, Via Cinthia 6, 80126, Napoli, Italy\\
$^{40}$ Dipartimento di Fisica e Astronomia "Augusto Righi" - Alma Mater Studiorum Universit\'a di Bologna, Viale Berti Pichat 6/2, 40127 Bologna, Italy\\
$^{41}$ Institut national de physique nucl\'eaire et de physique des particules, 3 rue Michel-Ange, 75794 Paris C\'edex 16, France\\
$^{42}$ Institute for Astronomy, University of Edinburgh, Royal Observatory, Blackford Hill, Edinburgh EH9 3HJ, UK\\
$^{43}$ European Space Agency/ESRIN, Largo Galileo Galilei 1, 00044 Frascati, Roma, Italy\\
$^{44}$ ESAC/ESA, Camino Bajo del Castillo, s/n., Urb. Villafranca del Castillo, 28692 Villanueva de la Ca\~nada, Madrid, Spain\\
$^{45}$ University of Lyon, Univ Claude Bernard Lyon 1, CNRS/IN2P3, IP2I Lyon, UMR 5822, 69622 Villeurbanne, France\\
$^{46}$ Institute of Physics, Laboratory of Astrophysics, Ecole Polytechnique F\'{e}d\'{e}rale de Lausanne (EPFL), Observatoire de Sauverny, 1290 Versoix, Switzerland\\
$^{47}$ Mullard Space Science Laboratory, University College London, Holmbury St Mary, Dorking, Surrey RH5 6NT, UK\\
$^{48}$ Department of Astronomy, University of Geneva, ch. d'Ecogia 16, 1290 Versoix, Switzerland\\
$^{49}$ Instituto de Astrof\'isica e Ci\^encias do Espa\c{c}o, Faculdade de Ci\^encias, Universidade de Lisboa, Campo Grande, 1749-016 Lisboa, Portugal\\
$^{50}$ Departamento de F\'isica, Faculdade de Ci\^encias, Universidade de Lisboa, Edif\'icio C8, Campo Grande, PT1749-016 Lisboa, Portugal\\
$^{51}$ INFN-Padova, Via Marzolo 8, 35131 Padova, Italy\\
$^{52}$ INAF-Osservatorio Astronomico di Padova, Via dell'Osservatorio 5, 35122 Padova, Italy\\
$^{53}$ Max Planck Institute for Extraterrestrial Physics, Giessenbachstr. 1, 85748 Garching, Germany\\
$^{54}$ University Observatory, Faculty of Physics, Ludwig-Maximilians-Universit{\"a}t, Scheinerstr. 1, 81679 Munich, Germany\\
$^{55}$ Institute of Theoretical Astrophysics, University of Oslo, P.O. Box 1029 Blindern, 0315 Oslo, Norway\\
$^{56}$ von Hoerner \& Sulger GmbH, Schlo{\ss}Platz 8, 68723 Schwetzingen, Germany\\
$^{57}$ Technical University of Denmark, Elektrovej 327, 2800 Kgs. Lyngby, Denmark\\
$^{58}$ Cosmic Dawn Center (DAWN), Denmark\\
$^{59}$ Max-Planck-Institut f\"ur Astronomie, K\"onigstuhl 17, 69117 Heidelberg, Germany\\
$^{60}$ Universit\"ats-Sternwarte M\"unchen, Fakult\"at f\"ur Physik, Ludwig-Maximilians-Universit\"at M\"unchen, Scheinerstrasse 1, 81679 M\"unchen, Germany\\
$^{61}$ Universit\'e de Gen\`eve, D\'epartement de Physique Th\'eorique and Centre for Astroparticle Physics, 24 quai Ernest-Ansermet, CH-1211 Gen\`eve 4, Switzerland\\
$^{62}$ Department of Physics, P.O. Box 64, 00014 University of Helsinki, Finland\\
$^{63}$ Helsinki Institute of Physics, Gustaf H{\"a}llstr{\"o}min katu 2, University of Helsinki, Helsinki, Finland\\
$^{64}$ NOVA optical infrared instrumentation group at ASTRON, Oude Hoogeveensedijk 4, 7991PD, Dwingeloo, The Netherlands\\
$^{65}$ Argelander-Institut f\"ur Astronomie, Universit\"at Bonn, Auf dem H\"ugel 71, 53121 Bonn, Germany\\
$^{66}$ Department of Physics, Institute for Computational Cosmology, Durham University, South Road, DH1 3LE, UK\\
$^{67}$ Universit\'e Paris Cit\'e, CNRS, Astroparticule et Cosmologie, 75013 Paris, France\\
$^{68}$ European Space Agency/ESTEC, Keplerlaan 1, 2201 AZ Noordwijk, The Netherlands\\
$^{69}$ Department of Physics and Astronomy, University of Aarhus, Ny Munkegade 120, DK-8000 Aarhus C, Denmark\\
$^{70}$ Centre for Astrophysics, University of Waterloo, Waterloo, Ontario N2L 3G1, Canada\\
$^{71}$ Department of Physics and Astronomy, University of Waterloo, Waterloo, Ontario N2L 3G1, Canada\\
$^{72}$ Perimeter Institute for Theoretical Physics, Waterloo, Ontario N2L 2Y5, Canada\\
$^{73}$ AIM, CEA, CNRS, Universit\'{e} Paris-Saclay, Universit\'{e} de Paris, 91191 Gif-sur-Yvette, France\\
$^{74}$ Universit\'e Paris-Saclay, Universit\'e Paris Cit\'e, CEA, CNRS, AIM, 91191, Gif-sur-Yvette, France\\
$^{75}$ Space Science Data Center, Italian Space Agency, via del Politecnico snc, 00133 Roma, Italy\\
$^{76}$ Institute of Space Science, Str. Atomistilor, nr. 409 M\u{a}gurele, Ilfov, 077125, Romania\\
$^{77}$ Dipartimento di Fisica e Astronomia "G.Galilei", Universit\'a di Padova, Via Marzolo 8, 35131 Padova, Italy\\
$^{78}$ Dipartimento di Fisica e Astronomia, Universit\'a di Bologna, Via Gobetti 93/2, 40129 Bologna, Italy\\
$^{79}$ Aix-Marseille Universit\'e, CNRS/IN2P3, CPPM, Marseille, France\\
$^{80}$ Institut d'Estudis Espacials de Catalunya (IEEC), Carrer Gran Capit\'a 2-4, 08034 Barcelona, Spain\\
$^{81}$ Institut de Ciencies de l'Espai (IEEC-CSIC), Campus UAB, Carrer de Can Magrans, s/n Cerdanyola del Vall\'es, 08193 Barcelona, Spain\\
$^{82}$ Aix-Marseille Universit\'e, CNRS, CNES, LAM, Marseille, France\\
$^{83}$ Centro de Investigaciones Energ\'eticas, Medioambientales y Tecnol\'ogicas (CIEMAT), Avenida Complutense 40, 28040 Madrid, Spain\\
$^{84}$ Instituto de Astrof\'isica e Ci\^encias do Espa\c{c}o, Faculdade de Ci\^encias, Universidade de Lisboa, Tapada da Ajuda, 1349-018 Lisboa, Portugal\\
$^{85}$ Universidad Polit\'ecnica de Cartagena, Departamento de Electr\'onica y Tecnolog\'ia de Computadoras,  Plaza del Hospital 1, 30202 Cartagena, Spain\\
$^{86}$ Kapteyn Astronomical Institute, University of Groningen, PO Box 800, 9700 AV Groningen, The Netherlands\\
$^{87}$ INFN-Bologna, Via Irnerio 46, 40126 Bologna, Italy\\
$^{88}$ Infrared Processing and Analysis Center, California Institute of Technology, Pasadena, CA 91125, USA\\
$^{89}$ Institut d'Astrophysique de Paris, 98bis Boulevard Arago, 75014, Paris, France\\
$^{90}$ Junia, EPA department, 41 Bd Vauban, 59800 Lille, France}

\date{}
\authorrunning{Casas et al.}
\abstract
{The \Euclid mission of the European Space Agency will perform a survey of weak lensing cosmic shear and galaxy clustering in order to constrain cosmological models and fundamental physics.}
{We expand and adjust the mock \Euclid likelihoods of the \montepython{} software in order to match the exact recipes used in previous \Euclid Fisher matrix forecasts for several probes: weak lensing cosmic shear, photometric galaxy clustering, the cross-correlation between the latter observables, and spectroscopic galaxy clustering. We also establish which precision settings are required when running the Einstein--Boltzmann solvers \class{} and \camb{} in the context of \Euclid.}
{For the minimal cosmological model, extended to include dynamical dark energy, we perform Fisher matrix forecasts based directly on a numerical evaluation of second derivatives of the likelihood with respect to model parameters. We compare our results with those of other forecasting methods and tools.}
{We show that such \montepython{} forecasts agree very well with previous Fisher forecasts published by the Euclid Collaboration, and also, with new forecasts produced by the \cosmicfish{} code, now interfaced directly with the two Einstein--Boltzmann solvers \camb{} and \class. Moreover, to establish the validity of the Gaussian approximation, we show that the Fisher matrix marginal error contours coincide with the credible regions obtained when running Monte Carlo Markov Chains with \montepython{} while using the exact same mock likelihoods.}
{The new \Euclid forecast pipelines presented here are ready for use with additional cosmological parameters, in order to explore extended cosmological models.}

\maketitle

\section{Introduction}
Forecasts for large-scale structure surveys are useful, first, for predicting the sensitivity of future experiments to cosmological models and parameters, and second, for paving the way to the analysis of real data. After the publication of many independent \Euclid-like forecasts, an effort was undertaken within the Inter-Science Taskforce for Forecasting (known as IST:F) working group of the Euclid Collaboration to compare several forecasting pipelines and validate them across each other. This has led to the publication of \citet[][`Euclid preparation: VII', hereafter EP:VII]{Blanchard:2019oqi}.

\istfisher\ presents a comparison of Fisher forecasts, based on the calculation  of the Fisher information matrix. This method provides a good approximation to the true experimental sensitivity as long as the posterior is nearly Gaussian, that is, as long as the likelihood is a nearly Gaussian function of model parameters (for fixed fiducial data and assuming flat priors on model parameters). Fisher matrices can be tricky to compute because they involve a calculation of derivatives with a finite difference method. The results may depend on the choice of algorithm and stepsizes because of two factors: (i) numerical noise in the theory codes and Fisher matrix codes lead to unstable derivatives in the small stepsize limit, and (ii) the posterior may deviate from a multivariate Gaussian. \istfisher\ shows how to mitigate these issues and obtains a very good level of agreement between several Fisher matrix codes. We choose one of them, \cosmicfish{} \citep{Raveri:2016leq}, as a representative case of \istfisher\ codes. In this work, we will validate a handful of new pipelines by comparing them directly with \cosmicfish. 

In order to compare different Fisher matrix codes in nearly ideal conditions, one can make a strategical choice. First, compute theory predictions in advance at various points in parameter space using an Einstein--Boltzmann Solver (EBS), and store the results in files. Secondly, let different Fisher matrix codes read this data and compute the Fisher matrix with their own algorithm. This strategy is the one adopted in \istfisher, with theory predictions usually computed by the EBS \camb{} \citep{Lewis:1999bs}.\footnote{This was not the case for all the algorithms used in \istfisher: for the photometric probe, two of the pipelines could call directly the EBS instead of storing its results in files.}

This approach is very well suited for the cross-comparison of several Fisher matrix codes, but not for testing the impact of different EBSs. One of the intermediate goals of this work is to achieve such a task. We will compare \cosmicfish{} results when the code reads some files produced either by \camb{} or \class{} \citep{Lesgourgues:2011re,Blas:2011rf}. We will find very good agreement between these two choices. We also provide a discussion of the settings that need to be imposed to each code in order to get stable, accurate, and mutually agreeing results. 

The strategy described above relies on the storage of the data containing theoretical predictions in some files. This was a rational attitude for the purpose of comparing different Fisher matrix codes. However, once Fisher matrix codes have been validated, the need to run EBSs separately and to fill up a directory with a substantial amount of data appears as relatively heavy. Ideally, one would like to call the EBS on the fly from the Fisher code, in order to save time and memory. Another intermediate goal of the current work is to implement this possibility in \cosmicfish. We will then validate some forecasts in which \cosmicfish{} calls either \camb{} or \class{} on the fly.

The Gaussian posterior approximation breaks in the case of parameters with asymmetric error bars, or with a posterior hitting the prior edge (e.g. for parameters that are always positive and whose best-fit value is close to zero), or in presence of a strong nonlinear degeneracy between parameters. In order to go beyond the Gaussian approximation, it is necessary to explore directly the full likelihood, instead of just its second derivatives computed at the best-fit point. This is often done using Bayesian inference algorithms -- such as, e.g., the Metropolis-Hastings algorithm -- involving Monte Carlo Markov Chains (MCMCs). Such MCMCs require a direct coding of the likelihood ${\cal L}$, unlike Fisher matrix codes like \cosmicfish{} that require the coding of its second derivatives, that is, of formulas in which the full likelihood itself does not appear explicitly.

\Euclid forecasts based on an MCMC approach were performed earlier in a few papers \citep{Audren:2012vy,Sprenger:2018tdb,Brinckmann:2018owf} using the framework of the \montepython{} Bayesian inference package \citep{Audren:2012wb,Brinckmann:2018cvx}. However, the \Euclid recipes used in these works were never accurately compared to those validated in \istfisher. 

An advantage of the \montepython{} package is that, once the mock likelihood describing an experiment has been implemented, it is possible to run \montepython{} either in Fisher mode or in MCMC mode. The former mode will estimate the Fisher matrix directly from the likelihood, evaluated at a few points in the vicinity of the best fit \citep[for details see][]{Brinckmann:2018cvx}. The MCMC mode allows for more reliable forecasts beyond the Gaussian approximation. It is important to stress that both methods rely on the numerical implementation of the likelihood formula in a single place in the code. Thus, once the likelihood has been validated in one mode, it can be considered as equally valid in the other mode. In our case, we can validate the \montepython{} \Euclid likelihoods against the forecasts of \istfisher\ by running \montepython{} (\MP) in Fisher mode (that we will call \MPFisher), and then, if needed, use the same likelihoods in the \montepython{} MCMC mode (that we call \MPMCMC). As such, \montepython{} is a valuable tool to transition from reliable Fisher forecasts to reliable MCMC forecasts. 

One of the main goals of this paper is actually to validate the \MPFisher{} \Euclid pipeline against \istfisher\ pipelines, for the same cosmological model as in \istfisher, that is, the minimal $\Lambda$CDM model extended to dark energy with two equation-of-state parameters $w_0$ and $w_a$ -- usually referred as the Chevallier--Polarski--Linder (CPL) model \citep{Chevallier:2000qy,Linder:2002et}. This validation will establish that the \montepython{} \Euclid likelihoods contain exactly the same modelling of \Euclid data as \istfisher, and can thus be used for robust forecasts in both \MPFisher{} and \MPMCMC{} modes. Note that \montepython{} can be readily used with all the free parameters of all the cosmological models implemented in the \class{} code. Thus, this validation is an important step for cosmologists who want to produce robust \Euclid forecasts in essentially whatever extended cosmological model, with or without the Fisher approximation.

However, we should stress that the \montepython{} mock \Euclid likelihoods used within this paper should not be confused with the official \Euclid likelihood, meant to be used with real data, that is currently under development within the Euclid Collaboration. At this stage, the \montepython{} likelihoods are only meant for forecasting purposes. They do not account for the state of the art in the modelling of theoretical and instrumental effects within the Euclid Collaboration.

In \cref{sec:fisher_gen}, we discuss the general analytic form of the likelihood and of the Fisher matrix assumed in \istfisher\ codes -- such as  \cosmicfish{} -- and in \montepython. These expressions rely on the same {\it observable power spectra} describing the data at the level of two-point statistics. In \cref{sec:lkl}, we provide more details  on the calculation of these observable power spectra, taking into account nonlinear corrections and instrumental errors. In \cref{sec:validation}, we compare five ways to compute the Fisher matrix with either \cosmicfish{} or \montepython. In \cref{sec:forecast_results}, we summarise our results for the sensitivity of \Euclid to the parameters of the $\Lambda$CDM+$\{w_0, w_a\}$ model, showing also the results of a \montepython{} MCMC forecast for comparison. In \cref{sec:acc}, we discuss the importance of correctly setting the input parameters of \camb{} or \class{} in order to get consistent and robust results -- showing in particular that precision settings need to be carefully handled in \camb. We present our conclusions in \cref{sec:conclusions}.

\section{Likelihood-based Fisher matrices\label{sec:fisher_gen}}

Forecasts for cosmological surveys are often based on the Fisher matrix formalism. In a Bayesian context, the Fisher matrix describes the curvature of the logarithm of the likelihood ${\cal L}$ in the vicinity of the best fit,\footnote{In the context of forecasts, the best fit is simply the assumed fiducial model.} 
\begin{equation}\label{eq:def_fisher}
F_{\alpha \beta}= - \partial_\alpha \partial_\beta \ln {\cal L} \big|_\text{best fit}\;,
\end{equation}
where $\{\alpha, \beta\}$ are the model (cosmological or nuisance) parameter indices. The Fisher matrix provides an accurate representation of the true likelihood only if the latter is close to being Gaussian, but it is fast to evaluate compared to a full MCMC forecast. In the presence of strong degeneracies among the parameters and highly non-Gaussian likelihoods, the Fisher matrix results may depend heavily on the steps used to evaluate numerical derivatives, and may differ significantly from MCMC results. However, in the case of a sufficiently Gaussian posterior, the inverse of the Fisher matrix can be used to estimate confidence ellipses for pairs of parameters and confidence intervals for individual parameters. The present work will confirm that for forecasts on the sensitivity of \Euclid to the parameters of the $\Lambda$CDM+$\{w_0, w_a\}$ model, the Fisher formalism is applicable. 

\subsection{Photometric surveys}
In a usual analysis of the photometric surveys of weak lensing (WL) and galaxy clustering (GCph), the observed\footnote{In typical forecasts there is no actual observed data. Instead, the fiducial model (referred here as ``fid'') is used to emulate a possible observation.} signal is decomposed into spherical harmonic coefficients $a_i^\mathrm{fid}(\ell,m)$ in each redshift bin $i$. In many analyses, these coefficients are modeled to be Gaussian distributed with a covariance matrix $C_{ij}^\mathrm{th}(\ell)$. In a realistic universe, this only captures a part of the information contained within the signal, but within this work we are not concerned with the analysis of other higher order statistics (such as for example the bi- or trispectrum, void or peak counts). Furthermore, it is often presumed that a reduced analyzed fraction of the sky $f_\mathrm{sky}<1$ causes a corresponding proportional decrease in the covariance matrix. This leads to the likelihood
\begin{equation}
    \mathcal{L} = \mathcal{N} \prod_{\ell m} \frac{1}{\sqrt{\det C_\ell^\mathrm{th}}} \exp\left[- f_\mathrm{sky} \frac{1}{2}\sum_{ij} a_i^\mathrm{fid}(\ell,m)\,\, (C_{ij}^\mathrm{th})^{-1}(\ell) \,\,[a_j^\mathrm{fid}(\ell,m)]^*\right]\;,
\end{equation}
where the indices $\{i,j\}$ run over each bin of each probe, while $\mathcal{N}$ is the likelihood normalization factor. Using the fact that the observed covariance can be estimated via $C^\mathrm{fid}_{ij}(\ell) = \frac{1}{2\ell+1}\sum_m a^\mathrm{fid}_{i}(\ell,m) \,\, [a^\mathrm{fid}_{j}(\ell,m)]^*$, a few steps of algebra lead to
\begin{equation}\label{eq:photometric_general}
\small
    \chi^2 \equiv -2 \ln \frac{\mathcal{L}}{\mathcal{L}_\mathrm{max}} = f_\mathrm{sky}  \sum_\ell (2\ell+1) \left\{\ln\left[\frac{\det \tens{C}^\mathrm{th}(\ell)}{\det \tens{C}^\mathrm{fid}(\ell)}\right] + \mathrm{Tr}[(\tens{C}^\mathrm{th})^{-1}(\ell) \,\, \tens{C}^\mathrm{fid}(\ell)] - N\right\}\;,
\end{equation}
where $N$ is the size of each $\tens{C}(\ell)$ matrix. From here, two possible steps can be taken. Either one follows the definition of Eq.~\eqref{eq:def_fisher} to derive the Fisher matrix, or one derives a more compact expression of the likelihood.
The latter approach leads to
\begin{equation}
    \chi^2 = f_\mathrm{sky} \sum_\ell \left(2\ell+1\right) \left( \frac{d_\ell^\mathrm{mix}}{d_\ell^\mathrm{th}} + \ln \frac{d_\ell^\mathrm{th}}{d_\ell^\mathrm{fid}} - N \right)\;,
    \label{eq:LWL}
\end{equation}
with
\begin{align}
    d_\ell^\mathrm{th} &= \det \left[C_{ij}^\mathrm{th}(\ell) \right]\;, \\
    d_\ell^\mathrm{fid} &= \det \left[ C_{ij}^\mathrm{fid}(\ell) \right]\;, \\
    d_\ell^\mathrm{mix} &= \sum_k \det\left[
    \begin{cases}
    C_{ij}^\mathrm{th}(\ell) & \text{ if } j \neq k \\ C_{ij}^\mathrm{fid}(\ell) & \text{ if } j=k \end{cases}
    \right]\;.
\end{align}
In the last definition, $k$ is an index running over each bin of each probe, and thus over each column of the theory matrix $\tens{C}^\mathrm{th}$. In each term of the sum, the determinant is evaluated over a matrix such that the $k$-th column of the theory matrix $\tens{C}^\mathrm{th}$ has been substituted by one column of the fiducial matrix $\tens{C}^\mathrm{fid}$.

The likelihood of \cref{eq:photometric_general} can also be used to derive an expression for the Fisher matrix by differentiating twice with respect to an arbitrary pair of parameters of indices $\{\alpha,\beta\}$ and evaluating at the fiducial. Then we find a well-known compact expression for the Fisher matrix (\citealt{Carron:2012pw}; \istfisher)
\begin{equation}
\small
F_{\alpha \beta} = \frac{1}{2} f_\mathrm{sky} \sum_\ell (2\ell+1) \mathrm{Tr} \left\{ \left[\tens{C}^\mathrm{fid}(\ell)\right]^{-1} \left[ \left. \partial_\alpha \tens{C}^\mathrm{th}(\ell)\right|_\mathrm{fid}\right] \left[{\tens{C}^\mathrm{fid}(\ell)}\right]^{-1} \left[\left. \partial_\beta \tens{C}^\mathrm{th}(\ell)\right|_\mathrm{fid} \right]
\right\}\;,
\label{eq:fisherWL}
\end{equation}
where we omitted redshift bin indices for concision.
All derivatives are evaluated precisely at the fiducial model.\footnote{There exists another option: one may use a likelihood that is Gaussian in the $C(\ell)$'s instead of the $a_{\ell m}$'s. The  \istfisher\ paper uses both methods to compute the Fisher matrix forecasts, showing that they are equivalent for all our practical purposes. See Eq.~(142) and Eq.~(143) in \istfisher\ for details.}

\subsection{Spectroscopic surveys}

Similarly, for galaxy clustering data in a spectroscopic survey (GCs), the likelihood is commonly assumed to be Gaussian with respect to the observed galaxy power spectrum $P_\mathrm{obs}(k,\mu,z)$, such that $\chi^2 \equiv -2 \ln ({\mathcal{L}}/{\mathcal{L}_\mathrm{max}})$ reads
\begin{equation}\label{eq:LGC}
    \chi^2 \equiv -2 \ln \frac{\mathcal{L}}{\mathcal{L}_\mathrm{max}}
    = \sum_{i} \!\! \int k^2 \mathrm{d}k  \!\! \int \!\! \mathrm{d}\mu \,\, \frac{V_i^\mathrm{fid}}{8\pi^2} \left\{\frac{P_\mathrm{obs}^\mathrm{th}\left[k_\mathrm{obs}(k,\mu,z_i),\mu_\mathrm{obs}(\mu,z_i),z_i\right]-P_\mathrm{obs}^\mathrm{fid}\left(k,\mu,z_i\right)}{P_\mathrm{obs}^\mathrm{th}\left[k_\mathrm{obs}(k,\mu,z_i),\mu_\mathrm{obs}(\mu,z_i),z_i\right]} \right\}^2\!\!,
\end{equation}
where $k$ is the Fourier wavenumber, $\mu$ is the cosine of the angle between the Fourier wavevector and the line of sight, $z_i$ is the central redshift of bin $i$ and $V_i^\mathrm{fid}$ is the volume covered by this bin. In the previous expression, $P_\mathrm{obs}^\mathrm{th}$ is the observable power spectrum, given by the galaxy spectrum of the theoretical cosmology (the cosmology that one wants to confront to the data) plus shot noise, and $P_\mathrm{obs}^\mathrm{fid}$ is the power spectrum of the mock observation (set equal to the galaxy spectrum of the fiducial cosmology plus shot noise). The volume $V_i^\mathrm{fid}$ is evaluated for the latter cosmology. Importantly, Eq.~\eqref{eq:LGC} already contains the \AP{} effect, in which the wavenumber $k$ and angle $\mu$ for the observed power spectrum are distorted through the impact of the analysis pipeline that converts the observed angular positions $\vec \vartheta$ and redshifts $z$ of individual galaxies into a power spectrum through the use of some reference cosmology. This gives $z$-dependent relations between the value of $(k,\mu)$ in the reference cosmology and the observed values $(k_\mathrm{obs}, \mu_\mathrm{obs})$. We identify the reference cosmology and the fiducial cosmology in order to get simpler equations, since in this case $P_\mathrm{obs}^\mathrm{fid}$ does not receive additional \AP{} factors. We write the precise formulas for the \AP{} effect and the full observed power spectrum $P_\mathrm{obs}(k,\mu,z)$ further below in \cref{sec:spectroscopic}.

By taking again the second derivative of Eq.~\eqref{eq:LGC} and evaluating it at the fiducial cosmology, we find the well-known and compact expression for the Fisher matrix
\begin{equation}
    F_{\alpha \beta} = \sum_i \frac{1}{8\pi^2} \int \mathrm{d} \mu \int k^2 \mathrm{d}k ~ \partial_\alpha \mathrm{ln} (P^\mathrm{th}_\mathrm{obs})\big|_\mathrm{fid} \,\, \partial_\beta \mathrm{ln} (P^\mathrm{th}_\mathrm{obs})\big|_\mathrm{fid}  V_i^\mathrm{fid}\;,
    \label{eq:fisherGC}
\end{equation}
where the derivatives are evaluated precisely at the fiducial model, and the arguments of the observed power spectrum are again $P^\mathrm{th}_\mathrm{obs}[k_\mathrm{obs}(k,\mu,z_i),\mu_\mathrm{obs}(\mu,z_i),z_i]$.\footnote{Note that \istfisher\ uses an equivalent expression in which the spectrum that appears in Eq.~\eqref{eq:fisherGC} does not include shot noise, while here we do assume that $P^\mathrm{th}_\mathrm{obs}$ and $P^\mathrm{th}_\mathrm{obs}$ include shot noise. In the conventions of \istfisher, to compensate for this, $V_i^\mathrm{fid}$ needs to be replaced by
\begin{equation}
V_i^\mathrm{eff}(k,\mu,z_i)=V_i^\mathrm{fid} \left(\frac{n_i^\mathrm{fid} \,\,P_\mathrm{obs}^\mathrm{fid}\left(k,\mu,z_i\right)}{n_i^\mathrm{fid} \,\,P_\mathrm{obs}^\mathrm{fid}\left(k, \mu,z_i\right)+1}\right)^2\;,
\end{equation}
where $n_i^\mathrm{fid}$ is the galaxy number density in each redshift bin and the product $n_i^\mathrm{fid} \,\, P_\mathrm{obs}^\mathrm{fid}$ is dimensionless.}

\subsection{Likelihood derivatives}

A specific code can compute the Fisher matrix with various methods, always involving the calculation of derivatives (of cosmological observables or directly of the likelihood) with respect to cosmological and nuisance parameters. Usually these derivatives are computed using numerical methods, such as an $n$-point derivative stencil or more advanced methods (such as fitting a low-order polynomial curve through a set of $n$ points), but in several cases (such as bias parameters or shot noise) analytical formulas can be used instead. To compute the derivatives, a given code receives the matter power spectrum $P_m(k,z)$ computed by an EBS (either linear as in \cref{sec:spectroscopic}, or corrected for nonlinear clustering using a specific recipe as in \cref{sec:photometric}) evaluated at a few points in model parameter space. The spectrum is used to compute the cosmological observable (that is, the photometric harmonic spectra $C_{ij}^\mathrm{th}(\ell)$ or the observable galaxy power spectrum $P^\mathrm{th}_\mathrm{obs}$) or directly the likelihood ${\cal L}$, and the derivative can be inferred.

\subsubsection{The IST:F method}
\istfisher\ uses various codes to get the Fisher matrix directly from \cref{eq:fisherWL,eq:fisherGC}. The first-order derivatives $\partial_\alpha C^\mathrm{th}(\ell)|_\mathrm{fid}$ or $\partial_\alpha \ln (P^\mathrm{th}_{i,\mathrm{obs}})|_\mathrm{fid}$ are inferred from finite differences, with $C^\mathrm{th}(\ell)$ or $P^\mathrm{th}_{i,\mathrm{obs}}$ being computed at the fiducial point and in its vicinity by an EBS. The main purpose of \istfisher\ was to compare such codes. Each of them was reading the same input matter power spectra in some files produced by the EBS \camb. These codes differed through the detailed numerical implementation of the Fisher formula and through the algorithm used to compute first-order derivatives. The validation of these different codes was based on a comparison between the predicted {\it marginalised error} on each model parameter (that is, the 68\% confidence limit on a each parameter when all other parameters are unknown). A given code was validated when all marginalised errors lied within 10\% of the median computed across all the codes -- which means that a maximum deviation of 20\% on each marginalised error was allowed between pairs of codes.

Since our goal in this work is to validate a few additional pipelines, we need to compare our new results with at least one of the validated \istfisher\ codes (comparing with all of them would have been intractable). We choose \cosmicfish{} as our reference \istfisher\ code. \cosmicfish{} can be set to compute the first-order derivative of cosmological spectra either with a simple two-sided finite difference method, or with a more advanced SteM algorithm using ten different step sizes, which essentially fits a tangent to the curve as a function of the cosmological parameters, see Appendix B of \cite{Camera:2016owj} for further details.

\subsubsection{\MPFisher{} method}
This method relies on the \montepython{} {\it Euclid mock likelihoods}, which are simply the numerical implementation of Eq.~\eqref{eq:LWL} and Eq.~\eqref{eq:LGC} in the Python-based inference code \montepython{} \citep[\MP;][]{Audren:2012wb,Brinckmann:2018cvx}. These \Euclid mock likelihoods could easily be transposed to other samplers such as Cobaya. Simpler forms of these likelihoods have been presented in previous papers \citep{Audren:2012vy,Sprenger:2018tdb} under the names {\tt euclid\_pk} and {\tt euclid\_lensing}. Together with this paper, we release\footnote{The release will take place upon acceptance of this paper as an update of the main \montepython{} branch in the public repository.} an improved version of these likelihoods, named {\tt euclid\_spectroscopic} and {\tt euclid\_photometric}. These are designed to match the same recipes as \istfisher.

To derive the corresponding Fisher matrices, the \MP{} code does not use Eq.~\eqref{eq:fisherWL} or Eq.~\eqref{eq:fisherGC}, but computes the derivative directly from the likelihood (or $\chi^2$), using Eq.~\eqref{eq:def_fisher}. All second derivatives are evaluated with a two-sided finite difference method. \MP{} uses the EBS \class{} to compute the full $\chi^2$ at the fiducial point $\{\bar{\theta}_\alpha\}$ and in its vicinity. Second derivatives with respect to a given parameter $\theta_\alpha$ are inferred from the likelihood at the fiducial point and at two adjacent points with $\theta_\alpha = \{\bar{\theta}_\alpha \pm \Delta \theta_\alpha\}$. Cross derivatives with respect to two parameters $\{ \theta_\alpha, \theta_\beta \}$ are inferred from the likelihood at the fiducial point and at four adjacent points with $\theta_\alpha = \{\bar{\theta}_\alpha \pm \Delta \theta_\alpha\}$ and $\theta_\beta = \{\bar{\theta}_\beta \pm \Delta \theta_\beta\}$.

\istfisher\ and the \MPFisher{} method would give the same result for any given stepsize $\Delta \theta_\alpha$ if there was no numerical noise inherent to the codes and if the likelihood was a perfect multivariate Gaussian function of model parameters. However, neither of these criteria is fulfilled, leading to a possible dependence of the result on the chosen stepsizes. We will illustrate this issue in the result section, and conclude that it can be overcome for the purpose of \Euclid forecasts, at least in the case of the $\Lambda$CDM+$\{w_0, w_a\}$ model.

The \MP{} code was initially designed for Bayesian parameter inference with various MCMC algorithms, but it features an automatic calculation of the Fisher matrix through finite differences since the release of version 3 in 2018 \citep{Brinckmann:2018cvx}. The public code offers an advanced scheme for the choice of parameter stepsizes, such as a search by iteration of each stepsize such that the variation of each parameter from $\bar{\theta}_\alpha$ to $\bar{\theta}_\alpha + \Delta \theta_\alpha$ leads to an increase of $-\ln {\cal L}$ by a target value $\Delta \ln \mathcal{L}$ chosen by the user, up to some tolerance. In this paper, for the sake of simplicity and efficiency, we disabled these iterations. In each run, we directly fix the stepsizes in our input files.

Note that we have placed all our input files, scripts and results in a public GitHub repository \url{https://github.com/sabarish-vm/Euclid_w0wa.git} to make our results entirely reproducible.

\section{Likelihood recipes}  \label{sec:lkl}

\subsection{Photometric likelihood}\label{sec:photometric}

\subsubsection{General expression}

The detailed calculation leading to the expression of the photometric spectra and likelihood adopted here has been described in many previous papers, including \istfisher. Here, we just want to precise the set of relations and assumptions used by our \cosmicfish{} and \montepython{} pipelines, which match the previous settings adopted in \istfisher.

For the purpose of evaluating either the likelihood, see Eq.~\eqref{eq:LWL}, or directly of the Fisher matrix, see Eq.~\eqref{eq:fisherWL}, both \cosmicfish{} and \montepython{} require the calculation of the photometric spectra $C_l^{XY}$ in multipole space, where $X,Y=L$ for the cosmic shear data or $G$ for the photometric galaxy data. In the sub-Hubble (Newtonian) limit and using the Limber approximation, one can write these spectra under the generic form
\begin{equation}
    C_{ij}^{XY}(\ell) = \! \int^{z_\mathrm{max}}_{z_\mathrm{min}} \de z  \frac{W_{i}^{X}(z) W_{j}^{Y}(z)}{c^{-1} H(z)r^{2}(z)} P_\mathrm{m} \! \left[ \frac{\ell+1/2}{r(z)},z \right] +N_{ij}^{XY}(\ell) \;,
    \label{eq:Cl_WL}
\end{equation}
where $W_{i}^{X}(z)$ is the window function of the $X$ observable in the $i$th redshift bin, $H(z)$ is the Hubble rate at redshift $z$, $r(z)$ is the comoving distance to $z$, $P_\mathrm{m}(k,z)$ is the nonlinear matter power spectrum in Fourier space and $N_{ij}^{XY}(\ell)$ is the noise spectrum. In the codes, we adopt units of $\rm{Mpc}^{-1}$ for $H/c$, $\rm{Mpc}$ for $r$ and $\rm{Mpc}^{3}$ for $P_\mathrm{m}$, making the $C_{ij}^{XY}$ dimensionless. The integrals run over the redshift range $[z_\mathrm{min},z_\mathrm{max}]$ covered by the survey, and we distinguish between $N_\mathrm{bin}$ redshift bins.

In absence of intrinsic alignment effects (and assuming that the standard Poisson equation applies to sub-Hubble scales), the lensing window function in the $i$th bin would simply be given by 
\begin{equation}
    W_i^{\gamma}(z) = \frac{3}{2} \, c^{-2} H_0^2 \, \Omega_{\mathrm{m},0} \,  (1+z) \, r(z) \int_z^{z_\mathrm{max}}\de z^\prime \,
    n_i(z^\prime)
    \left[1- \frac{r(z)}{r(z^\prime)}\right]\;,
\end{equation}
where $n_i(z)$ is the observed galaxy density in each redshift bin. 

\begin{figure}[htbp]
    \centering
    \includegraphics[width=0.5\linewidth]{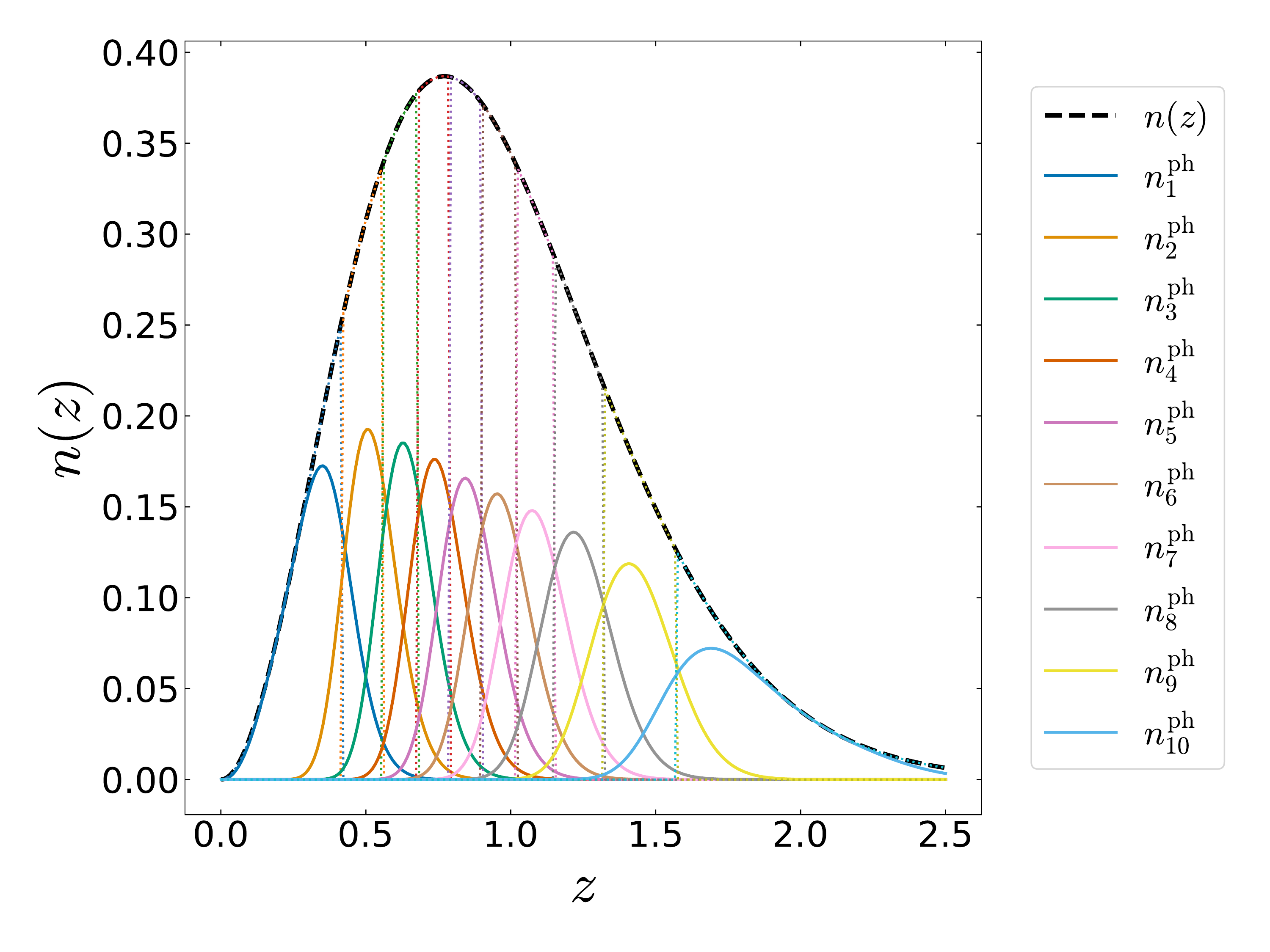}
    \caption[]{Underlying theoretical distribution of the number density of galaxies $n(z)$ (black dashed line), normalised to one, together with the $n_i(z)$ (solid colored lines) at each of the 10
    tomographic redshift bins (dotted vertical colored lines), normalised to $1/10$. The $n_i(z)$ are wide and overlap each other due to the photometric redshift errors. 
    }\label{fig:ni}
\end{figure}

In accordance with \istfisher, we model this density as some true underlying distribution $n(z)$, which is calibrated using a variety of different techniques \citep{vandenBusch:2020lur, Euclid:2022oea},
convolved with the photometric error $p_\mathrm{ph}(z_\mathrm{p}|z)$ of the experiment and normalized to unity,
\begin{equation}
    n_i(z) = \frac{\int_{z_i^-}^{z_i^+} \de z_\mathrm{p} \,\, n(z) \,\, p_\mathrm{ph}(z_\mathrm{p}|z)}{\int_{z_\mathrm{min}}^{z_\mathrm{max}}\de z \int_{z_i^-}^{z_i^+}\de z_\mathrm{p} \,\, n(z) \,\, p_\mathrm{ph}(z_\mathrm{p}|z)}\;, \label{eq:n_i}
\end{equation}
where $z_i^-$ and $z_i^+$ are the edges of the $i$th redshift bin. As in \istfisher, we assume that the galaxy distribution is given by\footnote{Notice that a more realistic photometric distribution based on the Euclid Consortium Flagship was recently published in \cite{Euclid:2021osj}. However, for the sake of validation, we stick to the analytic formula used in \istfisher.}
\begin{equation}
n(z) = \left(\frac{z}{z_0}\right)^2 \exp\left[-\left(\frac{z}{z_0}\right)^{1.5}\right]\;,
\end{equation}
with $z_0=z_\mathrm{mean}/\sqrt{2}$ and the mean redshift of the distribution, $z_\mathrm{mean}$, is given in \cref{tab:fixed_params_WL}. The photometric redshift error is modeled as \istfisher.
\begin{equation}
    p_\mathrm{ph}(z_\mathrm{p}|z) = \frac{1-f_{\rm out}}{\sqrt{2\pi}\sigma_{\rm b}(1+z)} \exp\left\{-\frac{1}{2}\left[\frac{z-c_{\rm b}z_\mathrm{p}-z_{\rm b}}{\sigma_{\rm b}(1+z)}\right]^2\right\} + \frac{f_{\rm out}}{\sqrt{2\pi}\sigma_0(1+z)} \exp\left\{-\frac{1}{2}\left[\frac{z-c_0z_\mathrm{p}-z_0}{\sigma_0(1+z)}\right]^2\right\}\;,
\end{equation}
where the second term accounts for a fraction $f_\mathrm{out}$ of catastrophic outliers. The parameters of this model are detailed in \cref{tab:fixed_params_WL}, while the observed galaxy density in each bin is shown in \cref{fig:ni}.

The contribution of intrinsic alignment effects to the total observed shear angular power spectrum can be modelled through a modification of the lensing window function. As a first step, one can assume the intrinsic alignment tracer to be biased with respect to the matter overdensity $\delta_\mathrm{m}$ as 
$\delta_\mathrm{IA}(k,z) = -\mathcal{A}_\mathrm{IA} \,\mathcal{C}_\mathrm{IA}\,\Omega_{\mathrm{m},0}\, \mathcal{F}_\mathrm{IA}(z)\,\delta_\mathrm{m}(k,z)/D(z)$, where $\mathcal{A}_\mathrm{IA}$ and $\mathcal{C}_\mathrm{IA}$ are two parameters, $D(z)$ is the linear growth factor of matter density fluctuations, and $\mathcal{F}_\mathrm{IA}(z)$ is a function of the redshift-dependent mean luminosity of galaxies $\langle L\rangle(z)$ in units of a characteristic luminosity $L_\star(z)$ \istfisher,
\begin{equation}
    \mathcal{F}_\mathrm{IA}(z) = (1+z)^{\eta_\mathrm{IA}} \left[\langle L\rangle(z) /L_\star (z)\right]^{\beta_\mathrm{IA}}\;.
\end{equation}
The parameters $\eta_\mathrm{IA}$ and $A_\mathrm{IA}$ are treated as nuisance parameters for the inference, while $\cal{C}_\mathrm{IA}$ and $\beta_\mathrm{IA}$ are fixed in our analysis. With such a model, one can account for intrinsic alignment by adding a new term to $W_i^{\gamma}(z)$. The total lensing window function then reads
\begin{equation}
    W_{i}^\mathrm{L}(z) = W_{i}^{\gamma}(z) - \mathcal{A}_\mathrm{IA} \mathcal{C}_\mathrm{IA}\Omega_{\mathrm{m},0} \frac{\mathcal{F}_\mathrm{IA}(z)}{D(z)} W_{i}^\mathrm{IA}(z) \;,
\end{equation}
with the definition
\begin{equation}
    W_i^\mathrm{IA}(z) = c^{-1} \, n_i(z)  H(z) \;.
\end{equation}
The values of the parameters entering these expressions are summarised in \cref{tab:fixed_params_WL}. The function $\langle L\rangle(z) /L_\star (z)$ is instead read in a file {\tt scaledmeanlum\_E2Sa.dat} provided on the public repository of the IST:F team.\footnote{\label{footn:site}\url{ https://github.com/euclidist-forecasting/fisher_for_public}}

The inclusion of photometric data on galaxy clustering involves the galaxy window function
\begin{equation}
    W_{i}^\mathrm{G}(z) = c^{-1} \, n_i(z) H(z)\, b_i(z) \;,
\end{equation}
where $n_i(z)$ is again the observed galaxy density in each redshift bin, still given by Eq.~\eqref{eq:n_i}, while $b_{i}(z)$ account for light-to-mass bias in the $i$th bin at redshift $z$. In principle, one could model bias as a unique continuous function of redshift, $b(z)$. However, in absence of a reliable bias model, the IST:F group treated the mean biases $b_i$ in each bin as nuisance parameters. Technically, this could be implemented in various ways:
\begin{itemize}
\item with a constant bias for each bin: for each $i$, $b_i(z)=b_i$ for $z_\mathrm{min} < z < z_\mathrm{max}$;
\item with a unique function across all bins, $b_{i}(z)=b(z)$ for $z_\mathrm{min} < z < z_\mathrm{max}$, where $b(z) $ is a step-like function taking the value $b_i$ within the range $z_i^- < z < z_i^+$.
\end{itemize} 
As a consequence of photometric redshift errors, the latter is not equivalent to the former because the functions $n_i(z)$ do not vanish outside of the range $[ z_i^- , z_i^+ ]$. Thus, in the second simplified model, the functions $n_i(z)$ feature discontinuities at each $z_i^+ = z_{i+1}^-$. Both options are implemented in our \cosmicfish{} and \montepython{} pipelines. However, since \istfisher\ adopted the second model, we will also stick to it.\footnote{This technical detail is important for calculating derivatives with respect to the $b_i$'s. Thus it is necessary to stick to the same convention in order to recover a similar Fisher matrix. In the codes, the second option (used here) is called {\tt binned}, while the first option is called {\tt binned\_constant}.}

Finally, we take into account the noise contribution to all auto-correlation spectra. Assuming a Poissonian distribution of galaxies, the noise spectra read 
\begin{equation}
    N_{ij}^\mathrm{LL}(\ell) = \frac{\sigma_\epsilon^2}{\Bar{n}_i}\delta_{ij}\;, \qquad N^\mathrm{GG}_{ij}(\ell) = \frac{1}{\Bar{n}_i} \delta_{ij}\;, \qquad
    N_{ij}^\mathrm{LG}(\ell)=N_{ij}^\mathrm{GL}(\ell)=0\;,
\end{equation}
where $\Bar{n}_i$ is the expected average number of galaxies per steradian in the given bin $i$, and $\sigma_\epsilon^2$ is is the variance of the observed ellipticities. The number $\Bar{n}_i$ is computed as the expected total number of galaxies per steradian, $n_{gal}$, divided by the number of bins.\footnote{We are assuming equally populated bins, as in \istfisher.} Like in \istfisher, we take a $\sigma_\epsilon$ of 0.21 per degree of freedom. Given the two degrees of freedom in the WL maps, this implies  $\sigma_\epsilon = 0.21 \sqrt{2} \simeq 0.30$. 

\begin{table}[]
    \centering
    \caption{Constants used in the Fisher and likelihood formulas. Only the maximum multipole values vary between the pessimistic and optimistic settings.}
    \begin{tabular}{c c |c}
        \hline
        Type & Name & Value (Pessimistic/Optimistic) \\
        \hline
        Redshift bins & $N_\mathrm{bin}$ & 10 \\
        Redshift bins & $z_\mathrm{min}$ & 0.001 \\
        Redshift bins & $z_\mathrm{max}$ & 2.5 \\
        Redshift bins & $z_1^+$, .., $z_5^+$ & 0.418, 0.560, 0.678, 0.789, 0.900 \\
        Redshift bins & $z_6^+$, .., $z_9^+$ & 1.019, 1.155, 1.324, 1.576 \\
        Redshift bins & $z_\mathrm{mean}$ & 0.9 \\
        \hline
        Photometric error& $c_0$ & 1.0 \\
        Photometric error& $c_{\rm b}$ & 1.0 \\
        Photometric error&$z_0$ & 0.1 \\
        Photometric error&$z_{\rm b}$ & 0.0 \\
        Photometric error&$\sigma_0$ & 0.05 \\
        Photometric error&$\sigma_{\rm b}$ & 0.05 \\
        Photometric error&$f_{\rm out}$ & 0.1 \\
        \hline
        Intrinsic alignment &$\mathcal{C}_\mathrm{IA}$ & 0.0134 \\ 
        Intrinsic alignment &$\beta_\mathrm{IA}$ & 2.17 \\ 
        \hline
        Noise &$\sigma_\epsilon$ & 0.3 \\
        Noise &$n_\mathrm{gal}$ & $30\ \mathrm{arcmin}^{-2}$ \\
        \hline
        Multipoles & $\ell_\mathrm{min}$ & 10 \\
        Multipoles & $\ell^\mathrm{WL}_\mathrm{max}$ & 1500 / 5000 \\
        Multipoles & $\ell^\mathrm{GCph}_\mathrm{max}$ & 750 / 1500 \\
        \hline
        Sky coverage & $f_\mathrm{sky}$ & 0.3636 \\
    \end{tabular}
    \label{tab:fixed_params_WL}
\end{table}

The lensing spectra, galaxy clustering spectra and cross-correlation spectra can be gathered in the  full cross-correlation matrix  
\begin{equation}
    \tens{C}^\mathrm{ph}(\ell) = \left[ \begin{array}{rr}
    C^\mathrm{LL}_{ij}(\ell) & C^\mathrm{GL}_{ij}(\ell) \\ 
    C^\mathrm{LG}_{ij}(\ell) & C^\mathrm{GG}_{ij}(\ell) \\
    \end{array}\right]\;.
    \label{eq:Cl_XC}
\end{equation}
The calculation of the likelihood, see Eq.~\eqref{eq:LWL}, or directly of the Fisher matrix, see Eq.~\eqref{eq:fisherWL}, relies on two matrices for each $\ell$, with elements $C^\mathrm{th}_{ij}(\ell)$ and $C^\mathrm{fid}_{ij}(\ell)$. These matrices are simply the $(2N_\mathrm{bin})\times(2N_\mathrm{bin})$ matrix of Eq.~\eqref{eq:Cl_XC}, computed either at an arbitrary point in parameter space in the case of $C^\mathrm{th}_{ij}(\ell)$, or at fiducial parameter values in the case of $C^\mathrm{fid}_{ij}(\ell)$. Eq.~\eqref{eq:LWL} and \eqref{eq:fisherWL} feature a sum over $\ell$, from $\ell_\mathrm{min}$ to some $\ell_\mathrm{max}$. Following \istfisher\ conventions, we choose a smaller value of the maximum multipole for the galaxy survey, $\ell_\mathrm{max}^\mathrm{GCph}$, than for the lensing survey, $\ell_\mathrm{max}^\mathrm{WL}=\ell_\mathrm{max}$. This choice allows us to stick to scales where galaxy bias can be modelled as approximately linear. The boundary values are given in \cref{tab:fixed_params_WL}. This means that for  $\ell > \ell_\mathrm{max}^\mathrm{GCph}$, $C^\mathrm{ph}_{ij}(\ell)$ reduces to the $N_\mathrm{bin}\times N_\mathrm{bin}$ matrix $C^\mathrm{LL}_{ij}(\ell)$. 

\subsubsection{Detailed numerical implementation}

The nonlinear matter power spectrum $P_\mathrm{m}(k,z)$ is obtained from either \class{} or \camb. Nonlinear corrections are computed within these codes using \texttt{Halofit} -- including \cite{Takahashi:2019hth} and \cite{Bird:2011rb} corrections, see \cref{app:nonlinear} for details. The nonlinear power spectrum is approximated as zero outside of the range $[k_\mathrm{min}, k_\mathrm{max}]=[0.001,50]\,h\,\mathrm{Mpc}^{-1}$. This range is sufficient to include most of the tails of the convolution kernels in Eq.~\eqref{eq:Cl_WL} even for the multipoles $\ell_\mathrm{min}=10$ and $\ell_\mathrm{max}=5000$.
The growth factor $D(z)$ is also extracted from  \class{} or \camb. 
In the case of \montepython, \class{} is called on-the-fly at every point in cosmological parameter space.
On the other hand, \cosmicfish{} can call either \class{} or \camb{} on-the-fly, but it also can read a set of pre-computed files with cosmological quantities, produced by any arbitrary EBS.

In the case of \montepython, which can also be used to perform MCMCs, computing the matrix elements $C^\mathrm{ph}_{ij}(\ell)$ for every single $\ell$ would be numerically expensive. Using the fact that these elements are smooth functions of $\ell$, we only compute them on a discrete grid of values of size 100 (with logarithmic spacing between $\ell_\mathrm{min}$ and $\ell_\mathrm{max}$), and then use a spline interpolation to get them at every integer $\ell$ within the range $[\ell_\mathrm{min}, \ell_\mathrm{max}]$. 
The log-likelihood is computed by summing over all integer values of $\ell$.
In the case of \cosmicfish, the angular spectra are computed on the same grid and do not need further interpolation, since the sum is evaluated only over the 100 $\ell$-bins.

\subsection{Spectroscopic likelihood}\label{sec:spectroscopic}

\subsubsection{General expression}

The details of our recipe for the spectroscopic likelihood have been described in many previous papers, including \istfisher. Here we just summarize briefly the set of relations and assumptions used by our \cosmicfish{} and \montepython{} pipelines, which match the prescription adopted by \istfisher. 

For the purpose of evaluating either the likelihood, see Eq.~\eqref{eq:LGC}, or directly of the Fisher matrix, see Eq.~\eqref{eq:fisherGC}, both \cosmicfish{} and \montepython{} require the calculation of the observed redshift-space galaxy power spectrum $P_\mathrm{obs}(k,\mu,z)$ at wavenumber $k$, angle cosine $\mu$ and redshift $z$. The calculation of this quantity starts from the evaluation of the linear real-space matter power spectrum $P_{\mathrm{m},\mathrm{lin}}(k,z)$, which can be computed with \camb{} or \class. The calculation sticks to scales where one can assume a linear relation between the galaxy and matter power spectra. Then, the real-space galaxy linear power spectrum reads
\begin{equation}
    P_\mathrm{gal,lin}(k,z) = b^2(z)P_{\mathrm{m},\mathrm{lin}}(k,z) \;,
\end{equation}
where $b(z)$ is the effective galaxy bias. We assume a constant bias $b_i$ in each redshift bin $i$ (with central redshift value $z_i$). Bin edges and fiducial bias values are specified in \cref{tab:redshift_params_GC}. To obtain the redshift-space galaxy linear power spectrum, one should further multiply $P_\mathrm{m,lin}$ by the Kaiser correction factor
\begin{equation}
     \left[b(z) + {f(z)} \mu^2 \right]^2 \;,
\end{equation}
where $\mu = \vec{k} \cdot \hat{\vec{r}}/k$ is the cosine of the angle between the wave-vector $\vec{k}$ and the line-of-sight direction $\hat{\vec{r}}$ and $f(z)$ is the scale-independent growth rate.\footnote{We recall that when the growth factor $D$ and growth rate $f$ are expressed as a function of the scale factor $a$, they are related to each other through $f(a) = \de\ln D(a)/\de\ln a$.} 
In absence of massive neutrinos (or modified gravity effects), the redshift dependence of $\sigma_8(z)$ can be modeled as 
\begin{equation}
    \sigma_8(z) = D(z)\sigma_8(z=0) \;.
\end{equation}
The Fingers-of-God effect can be accounted with an additional prefactor $F_\mathrm{FoG}(z)$, that we model as a Lorentzian \citep[see e.g.][]{2dFGRS:2004cmo},
\begin{equation}
    F_\mathrm{FoG}(z) = \frac{1}{1+[f(z) \, k \, \mu \, \sigma_p^\mathrm{fid}(z)]^2}\;,
    \label{eq:FOG}
\end{equation}
where the dispersion $\sigma_p^\mathrm{fid}$ is given by an integral over the linear spectrum of the fiducial model,
\begin{equation}
    \left[{\sigma_p^\mathrm{fid}}(z)\right]^2 = \frac{1}{6\pi^2} \int \de k\,P_\mathrm{m,lin}^\mathrm{fid}(k,z) \;. 
    \label{eq:sigma}
\end{equation}
In practice, for the latter integral, we use finite boundaries specified in the next section. 

Furthermore, baryon acoustic oscillations in the power spectrum are smoothed out by nonlinear matter clustering. This effect can be approximately modelled by replacing $P_\mathrm{m,lin}$ with the de-wiggled power spectrum
\begin{align}
    P_\mathrm{dw}(k,\mu,z)& \equiv P_\mathrm{m,lin}(k,z) \, \mathrm{e}^{-g_\mu k^2} + P_\mathrm{nw}(k,z)\left(1-\mathrm{e}^{-g_\mu k^2}\right) \;, \\
    g_\mu(k,\mu,z) &= \left[ \sigma_v^\mathrm{fid}(z)\right]^2  \left\{1-\mu^2+\mu^2[1+f^\mathrm{fid}(z)]^2 \right\}\;,
\end{align}
where in this case $f^\mathrm{fid}(z)$ is fixed to the fiducial cosmology. The no-wiggle power spectrum $P_\mathrm{nw}(k,z)$ is obtained by smoothing $P_\mathrm{m,lin}(k,z)$ as described in the next section. The variance of the displacement field $[\sigma_v^\mathrm{fid}]^2$ for the fiducial model is set equal to the quantity $[\sigma_p^\mathrm{fid}]^2$ defined in Eq.~\eqref{eq:sigma}.

In \istfisher, two cases were considered. In the first one, $\sigma_v$ and $\sigma_p$ were treated as free parameters in each redshift bin, with fiducial values calculated at the $z_\mathrm{mean}$ of the survey and rescaled by the growth factor $D(z)$. These parameters were then marginalized over. In the second one, in each redshift bin, these parameters were fixed to their fiducial value (computed with the power spectrum of the fiducial cosmology). The latter case assumes a better knowledge of nonlinear corrections and leads to more optimistic constraints. In this paper, we always stick to the second option and keep these variables fixed in both our pessimistic and optimistic approaches.

Additionally, spectroscopic redshift errors further suppress this power spectrum by an overall factor
\begin{equation}
    F_z(k,\mu,z) = \exp[-k^2 \mu^2 \sigma_r^2(z)] \;,
\end{equation}
where $\sigma_r(z)$ is the comoving distance error, which depends on the linear scaling of the redshift error $\sigma_{0,z}$:
\begin{equation}
    \sigma_r(z) = \frac{c}{H(z)}(1+z)\sigma_{0,z}\;.
\end{equation}
The value of $\sigma_{0,z}$ is given in \cref{tab:fixed_params_GC}.

Next, the \AP{} effect introduces a change of the observed wavenumbers and angles with respect to the reference cosmology used for the analysis (identified here to the fiducial model). In particular, distances parallel to the line of sight and orthogonal to it are modified respectively by
\begin{equation}
    k_{\perp,\mathrm{obs}} = k_\perp \,\, q_\perp \,\, \frac{h}{h^\mathrm{fid}}\;, \qquad \qquad
    k_{\parallel,\mathrm{obs}} = k_\parallel \,\, q_\parallel \,\, \frac{h}{h^\mathrm{fid}}\;,
\end{equation}
where
\begin{equation}
    q_\perp =  \frac{D^\mathrm{fid}_A(z)}{D_A(z)}\;, \qquad \qquad q_\parallel =  \frac{H(z)}{H^\mathrm{fid}(z)}\;. 
\end{equation}
Then, using $k^2 = k_\perp^2 + k_\parallel^2$ and $\mu = k_\parallel/k$ for both the fiducial wavevector (with no subscript) and the observed one (with subscript `obs'), we can express the observed wavevector components as a function of the fiducial ones,
\begin{align}
k_\mathrm{obs} &= k \,\, q_\perp \,\, \frac{h}{h^\mathrm{fid}} \left[ 1+\mu^2 \left( \frac{q_\parallel^2}{q_\perp^2} - 1\right) \right]^{1/2},\\
\mu_\mathrm{obs} &= \mu \,\, \frac{q_\parallel}{q_\perp} \, \left[ 1+\mu^2 \left( \frac{q_\parallel^2}{q_\perp^2} - 1\right) \right]^{-1/2}.
\end{align}
This shows explicitly that $\mu_\mathrm{obs}$ is a function of $(\mu,z)$ while $k_\mathrm{obs}$ is a function of $(k,\mu,z)$.
Additionally, the \AP{} effect implies that the overall power spectrum is multiplied by a factor of $q_\perp^2 q_\parallel$.

Finally, we also include a shot noise parameter $P_\mathrm{s}(z)$:
\begin{equation}
    P_\mathrm{s}(z_i) = \frac{1}{n_i^\mathrm{fid}} + p_\mathrm{s}(z_i)\;,
\end{equation}
where the residual shot noise $p_\mathrm{s}(z_i)$ is treated as a nuisance parameter with a fiducial value of 0 in each redshift bin, and $n_i^\mathrm{fid}$ is the galaxy number density in each bin given in \cref{tab:redshift_params_GC}.

In summary, the observed galaxy power spectrum in each bin $i$ reads (\istfisher) 
\begin{equation}
    P_\mathrm{obs}(k_\mathrm{obs}, \mu_\mathrm{obs}, z_i) = q_\perp^2 \,\,q_\parallel \,\frac{[b_i \, \sigma_8(z_i)+f(z_i)\, \sigma_8(z_i) \, \mu_\mathrm{obs}^2]^2}{1+[f^\mathrm{fid}(z_i)\,k_\mathrm{obs}\,\mu_\mathrm{obs}\,\sigma_p^\mathrm{fid}(z_i)]^2}
    \frac{P_\mathrm{dw}(k_\mathrm{obs},\mu_\mathrm{obs},z_i)}{\sigma_8^2(z_i)} \,F_z(k_\mathrm{obs},\mu_\mathrm{obs},z_i)+ P_\mathrm{s}(z_i)\;,
\end{equation}
where we omitted the arguments of the functions $k_\mathrm{obs}(k,\mu,z)$ and $\mu_\mathrm{obs}(\mu,z)$.
In each redshift bin, the product $b_i \sigma_8(z_i)$ is actually treated as an independent nuisance parameter, which means that it is kept fixed when varying the cosmological parameter $\sigma_8$. Together with the values of $p_\mathrm{s}(z_i)$, the spectroscopic case features a total of 8 nuisance parameters. The fiducial value of each $b_i \sigma_8(z_i)$ is inferred from the $b_i^\mathrm{fid}$ value reported in \cref{tab:redshift_params_GC} multiplied by $\sigma_8^\mathrm{fid}(z_i)$ (see \cref{tab:pcosmo_input_GC} for exact values).

To evaluate the likelihood (resp. the Fisher matrix), one should insert this expression into Eq.~\eqref{eq:LGC} (resp. into Eq.~\ref{eq:fisherGC}), which also depends on the volume $V_i^\mathrm{fid}$ and $n_i^\mathrm{fid}$ in each bin, whose values are listed in \cref{tab:redshift_params_GC}. Finally, one should integrate over $k$ from $k_\mathrm{min}$ to $k_\mathrm{max}$ (whose values are given in \cref{tab:fixed_params_GC} for the pessimistic and an optimistic case) and $\mu$ from $-1$ to $+1$.

\begin{table}[]
    \centering
    \caption{Redshift-dependent survey specifications evaluated at the central redshift $z_i$ of each redshift bin. Quantities with units $h\,\mathrm{Mpc}^{-1}$ are converted into $\mathrm{Mpc}^{-1}$ always using the unique fiducial value $h^\mathrm{fid}$.}
    \begin{tabular}{c c c|c|c|c}
        \multicolumn{3}{c|}{redshift bin} & survey volume & galaxy number density & galaxy bias \\
        $z_\mathrm{min}$ & $z_i$ & $z_\mathrm{max}$ & $V_i^\mathrm{fid}\,[\mathrm{Gpc}^3/(h^\mathrm{fid})^3]$ & $n_i^\mathrm{fid}\,[10^{-4}\,(h^\mathrm{fid})^3/\mathrm{Mpc}^3]$ & $b_i^\mathrm{fid}$ \\
        \hline
        0.90 & 1.00 & 1.10 & 7.94 & 6.86 & 1.46 \\
        1.10 & 1.20 & 1.30 & 9.15 & 5.58 & 1.61 \\
        1.30 & 1.40 & 1.50 & 10.05 & 4.21 & 1.75 \\
        1.50 & 1.65 & 1.80 & 16.22 & 2.61 & 1.90 \\
    \end{tabular}
    \label{tab:redshift_params_GC}
\end{table}

\begin{table}[]
    \centering
    \caption{Specifications used in the spectroscopic likelihood and Fisher formulas.}
    \begin{tabular}{c c |c}
        \hline
        Type & Name & Value (Pessimistic/Optimistic) \\
        \hline
        Spectroscopic error& $\sigma_{0,z}$ & 0.001 \\
        \hline
        Wavenumber & $k_\mathrm{min}$ & 0.001 $h^\mathrm{fid}/\mathrm{Mpc}$ \\
        Wavenumber & $k_\mathrm{max}$ & (0.25/0.30) $h^\mathrm{fid}/\mathrm{Mpc}$ \\
    \end{tabular}
    \label{tab:fixed_params_GC}
\end{table}

\subsubsection{Detailed numerical implementation}

For the sake of reproducibility, we provide here some details on the discretisation schemes, interpolation routines and integration algorithms used in our \cosmicfish{} and \montepython{} pipelines. These settings ensure some well-converged calculations, such that changing them slightly would have a negligible impact on our results.

In our \cosmicfish{} or \montepython{} algorithms, the linear matter power spectrum $P_\mathrm{m,lin}(k,z)$ is returned by \camb{} or \class{} on a logarithmically-spaced grid of 1055 discrete values of $k$ ranging from $k_\mathrm{min}=3.528 \times 10^{-5}$Mpc$^{-1}$ to $k_\mathrm{max}=50.79\,$Mpc$^{-1}$, corresponding to a logarithmic stepsize of $\de \ln k = 0.01354$.

The value of $\sigma_8(z=0)$ is read from \camb{} or \class. The growth factor $D(z)$ and growth rate $f(z)$ are also extracted from \camb{} or \class. 
Then $\sigma_8(z_i)$ is approximated as $\sigma_8(0) \, D(z_i)/D(0)$.

The no-wiggle power spectrum $P_\mathrm{nw}(k,z)$ is inferred from $P_\mathrm{m,lin}(k,z)$ using a Savitzky-Golay filter of order 3 on the same grid of 1055 points evenly spaced in $(\ln k)$ \citep[see e.g.][]{Boyle:2017lzt}. The filter's window length is set to 101 points, which corresponds to a smoothing over $\Delta \ln k=100 \, \ln(1+\de \ln k)=1.359$.

The algorithm then builds interpolators for $P_\mathrm{m,lin}(k,z)$ and $P_\mathrm{nw}(k,z)$ (using the \texttt{scipy} \texttt{CubicSpline} algorithm in \montepython{} and the order-3 \texttt{RectBivariate} spline algorithm in \cosmicfish). The final likelihood requires the evaluation of several quantities -- including these two spectra -- on a three-dimensional grid $(k_i,\mu_j,z_k)$, or, in the case of theoretical power spectra, $(k_{\mathrm{obs},i},\mu_{\mathrm{obs},j},z_k)= [k_\mathrm{obs}(k_i, \mu_j, z_k), \mu_\mathrm{obs}(\mu_j,z_k), z_k]$. In this grid,
\begin{itemize}
\item for \montepython{} (\cosmicfish) $k_i$ takes 500 (2048) discrete values of $k$ between some $k_\mathrm{min}$ and $k_\mathrm{max}$ (whose values are reported in \cref{tab:fixed_params_GC}) with even logarithmic spacing,
\item for \montepython{} (\cosmicfish) $\mu_j$ takes 9 (128) discrete values evenly spaced between $-1$ and $+1$,
\item $z_k$ runs over the 4 central redshift values of the 4 redshift bins reported in \cref{tab:fixed_params_GC}.
\end{itemize}

We have tested that having only 9 values for the grid in $\mu$ is enough for our purposes in both \montepython{} and \cosmicfish{} when the integral is performed using the Simpson algorithm of the Python package \texttt{scipy} \citep{2020SciPy-NMeth}.
However, for \cosmicfish{} we kept the grid of 128 values in $\mu$, since in general it is a much more time-efficient code.
In both codes, the final integral over $k$ is also performed using the Simpson algorithm.

\section{Validation of the forecast pipelines\label{sec:validation}}

\subsection{Methodology\label{sec:val_method}}

In this work, for each survey and each settings, we will compare five Fisher matrices corresponding to the following five cases:
\begin{enumerate}
\item \CFextCAMB: With this method, \cosmicfish{} (\CF) reads the information (spectra, distances...) in files produced in advance (externally) by \camb. This is one of the methods used in \istfisher. Thus, we include this case in order to cross-check that nothing significant has changed within the \cosmicfish{} code such that, a couple of years later, it still agrees with the results of \istfisher.
\item \CFintCAMB{} {\it (New)}: \cosmicfish{} calls \camb{} internally to extract all relevant information (spectra, distances...) on the fly: this is a more efficient approach that we also want to validate here. The comparison with the first case will prove that we integrated \camb{} within \cosmicfish{} in a correct way, with a proper use of the \camb{} Python wrapper.
\item \CFextCLASS{} {\it (New)}: \cosmicfish{} reads the relevant information in files similar to those of the first method, but produced in advance by \class. The comparison of this method with the first one will prove that the theoretical predictions from \camb{} and \class{} agree within the sensitivity of \Euclid.
\item \CFintCLASS{} {\it (New)}: \cosmicfish{} calls \class{} internally to extract all relevant information on the fly: this is again a more efficient approach than the external one. The comparison with the third case will prove that we integrated \class{} within \cosmicfish{} in a correct way, with a proper use of the \class{} Python wrapper.
\item \MPFisher{} {\it (New)}: \montepython{} (\MP) runs in Fisher mode and extracts all cosmological information on the fly from \class. The primary goal of this paper is to validate this pipeline against that of \istfisher.
\end{enumerate}

We will usually treat our results for the \CFextCAMB{} case as the ``reference result'' corresponding to the recipes of \istfisher. In order to show that this choice is justified, we will also directly compare the above five Fisher matrices with a second reference Fisher matrix from \istfisher. In the photometric case, our second reference will be the Fisher matrix posted on the online repository {\tt fisher\_for\_public}.\textsuperscript{\ref{footn:site}}\,
In the spectroscopic case, the second reference will be provided by \texttt{SoapFish} (\texttt{SF}), which is another of the codes validated by the IST:F group. Like \CFextCAMB, the \texttt{SF} pipeline relies on external files produced by \camb.

Finally, we can also estimate the survey sensitivity to cosmological parameter using the \MPMCMC{} method, that is, fitting some fiducial \Euclid data with our \Euclid mock likelihoods (which are the very same likelihoods as in the \MPFisher{} method), while exploring the parameter space with MCMCs. The focus of this paper is not on the comparison between Fisher and MCMC results. Still, knowing the MCMC results is useful in order to cross-check our Fisher results, evaluate the level of Gaussianity of our likelihood with respect to model parameters, or get insight on parameter degeneracies.

In the following section, we will compare the results given by all these methods with a seven-parameter cosmology ($\Lambda$CDM+$\{w_0,w_a\}$). Fiducial values of cosmological and nuisance parameters are summarised in \cref{app:fiducial}. We will prove that all our Fisher pipelines agree with each other within at most 10\%, which is the validation threshold set by the IST:F team.

Since for each case (photometric/spectroscopic survey with pessimistic/optimistic settings), we have five ways to compute the Fisher matrix, we can perform ten comparisons between pairs of matrices. The comparison can be made at the level of:
\begin{itemize}
\item unmarginalised errors, that is, the error on one parameter when all other parameters are kept fixed. This only involves the diagonal coefficients of the Fisher matrices.
\item marginalised errors, that is, the error on one parameter when all the others are unknown (and only constrained by the experiment). This involves the diagonal coefficients of the {\it inverse} Fisher matrices. As such, it depends on all coefficients in the Fisher matrices, or in other word, on the Fisher estimate of correlations between all pairs of parameters.
\end{itemize}
The most important quantities, which we use for validation, are of course the marginalised errors. Still, the knowledge of the unmarginalised ones is often useful. We will report both in what follows.

\subsection{Photometric likelihood}

\subsubsection{Pessimistic setting}

Table~\ref{fig:comparison_table_photo_pess} contains the most relevant information for the photometric survey with pessimistic settings, that is, the biggest discrepancy between marginalised errors, computed across all cosmological and nuisance parameters, in each of the 10 comparisons that can be made. The result is presented as a five-by-five symmetric matrix (with no information along the diagonal, since each method agrees with itself). In a nutshell, since the worst difference is at the level of 2\%, and thus well below the 10\% threshold set by the IST:F group, all five methods are validated -- that is, are in sufficiently good agreement with \istfisher\ results. We provide a more detailed discussion below.

\begin{table}
    \centering
     \caption{For the photometric survey with pessimistic settings, and for each pair of Fisher matrices obtained with different methods, largest percentage difference between the predicted marginalised error on each parameter (across all cosmological and nuisance parameters).}
    \includegraphics[width=0.65\textwidth]{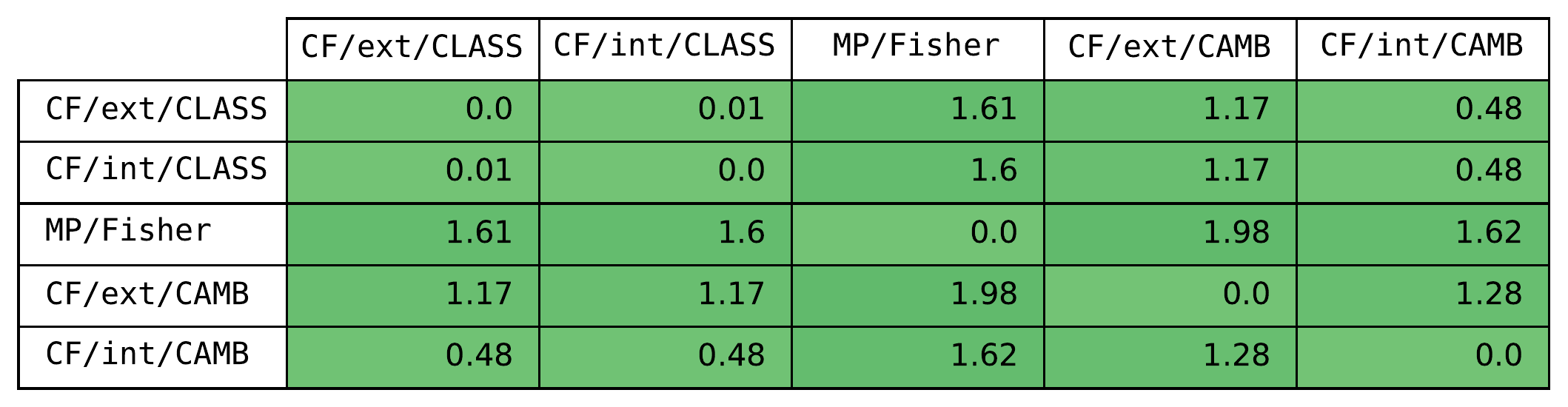}
    \label{fig:comparison_table_photo_pess}
\end{table}    

\begin{figure}[h]
    \centering
    \includegraphics[width=0.90\textwidth]{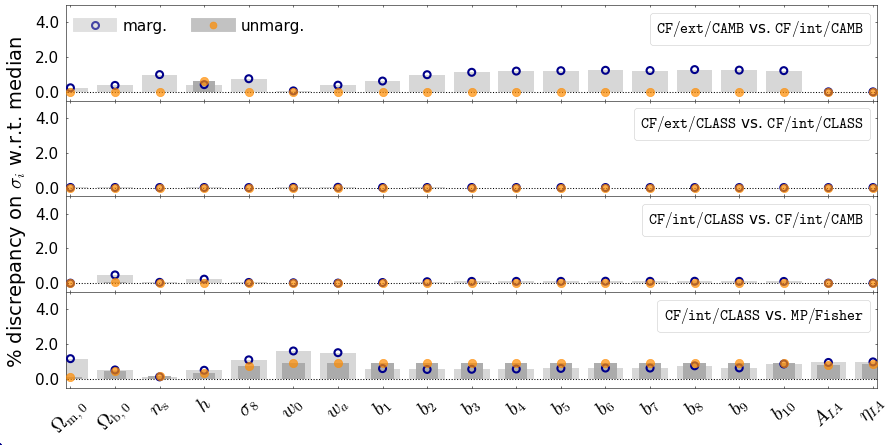}
    \caption{For the photometric survey with pessimistic settings, and for selected pair of Fisher matrices obtained with different methods, comparison of each Fisher marginalised (blue dot/light grey) and unmarginalised (orange dot/dark grey) error on each cosmological and nuisance parameters, for:
    {\it (First)} \CFextCAMB{} versus \CFintCAMB;
    {\it (Second)} \CFextCLASS{} versus \CFintCLASS;
    {\it (Third)} \CFintCLASS{} versus \CFintCAMB;
    {\it (Fourth)} \CFintCLASS{} versus \MPFisher. Each forecasted error $\sigma_i$ for each case and parameter is compared to the median of the two cases (see footnote \ref{foot:median}). The discrepancy is expressed in percent.
    \label{fig:comparison_errors_photo_pess}
    }
\end{figure}

\begin{figure}[h]
    \centering
    \includegraphics[width=0.75\textwidth]{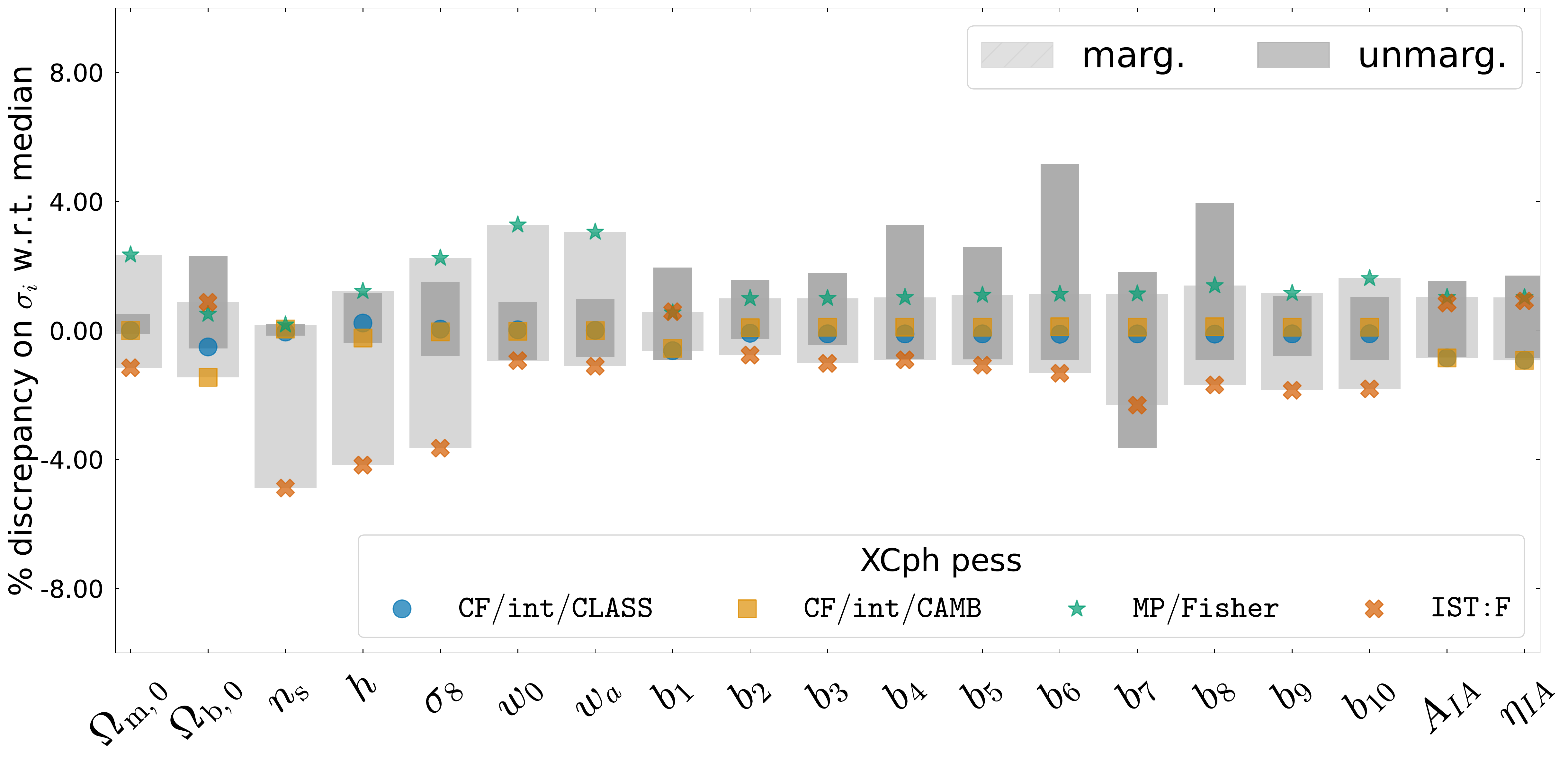}
    \caption{For the photometric survey with pessimistic settings, comparison of each Fisher marginalised (light grey) and unmarginalised (dark grey) errors on the cosmological and nuisance parameters, for:
     \CFintCLASS{} (blue circles) versus \CFintCAMB{} (orange squares) versus \MPFisher{} (green stars) versus the public IST:F results (red crosses).
     Each forecasted error $\sigma_i$ for each case and parameter is compared to the median of the four cases, and the discrepancy is expressed in percent.
    \label{fig:4-comparison_errors_photo_pess}
    }
\end{figure}

To begin, we observe that the two methods that use \cosmicfish{} combined with \camb{} are highly consistent with each other. The largest difference across all errors is only of 1.2\%. Individual marginalised and unmarginalised errors for these two cases are compared in Fig.~\ref{fig:comparison_errors_photo_pess} (first panel). In such plots, sticking to the plotting conventions of previous papers like \istfisher, we show the percentage discrepancy of each marginalised or unmarginalised error $\sigma_i$ with respect to the median.\footnote{In Fig.~\ref{fig:comparison_errors_photo_pess}, each panel presents a comparison between two cases only. Thus the median is just the average of the two errors, and the plotted discrepancy stands for one half of the relative difference between the two errors. In other figures like Fig.~\ref{fig:4-comparison_errors_photo_pess}, the median is instead computed across four different cases.\label{foot:median}}
This comparison proves that the integration of \camb{} within \cosmicfish{} has been done consistently. Note that \CFextCAMB{} uses only the \camb{} Fortran code, while \CFintCAMB{} uses also the \camb{} Python wrapper to extract quantities. The order $1\%$ differences found here can be attributed to details in the numerical algorithms used within this Python wrapper or within \cosmicfish, or different ways to interpret precision parameters in the \camb{} Fortran code and Python wrapper, see \cref{sec:acc} for details on accuracy settings. These differences are anyway hardly relevant and do not deserve further attention.

The two methods using \cosmicfish{} combined with \class{} are even more consistent, with a worst error of 0.02\%. Individual marginalised and unmarginalised errors for these two cases are compared in Fig.~\ref{fig:comparison_errors_photo_pess} (second panel). This proves that the integration of \class{} within \cosmicfish{} has also been done correctly. In the \class{} case, there is less room for differences, because the Python wrapper {\tt classy} is used by both the external and internal methods.\footnote{The extremely small (0.02\%) differences can be attributed to very small details in the sampling of functions, rounding errors when writing in files, etc. They are totally irrelevant at the \Euclid sensitivity level.}

At this point, we know that using \cosmicfish{} in external or internal mode makes no difference, and we can investigate the level of consistency of the  \camb{} and \class{} predictions. Comparing e.g. the \CFintCAMB{} and \CFintCLASS{} methods, we find again excellent agreement, with a worst error of 0.10\%. All individual marginalised and unmarginalised errors for these two cases are compared in Fig.~\ref{fig:comparison_errors_photo_pess} (third panel). Other comparisons between the \camb-based and \class-based methods are nearly as good, as can be checked from Table~\ref{fig:comparison_table_photo_pess}. This shows that using \camb{} or \class{} as our theory code makes no difference at the level of sensitivity of \Euclid. Note that, in order to reach such a conclusion, we had to enhance the settings of a few accuracy parameters in the two codes, and to ensure a very good match of the physical assumptions that they use, e.g. on the neutrino sector. These precise settings are detailed in \cref{app:EBS_settings} and commented in \cref{sec:acc}. We recommend using at least the precision settings discussed in \cref{sec:acc} in any forecast or real data analysis, for \Euclid or experiments with comparable sensitivity.

At this stage, we have validated all the methods involving \cosmicfish. We are only left with the comparison of the \cosmicfish{} versus \MPFisher{} results.

We see in Table~\ref{fig:comparison_table_photo_pess} and in the fourth panel of Fig.~\ref{fig:comparison_errors_photo_pess} that the worst discrepancy between the \cosmicfish{} and \montepython{} pipelines reaches about 2.0\%. Since this is much lower than the validation threshold, the \montepython{} Fisher pipeline is validated. We even performed a more demanding test: We compared all the pipelines presented in this paper to the final average results of \istfisher, available in the public repository {\tt fisher\_for\_public}.\textsuperscript{\ref{footn:site}}\,
Fig.~\ref{fig:4-comparison_errors_photo_pess} shows a final comparison between the Fisher marginalised and unmarginalised error from \cosmicfish, \montepython{} and \istfisher. We find a maximum deviation with respect to the mean of 5\%, which confirms validation. Note that \cosmicfish{} results are already known to agree with the results of \istfisher, since an older Fortran-based version of this code was one of the codes employed for the code-comparison project of \istfisher. However, these last tests demonstrates that the \cosmicfish{} pipeline used in this work is fully consistent with the older \cosmicfish{} implementation.

\begin{figure}[h]
 \centering
 \includegraphics[width=0.80\textwidth]{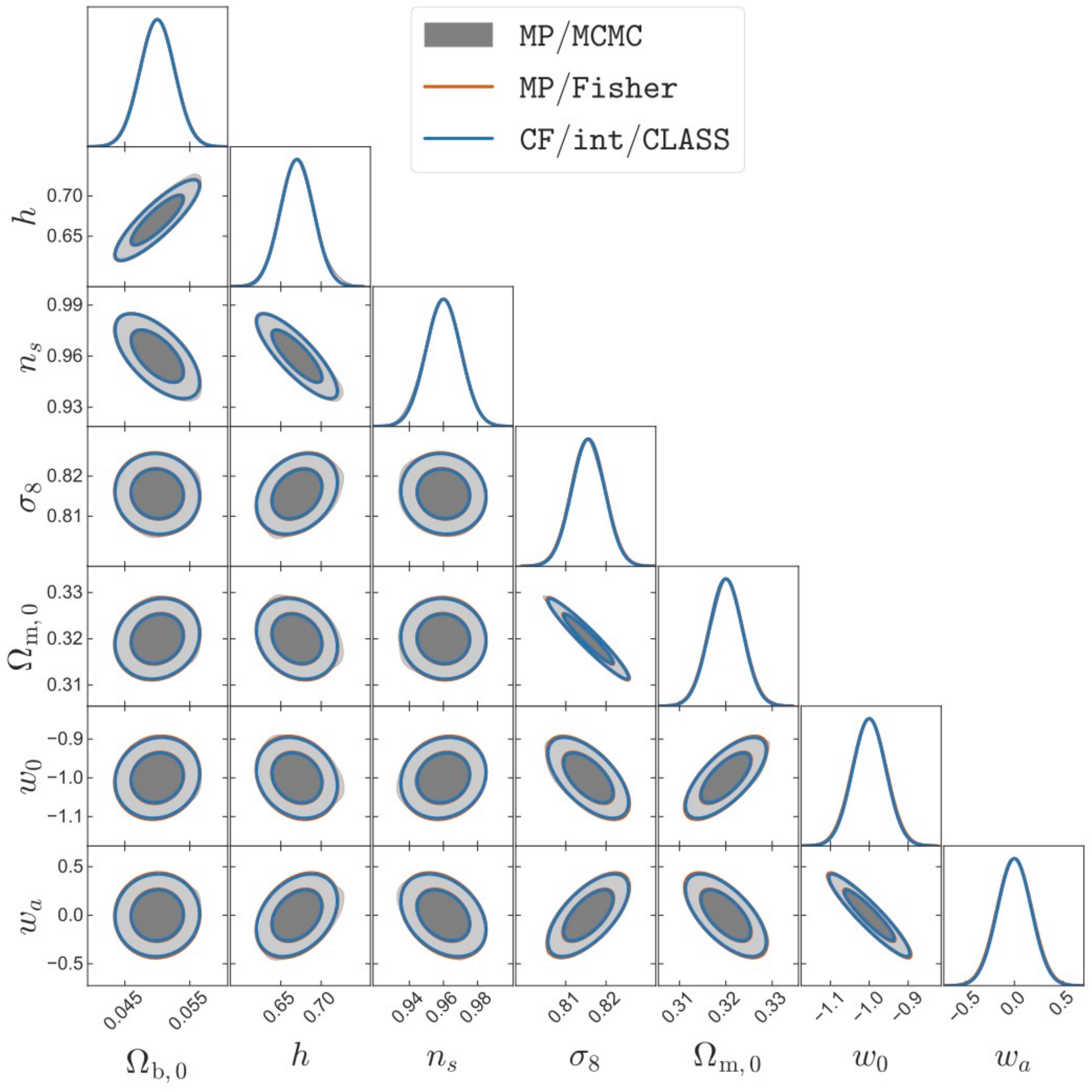}
 \caption{For the photometric survey with pessimistic settings, comparison of 1D posteriors and 2D contours (for 68\% and 95\% confidence level) from different methods: \MPMCMC{} (grey lines/contours), \MPFisher{} (orange lines), \CFintCLASS{} (blue lines). We only show here the cosmological parameters, but our contours involving nuisance parameters are shown in \cref{app:contours_nuisance}. Plotted using \texttt{GetDist}.
 \label{fig:contours_cosmo_photo_pess}}
\end{figure}  

Moreover, the visual comparison of Fisher ellipses with MCMC confidence contours, presented in Fig.~\ref{fig:contours_cosmo_photo_pess} for cosmological parameters (and in \cref{fig:contours_nuisance_photo_pess,fig:contours_cross_photo_pess} of \cref{app:contours_nuisance} for nuisance parameters) is enlightening. The MCMC contours turn out to be almost perfectly elliptical for all parameters. Thus, the approximation of a multivariate Gaussian likelihood, assumed in the Fisher method, is a good one. In all cases, the MCMC contours remain very close to the Fisher ellipses. This excludes the possibility that our Fisher results agree with each other accidentally, while being far from the actual confidence limits associated to the full likelihoods. It also confirms that, for each one of our Fisher methods, the numerical derivative step sizes have been chosen in a sensible way, that is, large enough to overcome numerical noise, and small enough to remain in the region where the likelihood is approximately Gaussian (see \cref{app:stepsizes_CF,app:stepsizes_MP} for details).

\FloatBarrier

\subsubsection{Optimistic setting}

\begin{figure}[h!]
    \centering
    \includegraphics[width=0.90\textwidth]{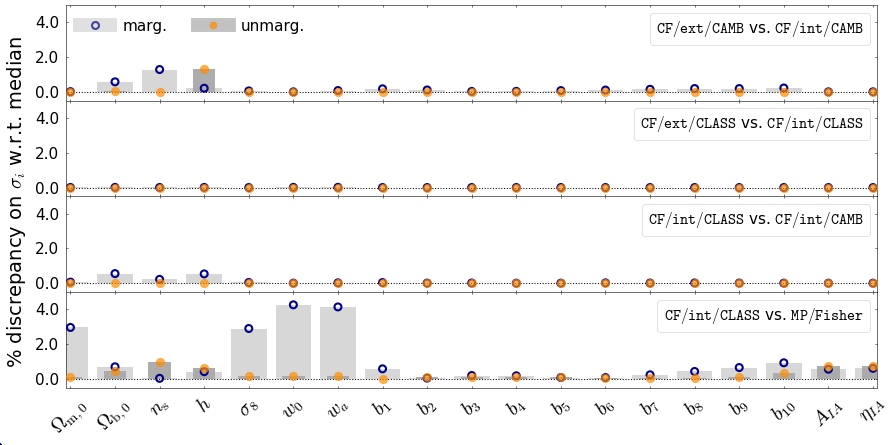}
    \caption{
    Same as Fig.~\ref{fig:comparison_errors_photo_pess} with optimistic settings.
    \label{fig:comparison_errors_photo_opt}
    }
\end{figure}

In this case, our results are summarised in Table~\ref{fig:comparison_table_photo_opt} and Fig.~\ref{fig:comparison_errors_photo_opt}. Qualitatively, all the conclusions reached for the pessimistic case also apply to this case. 

A priori, with optimistic settings, we may expect larger differences between our four \cosmicfish{} pipelines, because the likelihood is more sensitive to theoretical predictions for the nonlinear matter power spectrum. Indeed, the optimistic likelihood probes a larger range of $k$ values in the nonlinear power spectrum and has smaller observational errors. Thus, numerical errors are less likely to be masked by observational errors. However, we find that the difference between the four \cosmicfish{} pipelines (with internal/external  \camb/\class) increase only marginally. For instance, \CFintCAMB{} and \CFintCLASS{} still agree at an impressive 0.55\% level (instead of 0.11\% with pessimistic settings). The agreement between \CFintCAMB{} and \CFextCAMB{} is a bit worse (1.3\%), showing that the \camb{} Fortran code and the \camb{} Python wrapper handle accuracy settings differently, with a small but noticeable impact on small-scale predictions for the nonlinear power spectrum. The better agreement of the \CFintCAMB{} pipeline with the \class{} pipeline suggests that \camb{} is more accurate when called through the Python wrapper -- consistently with the fact that this implementation is the most recent one.\footnote{The small deviation of the \CFextCAMB{} pipeline with respect to the other ones may have to do with a small unexplained feature in the derivative of the nonlinear power spectrum with respect to $w_a$ observed in Fig.~\ref{fig:test_stepsizes} of \cref{app:stepsizes_CF}. We leave this issue for future investigation.}

\begin{figure}[h!]
    \centering
    \includegraphics[width=0.75\textwidth]{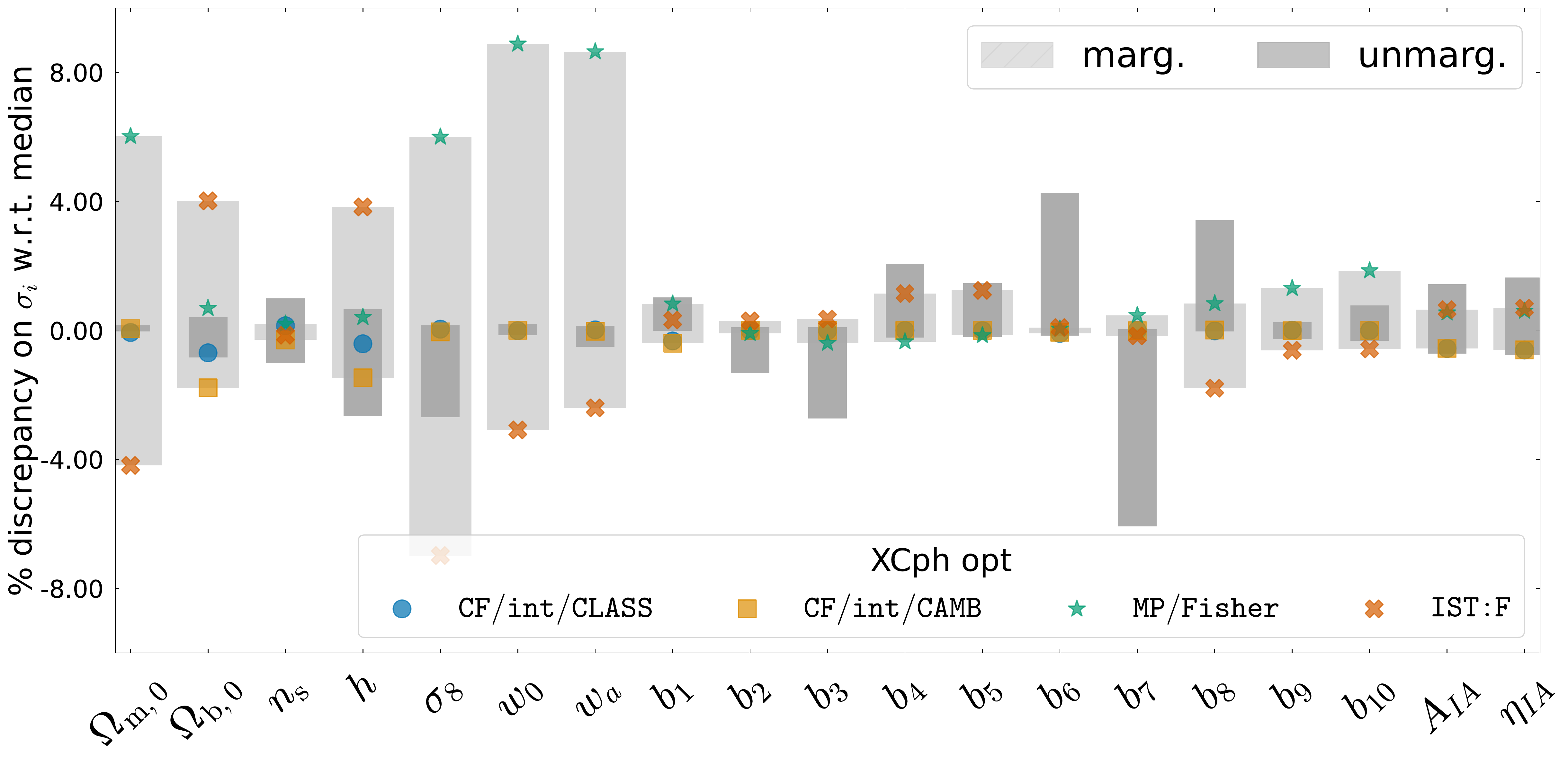}
    \caption{
     Same as Fig.~\ref{fig:4-comparison_errors_photo_pess} with optimistic settings.
    \label{fig:4-comparison_errors_photo_opt}
    }
\end{figure}

Besides, differences between \cosmicfish{} and \montepython{} also increase with respect to the pessimistic case, but only to the level of 4.3\%. Fig.~\ref{fig:comparison_errors_photo_opt} (bottom plot) shows that the difference is now dominated by the marginalised error on ($w_0$, $w_a$), that is, by the calculation of correlations between dark energy parameters and other parameters. However, the 10\% validation threshold is fulfilled. Even when we include the official \istfisher\ results in the comparison, as in Fig.~\ref{fig:4-comparison_errors_photo_opt}, the (un)marginalised errors differ from the median by at most 9\%.

\begin{figure}[h!]
    \centering
    \includegraphics[width=0.80\textwidth]{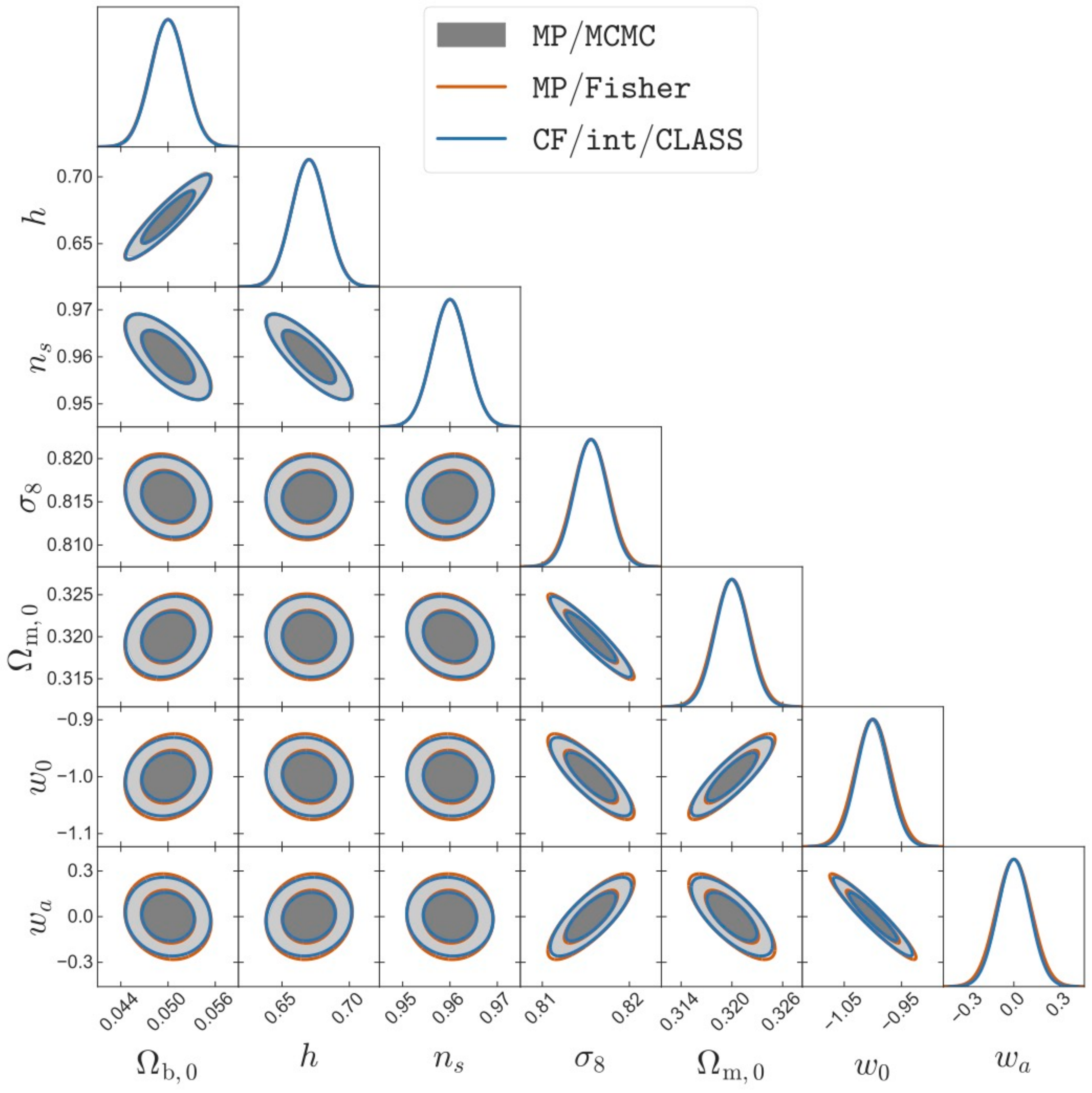}
    \caption{
    Same as \cref{fig:contours_cosmo_photo_pess} with optimistic settings.
    \label{fig:contours_cosmo_photo_opt}
    }
\end{figure}

In conclusion, our pipelines are also all validated in the photometric optimistic case. As a final check, the comparison between MCMC contours and Fisher ellipses in Fig.~\ref{fig:contours_cosmo_photo_opt} brings a final confirmation of the robustness of our forecasts against possible issues related to numerical noise or non-Gaussian posteriors (Fig.~\ref{fig:contours_cosmo_photo_opt} only includes cosmological parameters, but the plots for nuisance parameters are available in our public repository\footnote{In the GitHub repository \url{https://github.com/sabarish-vm/Euclid_w0wa.git}, these plots are located in {\tt plots/photometric/optimistic/WLxGCPh\_Opt\_nuisance.pdf} and {\tt plots/photometric/optimistic/WLxGCPh\_Opt\_cross.pdf}}).

\begin{table}[h!]
    \centering
    \caption{
    Same as Table~\ref{fig:comparison_table_photo_pess} with optimistic settings.}
    \includegraphics[width=0.65\textwidth]{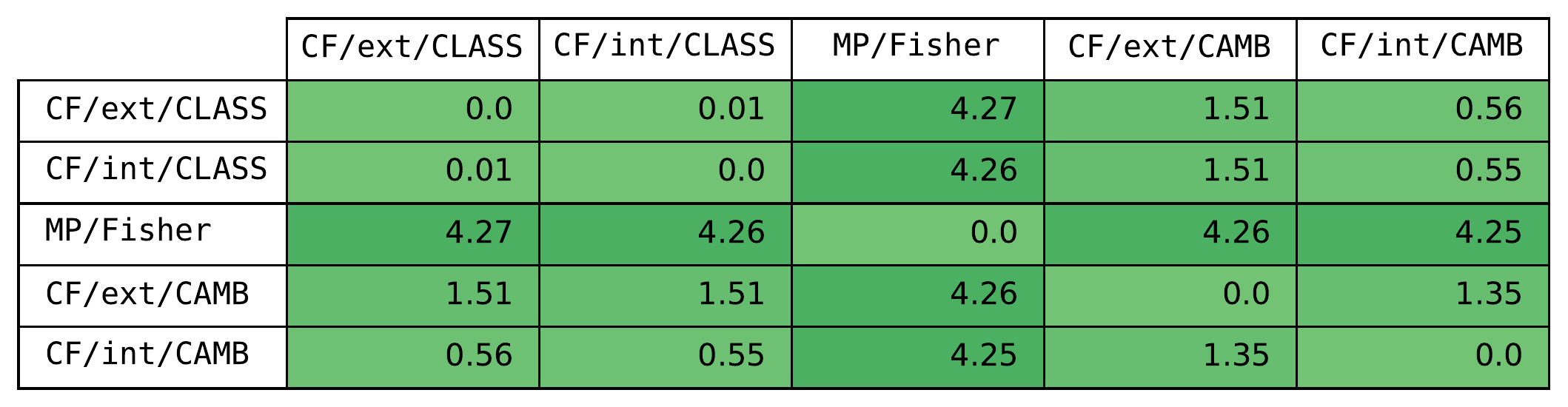}
    \label{fig:comparison_table_photo_opt}
\end{table}

\subsection{Spectroscopic likelihood}

\subsubsection{Pessimistic setting}

\begin{figure}[h]
    \centering
    \includegraphics[width=0.90\textwidth]{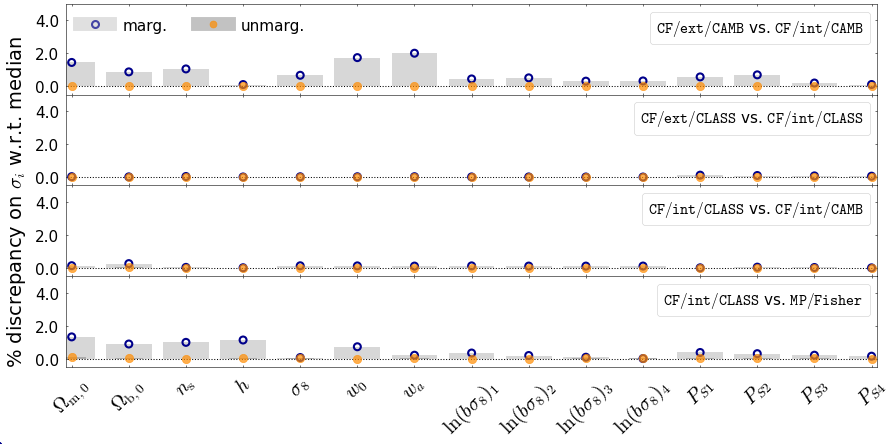}
    \caption{
    Same as Fig.~\ref{fig:comparison_errors_photo_pess} for the spectroscopic survey with pessimistic settings.
    \label{fig:comparison_errors_spectro_pess}
    }
\end{figure}

In this case, our results are summarised in Table~\ref{fig:comparison_table_spectro_pess} and Fig.~\ref{fig:comparison_errors_spectro_pess}. Qualitatively, there are no big differences between the conclusions to be drawn from the photometric and spectroscopic surveys.

\begin{table}[h]
    \centering
    \caption{
    Same as Table~\ref{fig:comparison_table_photo_pess} for the spectroscopic survey with pessimistic settings.}
    \includegraphics[width=0.65\textwidth]{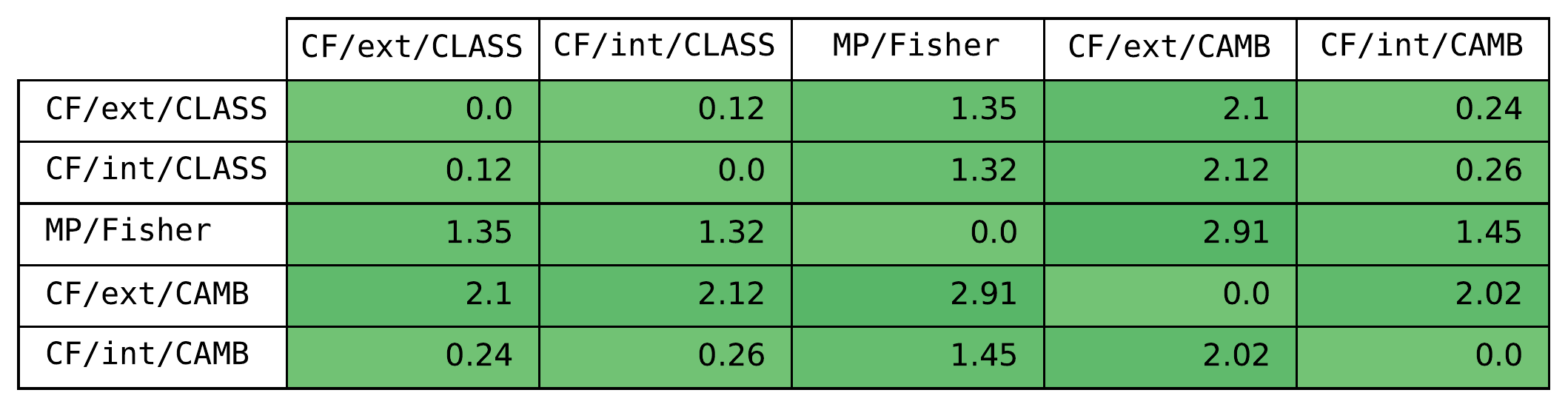}
    \label{fig:comparison_table_spectro_pess}
\end{table}

\begin{figure}[h]
    \centering
    \includegraphics[width=0.75\textwidth]{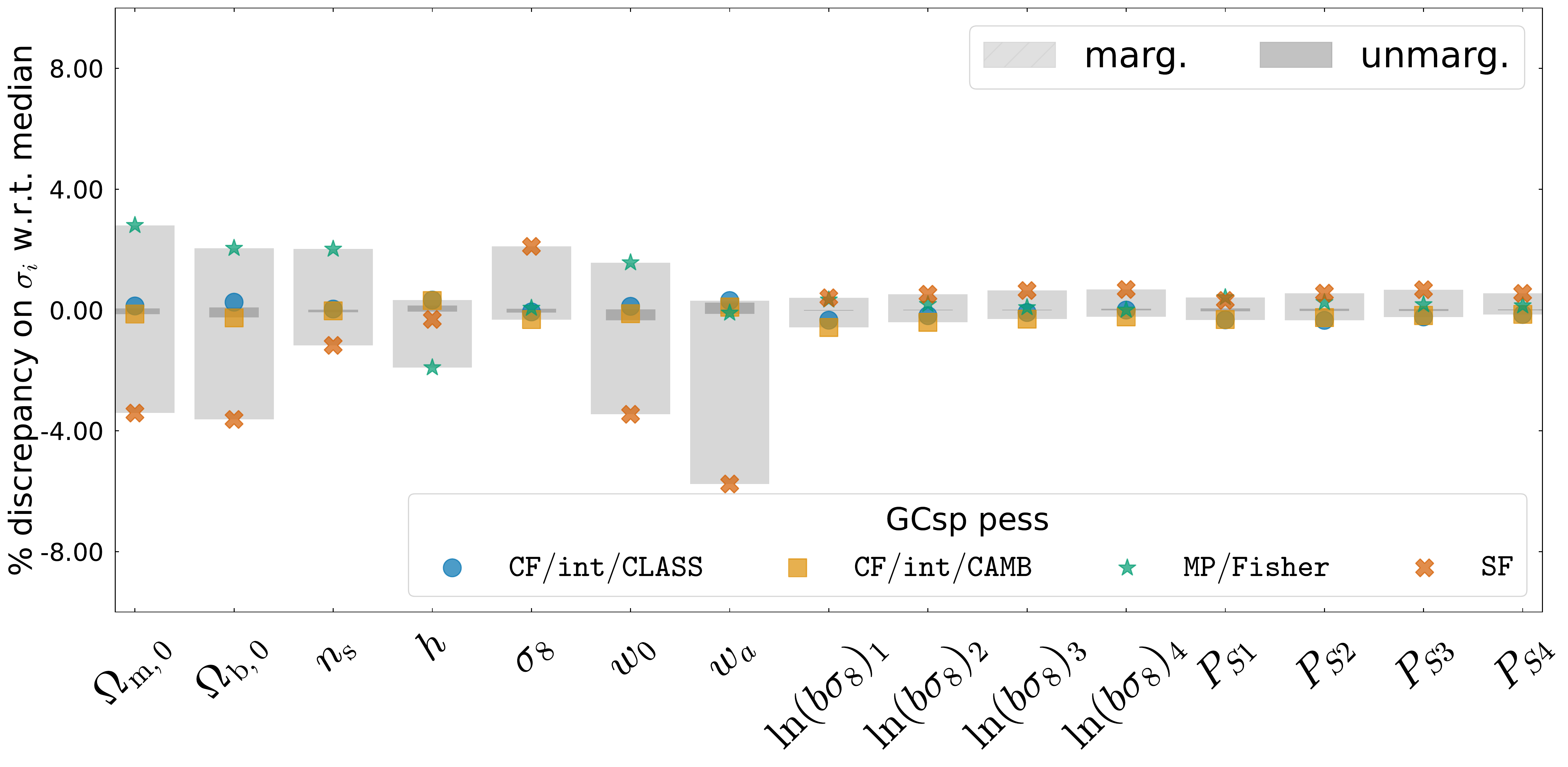}
    \caption{For the spectroscopic survey with pessimistic settings, comparison of each Fisher marginalised (light grey) and unmarginalised (dark grey) errors on the cosmological and nuisance parameters, for the cases of:
     \CFintCLASS{} (blue circles) versus \CFintCAMB{} (orange squares) versus \MPFisher{} (green stars) versus IST:F results from \texttt{SOAPFish} (\texttt{SF}, red crosses). Plotting conventions are the same as in Fig.~\ref{fig:4-comparison_errors_photo_pess}.
    \label{fig:4-comparison_errors_spectro_pes}
    }
\end{figure}  

As a matter of fact, the two pipelines using \cosmicfish{} combined with \camb{} are consistent with each other, with small differences in the errors of at most 2.1\% -- caused again by differences in the treatment of accuracy between the Fortran and Python methods for calling \camb. Once more, the pipelines using \cosmicfish{} combined with \class{} are even more consistent, with a largest differences of 0.12\%. When \cosmicfish{} switches from the \CFintCAMB{} method to the \CFintCLASS{} method, the errors remain unchanged up to 0.26\% differences.

\begin{figure}[htbp]
    \centering
    \includegraphics[width=0.80\textwidth]{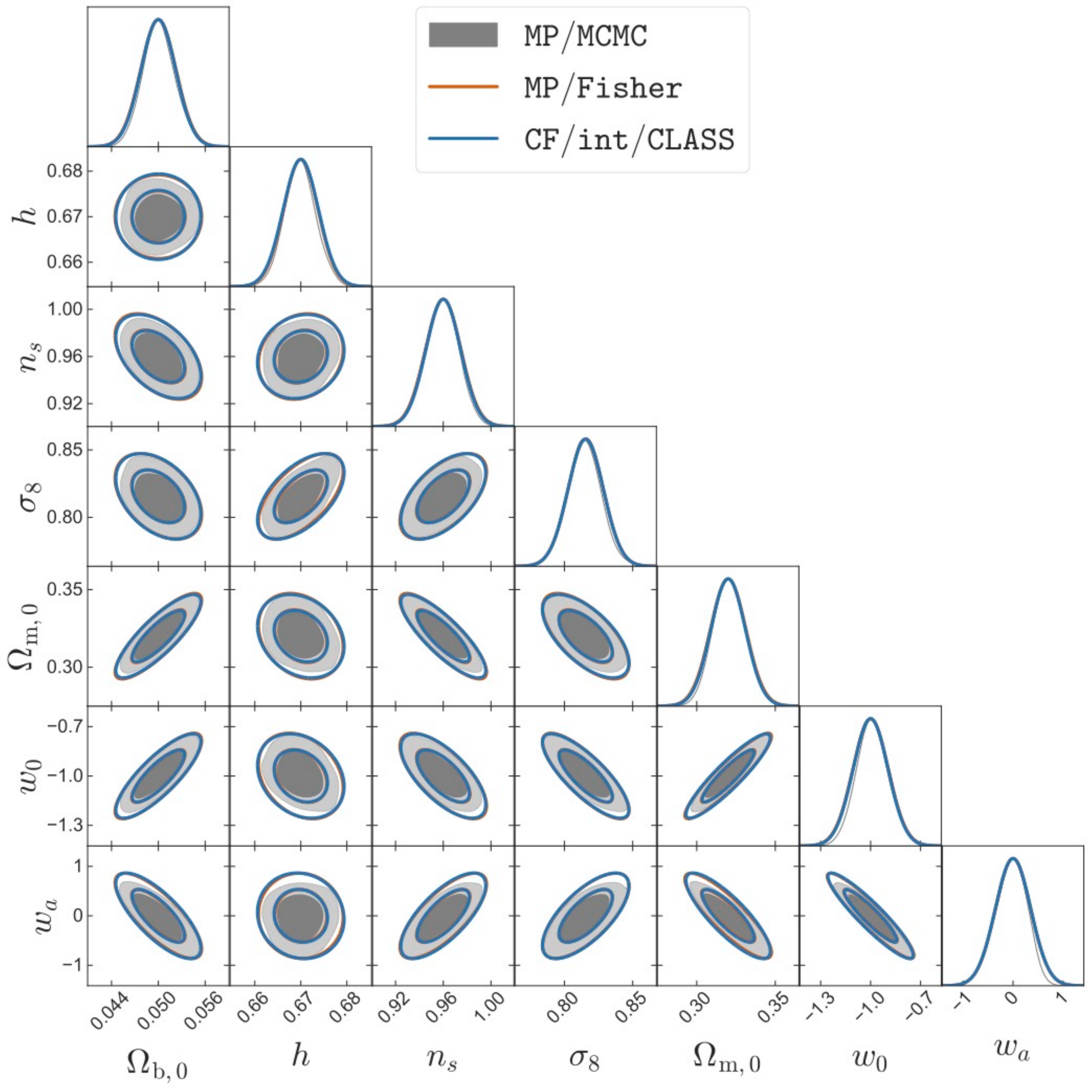}
    \caption{
    Same as Fig.~\ref{fig:contours_cosmo_photo_pess} for the spectroscopic survey with pessimistic settings.
    \label{fig:contours_cosmo_spectro_pess}
    }
\end{figure}  

The \MPFisher{} pipeline gets validated in the spectroscopic/optimistic case, with differences of at most 2.9\% with respect to \CFextCAMB{} and 1.4\% with respect to the other \cosmicfish{} methods. This is comparable to our results for the photometric probe. Our pipelines also pass successfully a more demanding test: when we add to the comparison set some previous \istfisher\ results, all errors remain within 5\% of the median, as shown in \cref{fig:4-comparison_errors_spectro_pes}. In this case, previous \istfisher\ results are represented by those of the \texttt{SOAPFish} code, as already mentioned in \cref{sec:val_method}.

Our Fisher forecasts agrees very well not only among each other, but also with the MCMC forecast shown for comparison in Fig.~\ref{fig:contours_cosmo_spectro_pess} for cosmological parameters (and Figs.~\ref{fig:contours_nuisance_spectro_opt} and \ref{fig:contours_cross_spectro_opt} of \cref{app:contours_nuisance} for nuisance parameters). Here, we can see by eye a small level of non-Gaussianity in the 1D and 2D posteriors. However, the Fisher ellipses appear to be excellent approximations to the actual confidence contours.

\FloatBarrier

\subsubsection{Optimistic setting}

\begin{figure}[htbp]
    \centering
    \includegraphics[width=0.75\textwidth]{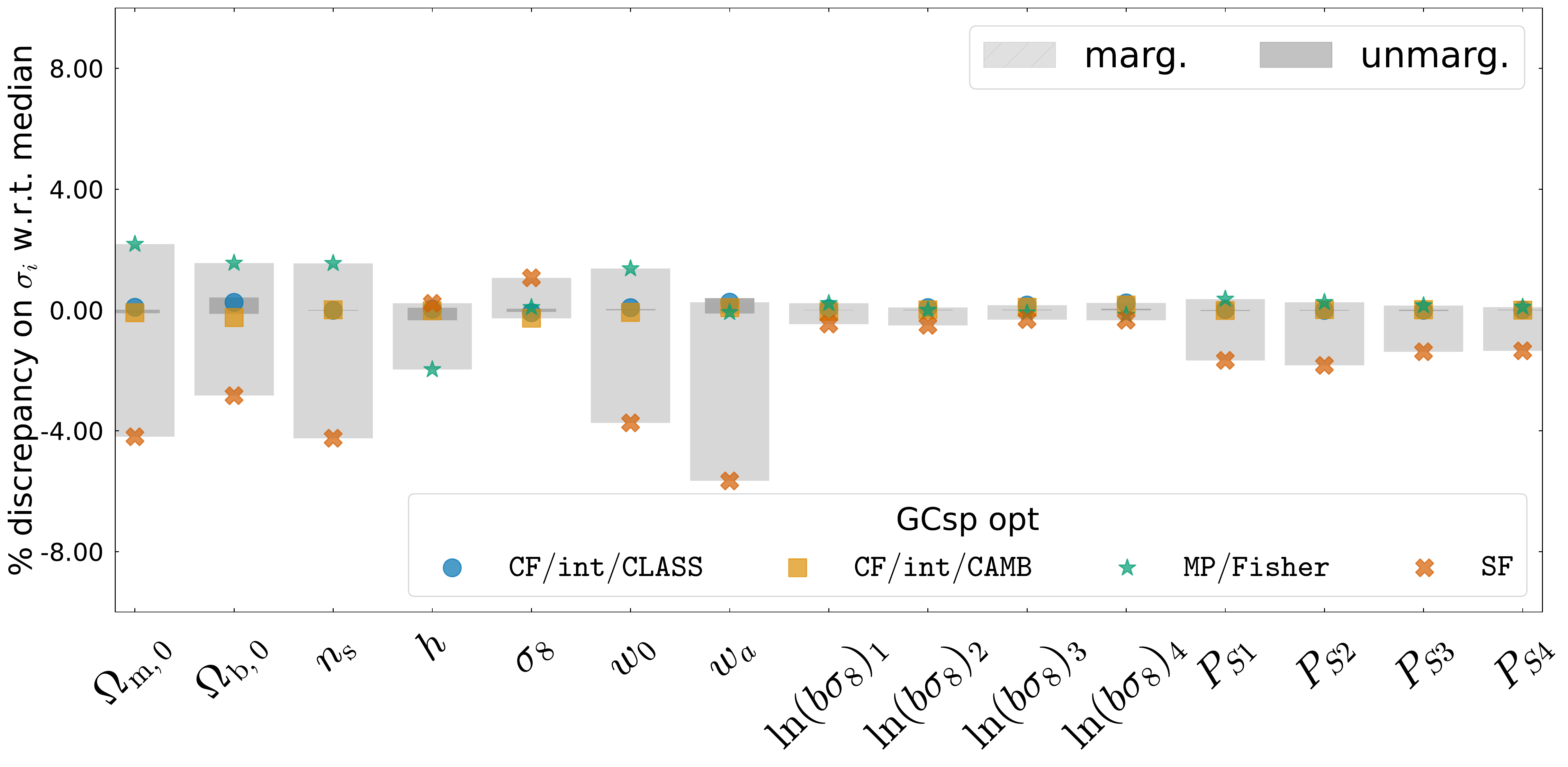}
    \caption{
     Same as Fig.~\ref{fig:4-comparison_errors_spectro_pes} with optimistic settings.
    \label{fig:4-comparison_errors_spectro_opt}
    }
\end{figure}

\begin{figure}[htbp]
    \centering
    \includegraphics[width=0.80\textwidth]{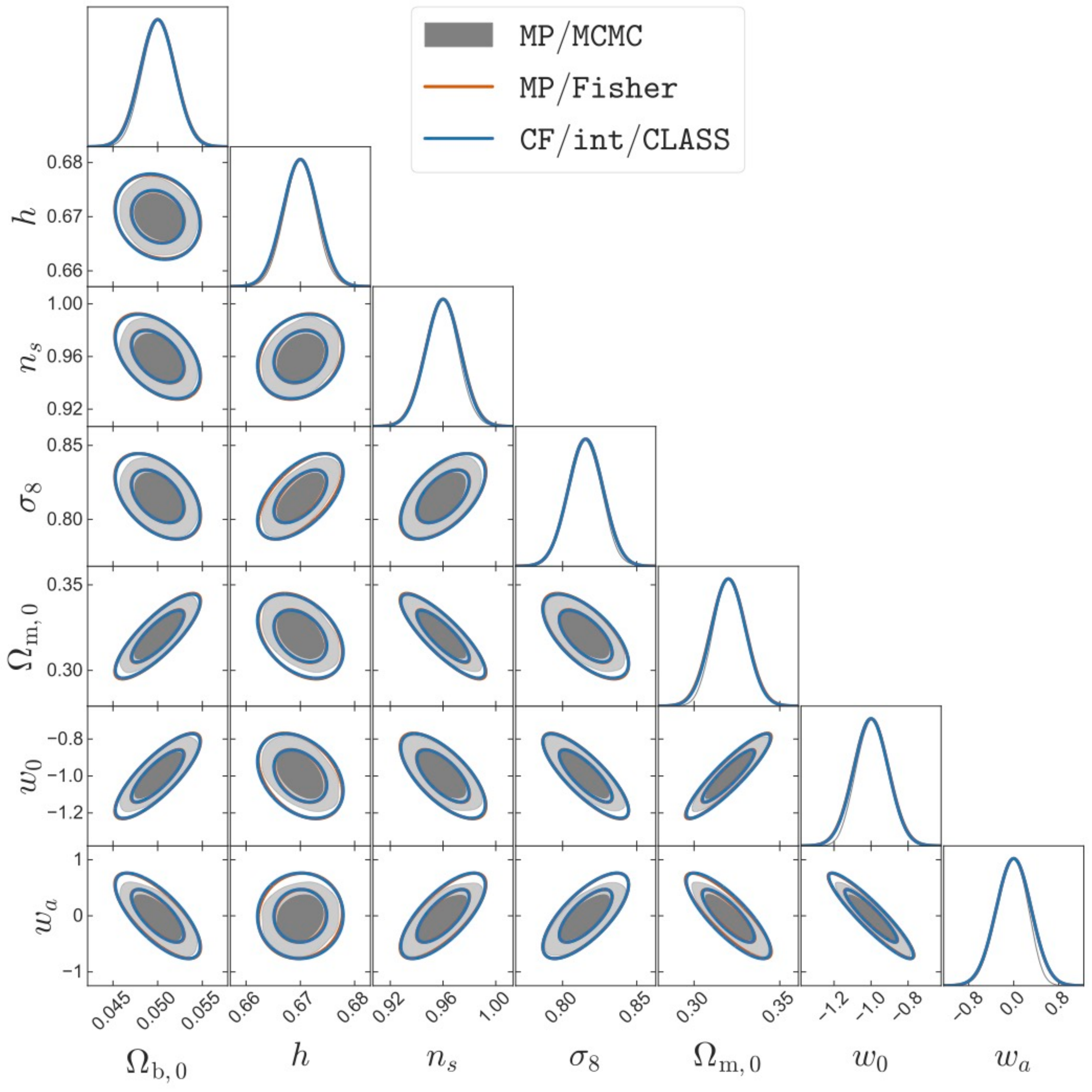}
    \caption{
    Same as Fig.~\ref{fig:contours_cosmo_photo_pess} for the spectroscopic survey with optimistic settings.
    \label{fig:contours_cosmo_spectro_opt}
    }
\end{figure}

This last case is very similar to the previous one. Switching to optimistic settings in the spectroscopic likelihood does not degrade nor improve significantly the comparison between the various approaches. The difference between all marginalised \cosmicfish{} and \montepython{} Fisher errors is below the 2.5\% level, see Table~\ref{fig:comparison_table_spectro_opt}. When adding \texttt{SOAPFish} to the comparison set, the difference increases to 6\%, see Fig.~\ref{fig:4-comparison_errors_spectro_opt}, still small enough to validate our \MPFisher{} pipeline. The very good agreement between all Fisher forecasts and an \MPMCMC{} forecast is shown for comparison in Fig.~\ref{fig:contours_cosmo_spectro_opt} for cosmological parameters 
(and in the repository\footnote{In the GitHub repository \url{https://github.com/sabarish-vm/Euclid_w0wa.git}, these plots are located in {\tt plots/spectroscopic/optimistic/GCsp\_Opt\_nuisance.pdf} and {\tt plots/spectroscopic/optimistic/GCsp\_Opt\_cross.pdf}} for nuisance parameters).

\begin{table}[htbp]
    \centering
    \caption{
    Same as Table~\ref{fig:comparison_table_photo_pess} for the spectroscopic survey with optimistic settings.}
    \includegraphics[width=0.65\textwidth]{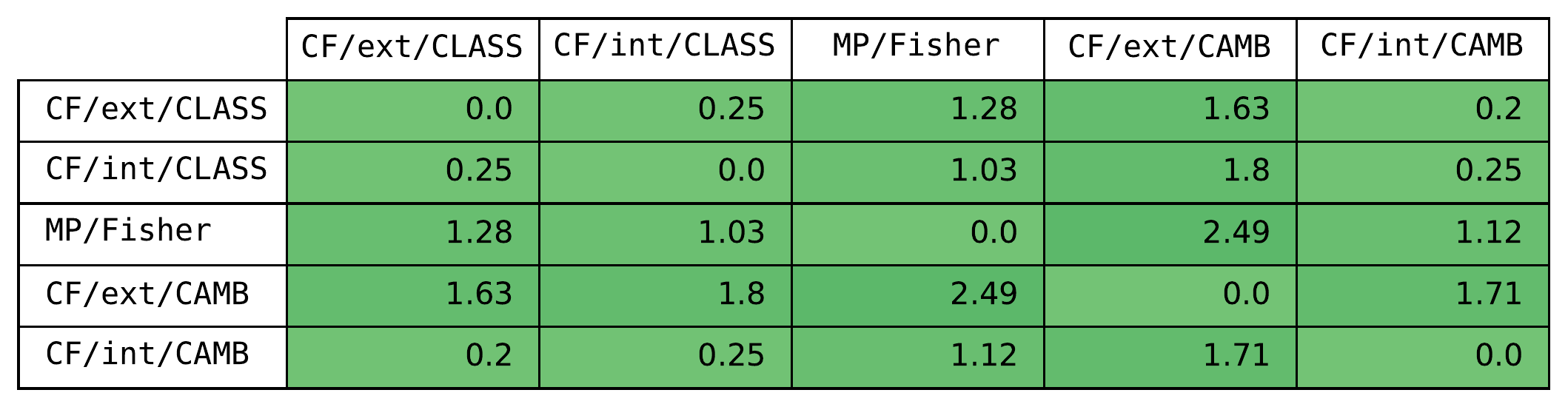}
    \label{fig:comparison_table_spectro_opt}
\end{table}

\begin{figure}[htbp]
    \centering
    \includegraphics[width=0.90\textwidth]{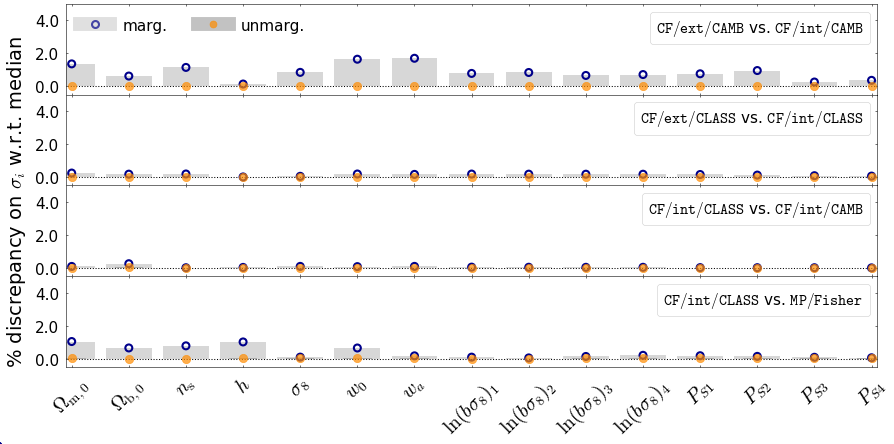}
    \caption{
    Same as Fig.~\ref{fig:comparison_errors_photo_pess} for the spectroscopic survey with optimistic settings.
    \label{fig:comparison_errors_spectro_opt}
    }
\end{figure}    

\FloatBarrier

\section{Forecast results\label{sec:forecast_results}}

\subsection{Photometric survey alone}

In \cref{tab:absolute_err_photo_pess,tab:absolute_err_photo_opt} we show the absolute 1-$\sigma$ errors on each cosmological parameters obtained for the photometric probe with our \cosmicfish, \MPFisher{} and \MPMCMC{} pipelines. We compare them with the mean values obtained by \istfisher. For \cosmicfish, we take here numbers from the \CFintCLASS{} case -- but we know from previous sections that other cases give essentially the same results. 

Previous tables showed the largest percentage difference between error bars across all cosmological and nuisance parameters. Here, by comparing different columns, one can check for individual parameters that these percentage differences are always very small -- even if the error bars of \istfisher{} are usually more optimistic than other errors by a tiny amount.

\begin{table}
\centering
\caption{For the photometric survey with pessimistic settings, marginalised 1-$\sigma$ errors on cosmological parameters found by: the average of \istfisher\ results, \CFintCLASS{} (the other versions of \cosmicfish{} give nearly the same results), \montepython{} in Fisher mode, \montepython{} in MCMC mode.}
\begin{tabular}{|c|c|c|c|c|}
\hline
 & IST:F & \CFintCLASS & \MPFisher & \MPMCMC \\
 \hline
 $\Omega_{\mathrm{b},0}$ & 0.0027 & 0.0027 & 0.0027 & 0.0026 \\
$h$ & 0.020 & 0.020 & 0.021 & 0.021 \\
$n_{\rm s}$ & 0.0097 & 0.010 & 0.010 & 0.011 \\
$\sigma_8$ & 0.0039 & 0.0041 & 0.0042 & 0.0044 \\
$\Omega_{\mathrm{m},0}$ & 0.0035 & 0.0036 & 0.0036 & 0.0038 \\
$w_0$ & 0.042 & 0.043 & 0.044 & 0.045 \\
$w_a$ & 0.17 & 0.17 & 0.18 & 0.18 \\
\hline
\end{tabular}
\label{tab:absolute_err_photo_pess}
\end{table}

\begin{table}
\centering
\caption{
Same as \cref{tab:absolute_err_photo_pess} for the photometric survey with optimistic settings.}
\begin{tabular}{|c|c|c|c|c|}
\hline
 & IST:F &  \CFintCLASS & \MPFisher & \MPMCMC \\
 \hline
$\Omega_{\mathrm{b},0}$ & 0.0023 & 0.0022 & 0.0022 & 0.0022 \\
$h$ & 0.014 & 0.013 & 0.013 & 0.013 \\
$n_{\rm s}$ & 0.0037 & 0.0037 & 0.0037 & 0.0038 \\
$\sigma_8$ & 0.0018 & 0.0019 & 0.0020 & 0.0020 \\
$\Omega_{\mathrm{m},0}$ & 0.0019 & 0.0020 & 0.0021 & 0.0020 \\
$w_0$ & 0.027 & 0.028 & 0.031 & 0.028 \\
$w_a$ & 0.10 & 0.11 & 0.11 & 0.11 \\
\hline
\end{tabular}
\label{tab:absolute_err_photo_opt}
\end{table}

\subsection{Spectroscopic survey alone}

In \cref{tab:absolute_err_spectro_pess,tab:absolute_err_spectro_opt}, we show the absolute 1-$\sigma$ errors on each cosmological parameters obtained with the same pipelines, but now for the spectroscopic probe. 
Note that there is a minor difference between our implementation and the \istfisher\ one regarding the nonlinear modelling. As a matter of fact, our pessimistic and optimistic cases agree in the $k_\mathrm{max}$ used for the probes, namely $0.25 h\,\mathrm{Mpc}^{-1}$ and $0.30 h\,\mathrm{Mpc}^{-1}$, respectively.
However, the pessimistic case of \istfisher\ is more conservative, since it marginalizes over the nonlinear parameters $\sigma_p$ and $\sigma_v$, while here we keep them both fixed at the fiducial cosmology, as already explained in \cref{sec:spectroscopic}.
Moreover, the recipe of \istfisher\ assumes a separation into shape- and redshift-dependent parameters for the observed galaxy power spectrum. In that recipe, the redshift-dependent parameters -- which are $H(z_i)$, $d_A(z_i)$, $f \sigma_8(z_i)$ and $b \sigma_8(z_i)$ -- are varied freely at each redshift bin $i$ and then projected onto the final cosmological parameter basis (or marginalized over in the case of the galaxy bias). This is again more conservative, firstly, because it assumes more freedom in the modelling of the observables, and secondly, because the marginalization over a larger parameter space degrades the constraints.
Since \istfisher\ did not publish a Fisher matrix using our direct full-shape method approach, we compare against the code \texttt{SoapFish}, which does contain the same implementation and is one of the validated codes of the IST:F group.

\begin{table}
\centering
\caption{
Same as \cref{tab:absolute_err_photo_pess} for the spectroscopic survey with pessimistic settings.}
\begin{tabular}{|c|c|c|c|c|}
\hline
 & \texttt{SoapFish} & \CFintCLASS & \MPFisher & \MPMCMC \\
 \hline
$\Omega_{\mathrm{b},0}$ & 0.0021 & 0.0022 & 0.0023 & 0.0019 \\
$h$ & 0.0038 & 0.0038 & 0.0038 & 0.0029 \\
$n_{\rm s}$ & 0.014 & 0.015 & 0.015 & 0.013 \\
$\sigma_8$ & 0.013 & 0.013 & 0.013 & 0.011 \\
$\Omega_{\mathrm{m},0}$ & 0.011 & 0.011 & 0.011 & 0.0095 \\
$w_0$ & 0.10 & 0.10 & 0.11 & 0.087 \\
$w_a$ & 0.33 & 0.35 & 0.35 & 0.28 \\
\hline
\end{tabular}
\label{tab:absolute_err_spectro_pess}
\end{table}

\begin{table}
\centering
\caption{
Same as \cref{tab:absolute_err_photo_pess} for the spectroscopic survey with optimistic settings.}
\begin{tabular}{|c|c|c|c|c|}
\hline
 & \texttt{SoapFish} &  \CFintCLASS & \MPFisher & \MPMCMC \\
 \hline
 $\Omega_{\mathrm{b},0}$ & 0.0019 & 0.0019 & 0.0020 & 0.0019 \\
$h$ & 0.0032 & 0.0032 & 0.0032 & 0.0029 \\
$n_{\rm s}$ & 0.013 & 0.013 & 0.013 & 0.013 \\
$\sigma_8$ & 0.012 & 0.012 & 0.012 & 0.011 \\
$\Omega_{\mathrm{m},0}$ & 0.0096 & 0.010 & 0.010 & 0.0095 \\
$w_0$ & 0.090 & 0.093 & 0.095 & 0.087 \\
$w_a$ & 0.29 & 0.31 & 0.31 & 0.28 \\
\hline
\end{tabular}
\label{tab:absolute_err_spectro_opt}
\end{table}

\subsection{Combined surveys}

We finally combine the photometric and spectroscopic probes, which amounts in summing up their Fisher matrices. We consider all four combinations of pessimistic or optimistic settings for each probe. The absolute 1-$\sigma$ errors on each cosmological parameters obtained with \cosmicfish{} are shown in \cref{tab:absolute_err_combined}. We take here numbers from the \CFintCLASS{} case -- but we cross-checked that other cases give essentially the same numbers. To visualise the way in which the combined probe lifts parameter degeneracies, the best is to compare the Fisher ellipses coming from individual and combined probes. These are shown in Fig.~\ref{fig:contours_cosmo_combined_opt_opt} for the optimistic-optimistic case, in Appendix~\ref{app:combined} for the pessimistic-pessimistic case, and in our repository for other cases.\footnote{In the GitHub repository \url{https://github.com/sabarish-vm/Euclid_w0wa.git}, these plots are located in {\tt plots/combined/combined\_pess\_opt\_cosmo.pdf} and {\tt plots/combined/combined\_opt\_pess\_cosmo.pdf}}

Note that, in principle, there is a correlation between data from the photometric and spectroscopic galaxy clustering probes at the redshifts where they overlap. In this section, we neglect this correlation for simplicity, which makes our forecasts slightly too optimistic. This differs from the `\istfisher\ pessimistic' approach, in which high-redshift data was removed from the photometric survey, leading instead to slightly too pessimistic results. 

The combination is particularly efficient for the determination of $\Omega_\mathrm{b,0}$ and $h$, for which the photometric and spectroscopic surveys probe different degeneracy directions. The combination reduces significantly the error on these parameters compared to each individual probe. For other parameters, the measurement is dominated by the photometric probe.

\begin{table}
\centering
\caption{For the combined photometric and spectroscopic surveys with various combinations of pessimistic and optimistic settings, marginalised 1$\sigma$ errors on cosmological parameters found by \CFintCLASS{} (the other versions of \cosmicfish{} give nearly the same results).}
\begin{tabular}{|c|c|c|c|c|}
\hline
 & \multicolumn{4}{c|}{combined photometric/spectroscopic}\\
 & pess/pess & pess/opt & opt/pess & opt/opt \\
 \hline
$\Omega_{\mathrm{b},0}$ & 0.00099 & 0.00095 & 0.00075 & 0.00072 \\
$h$ & 0.0016 & 0.0015 & 0.0012 & 0.0011 \\
$n_{\rm s}$ & 0.0047 & 0.0046 & 0.0019 & 0.0019 \\
$\sigma_8$ & 0.0033 & 0.0032 & 0.0018 & 0.0018 \\
$\Omega_{\mathrm{m},0}$ & 0.0031 & 0.0031 & 0.0019 & 0.0019 \\
$w_0$ & 0.036 & 0.035 & 0.026 & 0.025 \\
$w_a$ & 0.14 & 0.13 & 0.096 & 0.094 \\
\hline
\end{tabular}
\label{tab:absolute_err_combined} 
\end{table}

\begin{figure}[h]
    \centering
    \includegraphics[width=0.95\textwidth]{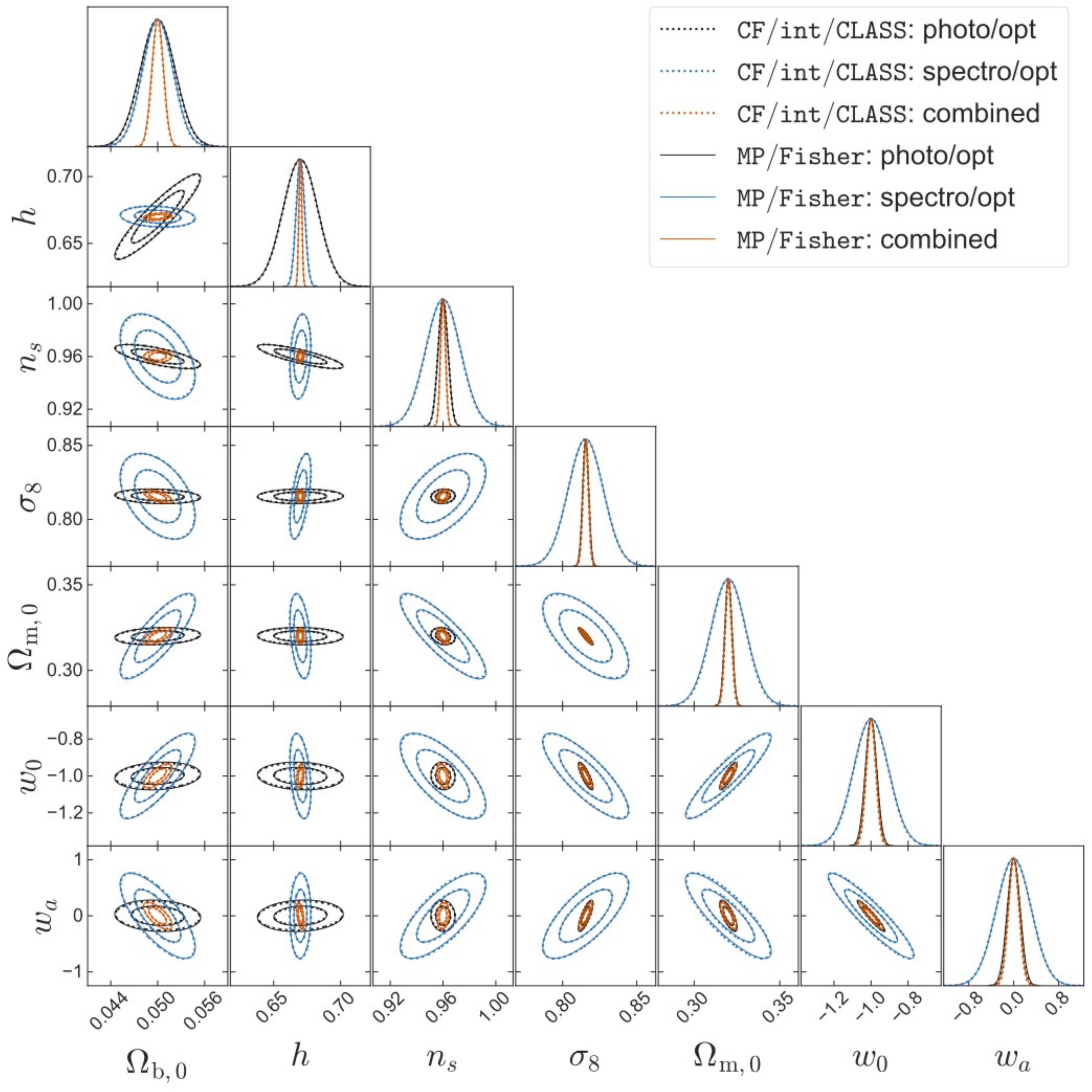}
    \caption{For the individual and combined photometric and spectroscopic surveys with optimistic settings, 1D posterior and 2D contours (for 68\% and 95\% confidence level) on cosmological parameters from  \CFintCLASS{} (dotted lines) and \MPFisher{} (solid lines).
    Plotted using \texttt{GetDist}.
    \label{fig:contours_cosmo_combined_opt_opt}
    }
\end{figure}

A traditional estimate of the constraining power of a survey dedicated to the study of dark energy is the dark energy Figure of Merit (DE FoM), which is defined as 
\begin{equation}
    \mathrm{FoM} = \sqrt{\det ( \tilde{F}_{w_0 w_a} )}\;,
\end{equation}
where $\tilde{F}_{w_0 w_a}$ is the $2 \times 2$ Fisher matrix after marginalizing over all other cosmological and nuisance parameters except $w_0$ and $w_a$.
For a Gaussian posterior (i.e. for elliptical Fisher contours) this quantity is related to the area $S$ of the $1\sigma$ probability contours by 
\begin{equation} \label{eq:fom-area}
    \mathrm{FoM} = \frac{2.3 \pi }{S}\;.
\end{equation}
For a Fisher matrix, the computation of the determinant is a trivial operation, while for MCMC chains, there are two options. The first option, and most common in the literature, is to compute the covariance matrix of the chains, which is a good approximation if the posterior is Gaussian: then, the covariance matrix is equivalent to the inverse of the Fisher matrix. However, since the covariance matrix is by definition symmetrical, computing the DE FoM from it might under- or over-estimate the constraining power of the probe in the $w_0$-$w_a$ parameter space. The difference would depend on the exact shape of the MCMC contours and how non-Gaussian they are.
The second option is to use Eq.~\eqref{eq:fom-area} to calculate the FoM from the enclosed area of the MCMC contour itself. The computation of such contours is a well-known problem in the literature. The reader can refer to \citet{Lewis:2019xzd} and \citet{Brinckmann:2018cvx} for more details on the estimation of the probability contours from a discrete set of Markov chains.

For the photometric probe, since our approach is exactly equivalent to the one of \istfisher, we also find exactly the same DE FoM. For the spectroscopic one, we have highlighted and proved in this work that while using the same likelihood recipe, we adopt a different parametrisation: we directly compute the Fisher matrix in the space of cosmological parameters, while \istfisher\ computed it with respect to a basis of phenomenological parameters, and then used a non-trivial transformation to go to the space of cosmological parameters. Thus we find a slightly different FoM, that we provide for completeness in \cref{tab:FoM_table} -- using either one of the \cosmicfish{} Fisher pipelines or the MCMC pipeline with an evaluation of the area enclosed by the marginalised contours.

\begin{table}[h]
\centering
\caption{Dark energy Figure of Merit for the spectroscopic survey with optimistic and pessimistic settings, found by: the determinant of the \cosmicfish{} Fisher matrix and the area of the \montepython{} MCMC contours.}
\begin{tabular}{|c|c|c|}
\hline
\multicolumn{3}{|c|}{dark energy Figure of Merit (FoM)}\\
\hline
 & \cosmicfish{} determinant
 & \multicolumn{1}{p{5cm}|}{ \montepython{} MCMC area} \\
 \hline
GCsp pessimistic & 69 & 76 \\
GCsp optimistic & 87 & 92  \\
\hline
\end{tabular} 
\label{tab:FoM_table}  
\end{table}

\section{Impact of accuracy settings in Einstein--Boltzmann solvers \label{sec:acc}}

\camb{} and \class{} have been proved to agree remarkably with each other when using very high accuracy settings \citep[see][]{Lesgourgues:2011rg}. The settings defined in \citet{Lesgourgues:2011rg} are however unpractical because they slow down both codes considerably. The question of properly choosing the default value of each precision parameter in \camb{} and \class{} has been extensively investigated and tested by various groups from 2011 to 2013, that is, during the first stages of the {\it Planck} data analysis. Since then, both codes are provided with default-precision settings designed to avoid biasing the results of MCMC analyses of, typically, {\it Planck} and SDSS/BOSS data, while keeping them as fast as possible.

Accuracy settings in \camb{} and \class{} deserve to be revisited in the present context, mainly for three reasons:
\begin{itemize}
\item In general, Fisher analyses require more precision than MCMC runs. Indeed, the calculation of numerical derivatives involves the comparison of power spectra obtained under very small variations of cosmological parameters. These variations should be dominated by physical effects rather than numerical errors. MCMC runs avoid such an issue because, in MCMCs, random numerical errors tend to be averaged out each time that results are marginalised over unwanted parameters. Thus, in this work, MCMC runs are always performed with default \class{} precision.
\item Among the Fisher methods used here, the \MPFisher{} method, which requires the calculation of second derivatives of the likelihood, tends to require even higher precision.
\item \Euclid will probe the matter power spectrum with much higher accuracy than SDSS/BOSS. Precision parameters that are particularly relevant for computing the matter power spectrum need to be tuned accordingly.
\end{itemize}

\subsection{Accuracy settings and fiducial spectra}
\label{sec:acc_fid}

In \cref{app:EBS_settings}, we detail the list of all fixed parameters passed to \camb{} and \class{} before each run. The list includes a few options concerning the setting of key precision parameters. Here we summarize these different settings and we show how they impact the accuracy of the power spectrum calculation for a given cosmological model -- in this section, we just chose to focus on the fiducial model.\footnote{In this section we will only compare the linear and nonlinear power spectra from \camb{} and \class. Internally, we also compared the functions $D(z)$, $f(z)$ and $\sigma_8(z)$. For the latter, we did not find any noticeable difference that would be worth reporting.} For more details, the reader can consult \cref{app:EBS_settings}.

For \camb, the two parameters {\tt accuracy\_boost} and {\tt l\_accuracy\_boost} play an essential role for the calculation of background, thermodynamical and perturbation quantities for each cosmology. In this work, we never consider values of these parameters below 2 (2 was the setting adopted in \istfisher). Additionally, in the context of Fisher forecasts, one must be aware of the role of a tolerance parameter in the version of \texttt{Halofit} implemented in \camb. This tolerance governs the accuracy with which the algorithm finds the scale of nonlinearity for each cosmology, using a bisection method. In \camb, this tolerance is hard-coded to $10^{-3}$. We replaced it by a new accuracy parameter passed in input, {\tt halofit\_tol\_sigma}, and we explored values down to $10^{-6}$ (we checked that even smaller values make no difference within \cosmicfish{} forecasts). We define three precision settings for \camb, which we call P1, P2, P3 in order of growing precision:

\camb{} (P1 / P2 / P3)
\begin{lstlisting}
accuracy_boost = 2 / 2 / 3
l_accuracy_boost = 2 / 2 / 3
halofit_tol_sigma = 1.e-3 / 1.e-6 / 1.e-6
\end{lstlisting}

Thus, (P1) corresponds to settings of \istfisher, (P2) shows the impact of decreasing the \texttt{Halofit} tolerance by 3 orders of magnitude, and (P3) stands for our most enhanced settings, used by default in the rest of this work.

For \class, we will consider the default-precision (DP) settings of the public code and compare them with high-precision (HP) settings. In order to define the latter, we recall that \class{} is downloaded together with an optional input file called {\tt pk\_ref.pre}, designed to push the code to extreme accuracy, and used in \cite{Lesgourgues:2011rg} to demonstrate that boosted versions of \camb{} and \class{} do agree at the 0.01\% level for the matter power spectrum and other observables. These settings are unpractical for forecasting purposes due to their CPU cost. Additionally, many parameters in {\tt pk\_ref.pre} are irrelevant for the calculation of the matter power spectrum. In \cref{app:HP}, we propose a more concise set of accuracy parameters that define our HP settings. We will show here that these HP settings lead to a very good convergence of the matter power spectrum without slowing \class{} down in an exaggerated way. Note that the tolerance parameter of \texttt{Halofit} is set by default to $10^{-6}$ within \class{} (like in our \camb{} P2 and P3 settings). We lower it to $10^{-8}$ in \class{} HP settings (this is lower than in \camb{} P2 and P3, because \class{} is used within the \MPFisher{} pipeline which, as argued before, requires extra precision).

\begin{figure}[h]
 \centering
 \includegraphics[width=0.95\textwidth]{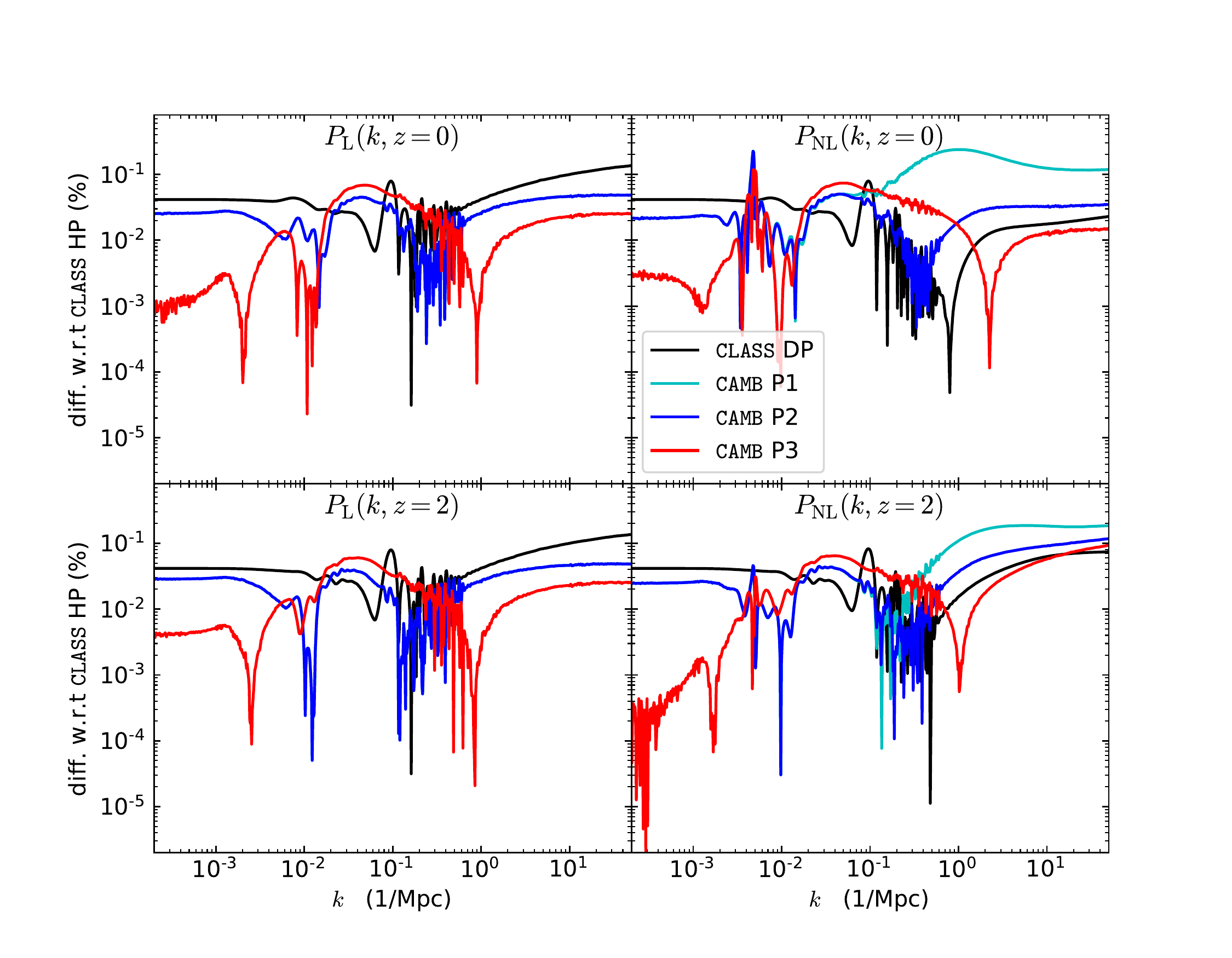}
 \caption{For the linear (left) and nonlinear (right) matter power spectrum of the fiducial model at redshift $z=0$ (top) and $z=2$ (bottom), percentage difference between the predictions of \class{} and \camb. Here, the reference is the \class{} high-precision (HP) result. We compare it with \class{} Default Precision (DP) and \camb{} with the precision settings defined as P1, P2, P3 in the text. In the linear case (left), the \camb{} P1 curve is always overdrawn by the P2 one, since the difference between P1 and P2 is only relevant at the level of the nonlinear spectrum.}
 \label{fig:fiducial_ratios}
\end{figure}  

In Fig.~\ref{fig:fiducial_ratios}, we compare the linear and nonlinear spectrum of the fiducial model at redshift $z=0$ and $z=2$, computed with each version of \camb{} and \class. Here, \class{} HP is always used as a reference.

First, we can compare the results with the highest settings: in Fig.~\ref{fig:fiducial_ratios}, the red curves show the ratio of the \camb{} P3 over \class{} HP spectra. We see that the relative difference is always below 0.07\% for the linear spectrum. This is not as small as the 0.01\% agreement found in \cite{Lesgourgues:2011rg}, because this reference relied on even higher precision settings, with accuracy boost parameters as large as (4, 8) in \camb, and settings from {\tt pk\_ref.pre} in \class. Nevertheless, a 0.07\% agreement is remarkable and definitely sufficient for the purpose of our forecasts, as demonstrated by the agreement between the various Fisher matrices derived in the previous sections. For the nonlinear power spectrum, the level of agreement is similar (excepted for a narrow spike of 0.2\% at $k \simeq 3.3\times 10^{-3}$Mpc$^{-1}$ coming from the transition from linear to nonlinear calculations in both codes -- but a feature at such a low values of $k$ is harmless for \Euclid predictions).

Next, we can investigate the impact of degrading precision in the two codes. With \class{} DP, the error remains below the 0.1\% level on large scales, but slightly exceeds 0.12\% on small scales for the linear power spectrum. This is anyhow a very small degradation compared to the sensitivity of \Euclid. As a matter of fact, we only observe very small differences between the  \cosmicfish{} results obtained with either \class{} DP or HP. However, the HP settings provide more stability against the choice of step sizes when computing the second derivatives of the likelihood with the \MPFisher{} approach, as shown in \cref{app:stepsizes_MP}. Thus, in the rest of this work, we decided to stick everywhere to \class{} HP settings.

Comparing the results from \camb{} P2 and P3 in Fig.~\ref{fig:fiducial_ratios}, we see that \camb{} P2 is a bit further away from \class{} HP than \camb{} P3 on small scales, but nothing dramatic. The results for P1 and P2 are identical at the level of the linear spectra, since these settings only differ at the level of one \texttt{Halofit} parameter. However, the nonlinear spectra from P1 and P2 differ significantly -- at the 0.25\% for our fiducial model. This shows that a \texttt{Halofit} tolerance of $10^{-3}$ is insufficient. It leads to a poorly converged bisection in \texttt{Halofit}, and thus, to a spurious step-like response of the nonlinear spectrum to small variations in the cosmological parameters. Thus, it compromises the calculation of numerical derivatives. We believe that the IST:F group overcame this difficulty by using advanced methods for the calculation of derivatives, based on multiple steps. With simple two-sided derivatives, it would be nearly impossible to obtain stable derivatives with P1 settings, as we shall see better in the next subsection.

As a small technical detail, we note that, in order to produce Fig.~\ref{fig:fiducial_ratios}, we adjusted two parameters in a slightly different way than in the rest of this paper:
\begin{itemize}
\item Instead of passing the same value of $\sigma_8$ to both codes, we passed the same value of $A_\mathrm{s}$, in order to be insensitive to very small differences in the $\sigma_8$-to-$A_\mathrm{s}$ conversion performed inside \camb{} or \class. This choice is relevant only for the sake of producing Fig.~\ref{fig:fiducial_ratios}. Indeed, in the rest of this work, we only want to compute the variation of the spectra in the vicinity of the fiducial model. Thus, a very small difference in the absolute value of one cosmological parameter ($A_\mathrm{s}$) is irrelevant. Therefore, in other sections, we can safely adopt the same input value of $\sigma_8$ in the two codes. 
\item In order to match the scale $k_*$ at which both codes start computing nonlinear corrections to the power spectrum, we adjusted manually the transition scale parameter in \class{} to its \camb{} value, that is, {\tt halofit\_min\_k\_nonlinear=3.3e-3}. Otherwise, nonlinear corrections are computed by \class{} for any $k>k_*^\mathrm{CLASS}=10^{-4}$Mpc$^{-1}$, and by \camb{} for any $k>k_*^\mathrm{CAMB}=3.3 \times 10^{-3}$Mpc$^{-1}$. This produces a small degradation of the \camb-\class{} agreement for $k_*^\mathrm{CLASS} < k<k_*^\mathrm{CAMB}$. However, \Euclid does not probe accurately such huge scales / tiny wavenumbers. Thus, in other parts of this work, we stick to the default value of the \class{} parameter {\tt halofit\_min\_k\_nonlinear}.
\end{itemize}

\subsection{Accuracy settings and first derivatives of the spectra}
\label{sec:acc_der}

\begin{figure}[h]
    \centering
    \includegraphics[width=0.95\textwidth]{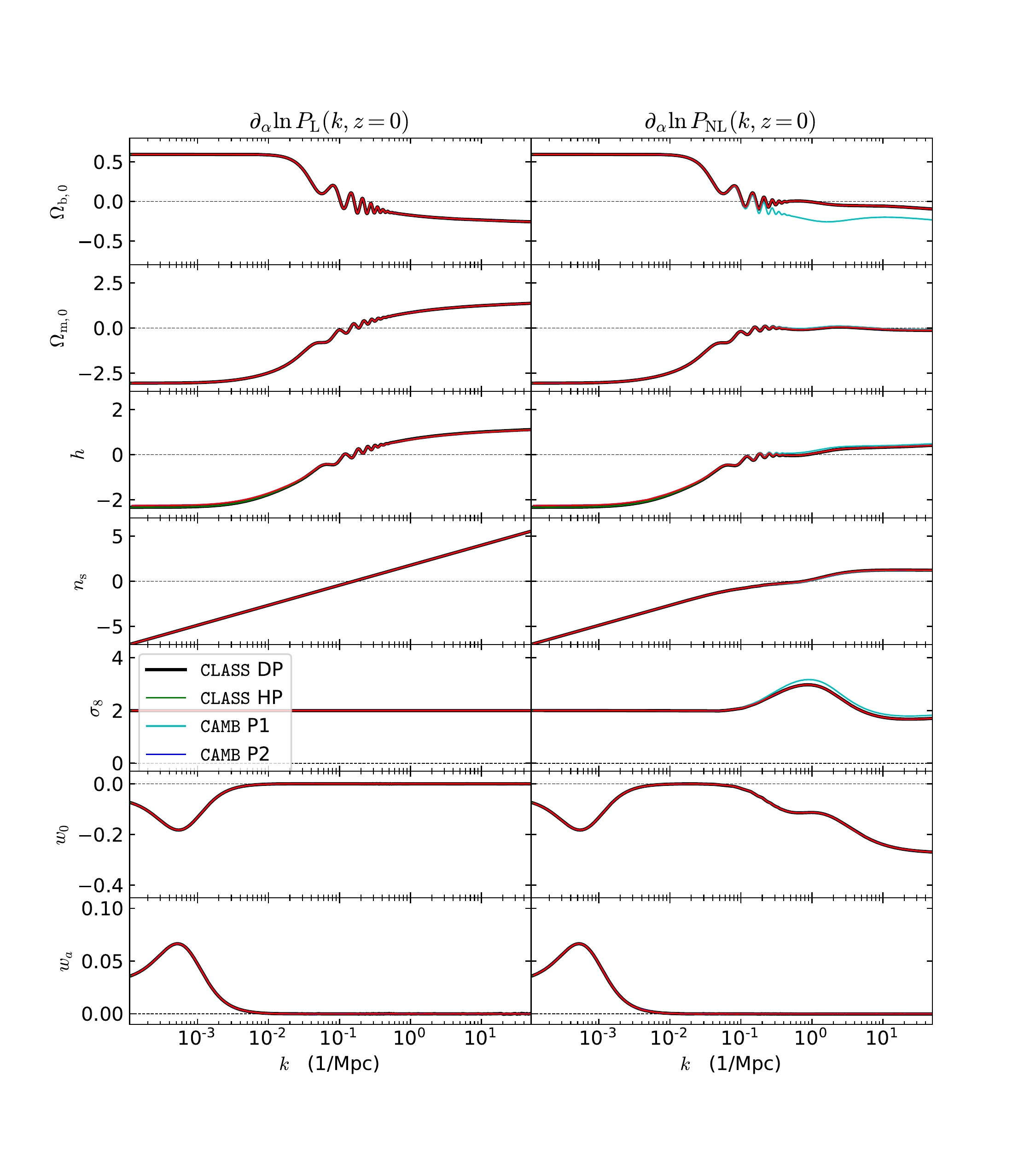}
    \caption{Derivative of the logarithm of the linear (left) and nonlinear (right) matter power spectrum with respect to each cosmological parameter in our basis, computed with \class{} (with either DP or HP) and \camb{} (with either P1, P2, or P3 precision). The derivatives are computed at $z=0$. In the linear case (left), the \camb{} P1 curve is always overdrawn by the P2 one, since the difference between P1 and P2 is only relevant at the level of the nonlinear spectrum. Moreover, P2 is always nearly overdrawn by P3.}
    \label{fig:derivatives}
\end{figure}  

\begin{figure}[h]
    \centering
    \includegraphics[width=0.95\textwidth]{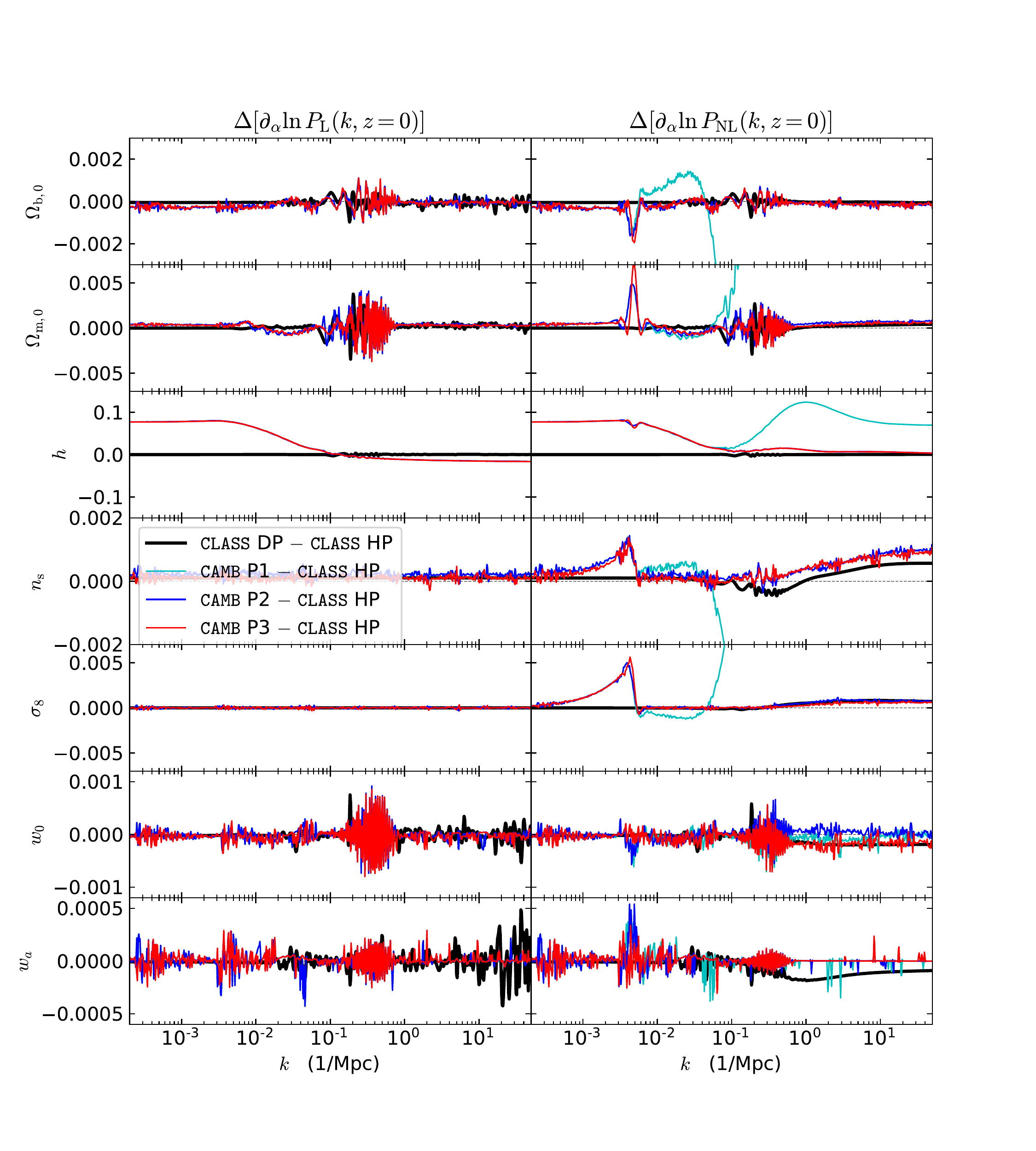}
    \caption{Same Fig.~\ref{fig:derivatives} but now for the {\it difference} between the derivative of the logarithm of the linear (left) and nonlinear (right) matter power spectrum computed with one of \class{} DP or \camb{} (with either P1, P2, or P3 precision) with the same derivative computed with \class{} HP. }
    \label{fig:derivative_errors}
\end{figure}  

We now bring our precision tests to the level of first derivative calculations. Since we use the Limber approximation for the photometric probe, the quantities of interest for both probes are the derivatives of the matter power spectrum. We will show here the derivatives $\partial_\alpha \ln P(k,z)$ of the linear and nonlinear power spectrum at $z=0$ for all cosmological parameters $\alpha$. Like in our \cosmicfish{} runs, we use two-sided derivatives with step sizes $\Delta p_\alpha$ set to 1\% of fiducial values (or 0.01 in the case of $w_0$ and $w_a$). Fig.~\ref{fig:derivatives} shows the derivatives $\partial_\alpha \ln P(k,z=0)$ as a function of $k$ computed with \class{} DP, HP, and \camb{} P1, P2, P3. To see better the difference between these  curves, we show in Fig.~\ref{fig:derivative_errors} the difference between the derivatives computed with either \class{} DP, \camb{} P1, \camb{} P2 or \camb{} P3 and those computed with \class{} HP.

For the linear power spectrum, all codes and all accuracy settings predict derivatives that are visually identical in Fig.~\ref{fig:derivatives}. Fig.~\ref{fig:derivative_errors} reveals tiny differences. By comparing the two figures, we see that differences are typically two order of magnitude smaller than the derivative themselves (the biggest discrepancy is a 5\% difference between the \camb{} and \class{} predictions for the derivative with respect to $h$). Such a high level of agreement on linear derivatives explains the very good match between the Fisher matrices obtained with \cosmicfish/\camb{} and \cosmicfish/\class{} for the spectroscopic probe.

At the level of the nonlinear power spectrum, we see in both figures that \camb{} P1 predictions are very discrepant -- especially for ($\Omega_\mathrm{b}$, $\sigma_8$), but also for ($\Omega_{\mathrm{m},0}$, $h$, $n_{\rm s}$). This is due to the glitches generated by insufficient settings for the \texttt{Halofit} tolerance parameter. However, the level of agreement between \class{} DP, HP and \camb{} P2, P3 is again excellent, typically of the same order of magnitude as for the linear spectrum -- which proves that \texttt{Halofit} is implemented consistently and accurately in the two codes.

From a glance at \cref{fig:derivatives,fig:derivative_errors}, it is difficult to assess whether the Fisher matrix of the \cosmicfish{} pipeline should be computed with DP or HP for \class{} and with P2 or P3 for \camb: these settings seem roughly equivalent at the level of these plots. However, tiny differences at the level of derivatives -- and thus of Fisher matrices -- can be amplified by the Fisher matrix inversion. Thus the only way to check the required level of accuracy is to compare the marginalised errors obtained with different precision levels. We performed such tests using the \CF/ext pipeline.

\begin{figure}[h]
    \centering
    \includegraphics[width=0.80\textwidth]{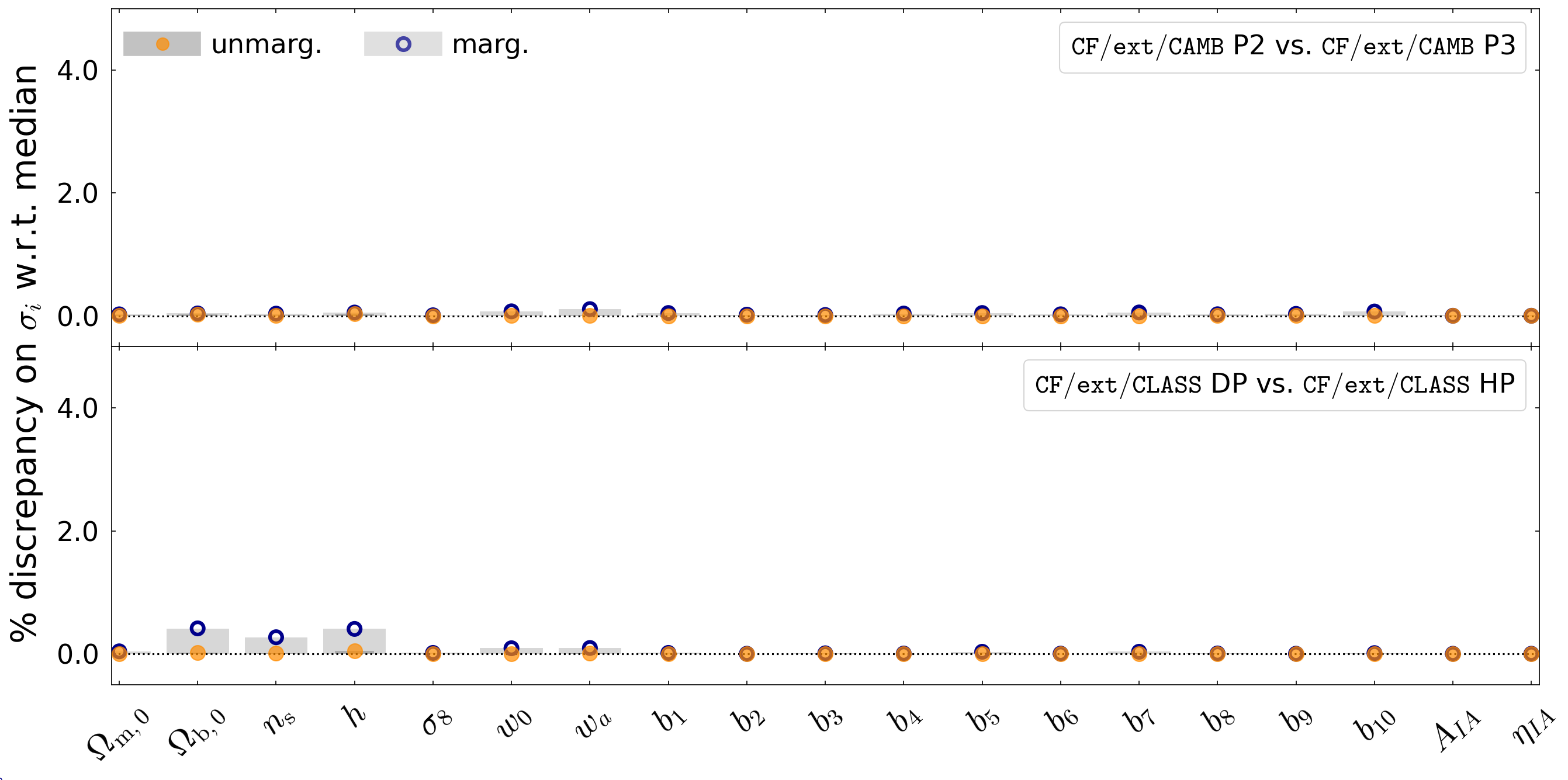}
    \includegraphics[width=0.80\textwidth]{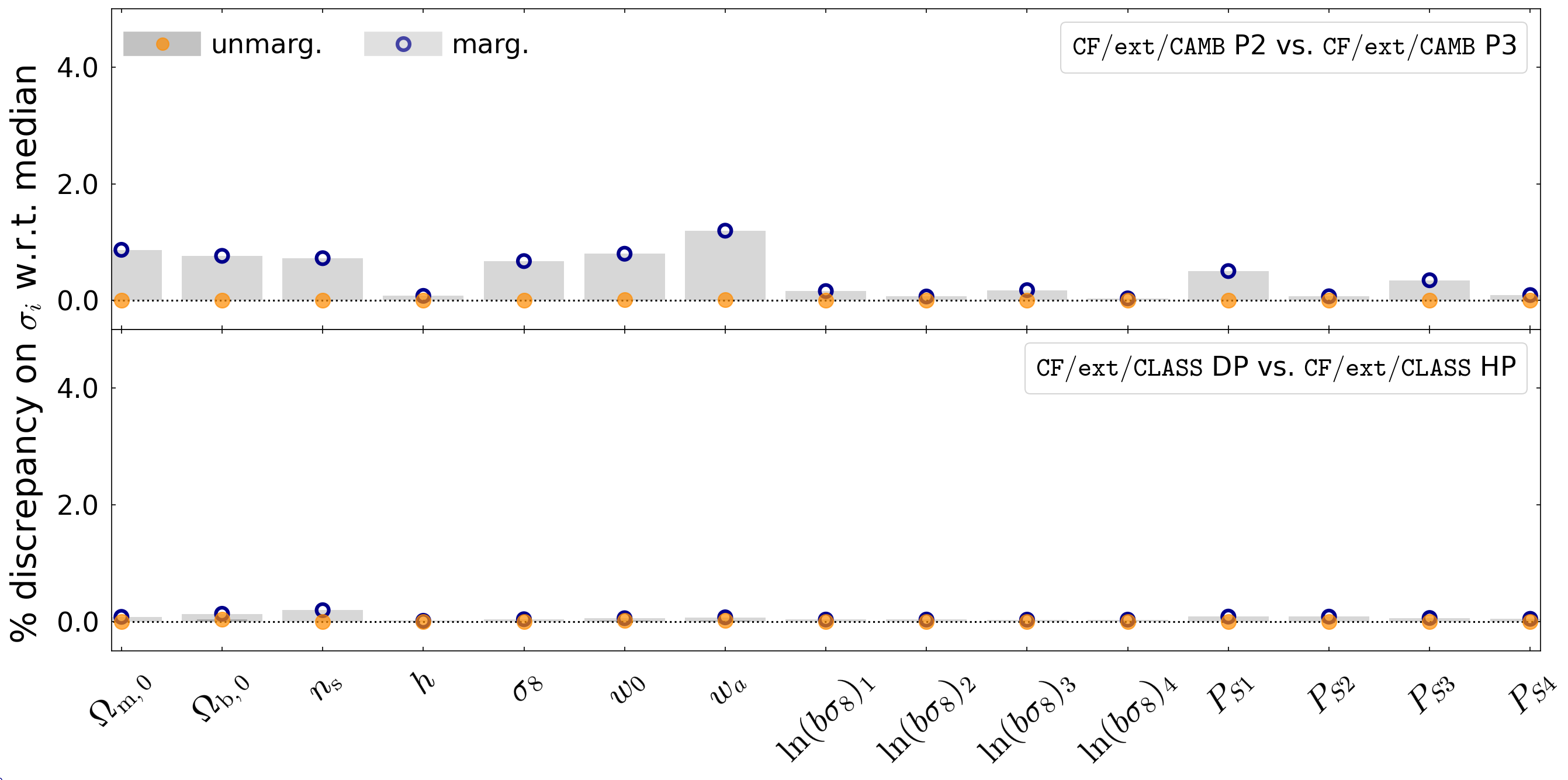}
    \caption{For the photometric survey (top panels) and spectroscopic survey (bottom panels) both with optimistic settings, comparison of the Fisher marginalised and unmarginalised errors on each cosmological and nuisance parameters for \camb{} P2 versus \camb{} P3 precision (first and third panels), or \class{} DP versus HP precision (second and fourth panels). 
    Plotting conventions are the same as in Fig.~\ref{fig:comparison_errors_photo_pess}.
    \label{fig:comparison_errors_camb_precision_CF}
    }
\end{figure}  

We show the difference between the error inferred from \camb{} P2 versus P3 or from from \class{} DP versus HP in Fig.~\ref{fig:comparison_errors_camb_precision_CF}. In each test, we consider the case of the photometric/optimistic probe (top panels) and spectroscopic/optimistic probe (bottom panels). Here we always use the {\tt CF/ext} pipeline. 

The largest differences appear between \camb{} P2 and \camb{} P3 for the spectroscopic probe: they are of the order of 1\% on marginalised errors. These differences are still small, which suggests that results obtained with P2 are already reasonable, while those of P3 are likely to be well converged. We checked that the results obtained with P3 are the closest ones to the average of other pipelines, and thus, as expected, the most accurate. This motivates our choice to stick to P3 precision in the rest of this work. We recall that the matrices of \istfisher\ were based on P1 settings, and that the significant error introduced by such settings was mitigated through the use of the SteM scheme for the calculation of derivatives.

Between \class{} DP and \class{} HP, we never find differences exceeding 0.5\%. Thus, running with HP is not strictly necessary, but it confers extra precision and robustness to the results. In all previous sections, our Fisher matrices were obtained with HP settings.

\subsection{Accuracy settings and second derivatives of the likelihood}
\label{sec:acc_der2}

The \MPFisher{} pipeline relies on the calculation of second derivatives of the likelihood, which potentially require even more precision than the evaluation of first derivatives. We can do various tests of the impact of accuracy settings on such second derivatives, that is, on the direct calculation of the Fisher matrix coefficient.

In \cref{app:stepsizes_MP}, we show how the prediction of a given Fisher matrix coefficient varies with the \class{} precision setting (DP versus HP) and the step size. We find that with small step sizes, the results can be affected by random numerical noise, while with large step sizes they pick up higher-order contributions related to the non-Gaussianity of the likelihood. Adopting HP settings offers an opportunity to decrease the step sizes without being dominated by numerical noise, and thus, to have a safe and robust calculation of Fisher matrix coefficients, closer to the Gaussian approximation. This supports our decision to stick to HP settings throughout this work -- while in the \cosmicfish{} pipeline, based on the calculation of first derivatives, \class{} DP settings would suffice, as shown in the last subsection.

We can also compare the marginalised and unmarginalised errors on all parameters obtained with the \MPFisher{} pipeline, using either \class{} DP or HP. For this test, we stick to the same small step sizes as in the rest of this work, that is, for the photometric probe, 10\% of the marginalised error on each parameter, and for the spectroscopic probe, 5\% of them (as explained in \cref{app:stepsizes_MP}). This test leads to tiny differences (1\%) on unmarginalised errors, but large errors on the marginalised one, reaching 11\% for the spectroscopic/optimistic probe and even much more for the photometric/optimistic probe. Thus we conclude that the use of HP is optional for the \cosmicfish{} pipeline and absolutely required for the \MPFisher{} pipeline. We reiterate that the \MPMCMC{} pipeline is the least sensitive to numerical errors and can be safely run in DP.

\section{Conclusions\label{sec:conclusions}}

For the purpose of \Euclid forecasts, we have thoroughly validated the \MPFisher{} pipeline against a bunch of \cosmicfish{} Fisher forecast pipelines and a few other previous results from \istfisher. Despite of the very different method through which \montepython{} and \cosmicfish{} compute Fisher matrices, we found a spectacular level of agreement between them. Typically, with pessimistic \Euclid settings, all errors agree at least at the 2\% level between the \MPFisher{} and \cosmicfish{} pipelines, for both the photometric and spectroscopic cases. Even when involving additional forecast pipelines from \istfisher, all error bars agree much better than the ``10\% with respect to the median'' threshold fixed by the latter group. We also showed that forecasts performed with \class{} or \camb{} agree very well with each other, especially when both codes are called through their Python wrappers (leading to sub-0.5\% differences between all errors computed with \CFintCAMB{} versus \CFintCLASS). Additionally, we proved that the Gaussian approximation on which all Fisher forecasts are relying is valid in the context of \Euclid and of the $\Lambda$CDM+$\{w_0, w_a\}$ model, since the Fisher ellipses overlap with the contours derived from an MCMC run using the same \montepython{} mock \Euclid likelihoods.

This validation step is interesting by itself, because it allows us to cross-check the impact and the self-consistency of several details in the recipes used for modelling the \Euclid photometric and spectroscopic probes, and by the same occasion, to validate some new ways of using the \cosmicfish{} code, with four different interfaces with the \camb{} and \class{} Einstein--Boltzmann solvers. In the present paper, we tried to document thoroughly all the physical assumptions and numerical methods used by these pipelines, reporting things exactly as they are implemented in the codes. We also reported our predictions for the sensitivity of \Euclid to the parameters of the $\Lambda$CDM+$\{w_0, w_a\}$ model when the photometric and spectroscopic probes are used either independently or in combination with each other.

However, in the context of preliminary work for \Euclid, the most important consequence of this work of validation is to pave the way to further robust forecasts. As a matter of fact:
\begin{itemize}
\item Any cosmological model implemented in the \class{} code can now be immediately used in a forecast: for this, one just needs to add to the input file of \montepython{} one line per free parameter. If this parameter is known by \class, there is nothing else to be done. Thanks to the new \CFintCAMB{} pipeline presented here, this is actually also true for any parameter or model implemented in \camb. However, the version of the mock likelihoods used in this paper relies on a few model-dependent assumptions (like the standard Poisson equation or the fact that galaxies trace the total matter power spectrum). These assumptions require some generalisation in the case e.g. of massive neutrinos or modified gravity models. A generalisation of the \montepython{} \Euclid mock likelihoods to the case of massive neutrinos will be released soon (together with a forthcoming publication).
\item Since \montepython{} calls the very same functions (describing the two \Euclid mock likelihoods) when computing the Fisher matrix or when running an MCMC forecast, and since \MPFisher{} forecasts have been validated against \istfisher\ forecasts, we automatically know that \MPMCMC{} forecasts can be trusted on equal footing with \istfisher\ forecasts, with the additional advantage of taking consistently into account a possible non-Gaussianity of the posteriors.
\end{itemize}

The \montepython{} \Euclid mock likelihoods should not be confused with the official \Euclid likelihood that is currently being developed within the collaboration. The latter is more ambitious and includes effects that are neglected here (such as e.g. super-sample covariance or a different modelling of nonlinear effects). Nonetheless, the \montepython{} \Euclid likelihoods provide a robust tool against which the official likelihood can be compared and tested. 

Another more general benefit from our analysis is that we have established and documented a list of accuracy settings for the EBSs \camb{} and \class{} that prove to be sufficient for the purpose of \Euclid forecasts, and thus, in principle, for the analysis of real \Euclid data.

\bibliographystyle{aa}
\bibliography{biblio}

\begin{acknowledgements}
\\
N.S.\ acknowledges support from the Maria de Maetzu fellowship grant: CEX2019-000918-M, financiado por MCIN/AEI/ 10.13039/501100011033.
N.F.\ is supported by the Italian Ministry of University and Research (MUR) through the Rita Levi Montalcini project ``Tests of gravity on cosmic scales'' with reference PGR19ILFGP. N.F.\ and F.P.\ also acknowledge the FCT project with ref. number PTDC/FIS-AST/0054/2021. 
S.V.\ acknowledges support funded by the Deutsche Forschungsgemeinschaft (DFG, German Research Foundation) under Germany's Excellence Strategy -- EXC-2121 ``Quantum Universe'' -- 390833306. E.B.\ has received funding from the European Union’s Horizon 2020 research and innovation program under the Marie Skłodowska-Curie grant agreement No 754496.
F.P.\ acknowledges partial support from the INFN grant InDark and the Departments of Excellence grant L.232/2016 of the Italian Ministry of University and Research (MUR).
St.C.\ acknowledges support from the `Departments of Excellence 2018-2022' Grant (L.\ 232/2016) awarded by the Italian Ministry of University and Research (\textsc{mur}).
Z.S.\ acknowledges funding from DFG project 456622116.
\AckEC
\end{acknowledgements}

\appendix

\newpage

\section{Input and precision settings in Einstein--Boltzmann solvers\label{app:EBS_settings}}

We list below the input parameter passed to either \camb{} or \class{} in our \Euclid forecasts. \Cref{sec:acc} discusses the relevance of precision parameter settings for such forecasts.

\subsection{Fiducial cosmology parameters\label{app:EBS_fiducial}}

We use the same fiducial value of cosmological parameters as in \istfisher:

\camb{}
\begin{lstlisting}
use_physical = T  # to pass big Omegas

sigma8 = 0.815584
hubble = 67
omega_baryon = 0.05
omega_m = 0.32
scalar_spectral_index(1) = 0.96
w = -1.0
wa = 0.0

dark_energy_model = PPF  # use PPF scheme in DE perturbations, like CLASS
helium_fraction = 0.2454006
temp_cmb = 2.7255 # relevant in neutrino mass conversion
reionization = F # Reio is irrelevant here. Switching it off gives a small
                   # speedup. Not possible when CMB included. 
\end{lstlisting}

We get the same fiducial cosmology in \class{} with the following settings:

\class{}
\begin{lstlisting}
sigma8 = 0.815584
h = 0.67
Omega_b = 0.05
Omega_m = 0.32 
n_s = 0.96
w0_fld=-1.
wa_fld=0.

Omega_Lambda=0 # setting the true cosmological constant to zero, in order 
                 # to activate the DE fluid (fld) with a CPL equation of state
YHe = 0.2454006 # Helium fraction
T_cmb = 2.7255 # relevant in neutrino mass conversion
reio_parametrization = reio_none               
\end{lstlisting}

\subsection{Neutrino settings\label{app:neutrino}}

This issue is actually tricky because, by default, \camb{} and \class{} model the details of the neutrino sector differently. The modelling used in \class{} aims at being closer to reality (that is, to the results of the most detailed studies of neutrino decoupling, such as \cite{Froustey:2020mcq,Bennett:2020zkv}). However, for the purpose of comparing our forecast pipelines -- rather than fitting real data -- the quest for realism is irrelevant: we just want the neutrino modelling to be reasonable and identical in the two codes. Then, it is easier to set up \class{} parameters in such way to mimic the \camb{} neutrino model.

The \camb{} neutrino model used here is the one corresponding to the option {\tt share\_delta\_neff = T}, explained in \url{https://cosmologist.info/notes/CAMB.pdf}. The relevant neutrino parameters that we pass to \camb{} are:

\camb
\begin{lstlisting}
share_delta_neff = T
num_mass_eigenstates = 1
massless_neutrinos = 2.046
massive neutrinos = 1
mnu=0.06
\end{lstlisting}

In \camb, these settings trigger several non-trivial operations, with the goal of having the same temperature shared by the three neutrino species. The true effective number of ultra-relativistic degrees of freedom (i.e. massless neutrinos) actually used in the \camb{} equations is redefined internally from 2.046 to a new number: 
\begin{equation}
N_\mathrm{ur} = 2 + \frac{2}{3} \times (2.046 - \mathrm{floor}[2.046]) \simeq 2.030666\, ,
\end{equation}
while the massive neutrinos are modelled as a perfect Fermi-Dirac species with a temperature $T_\nu$ and fractional density $\Omega_\nu$ computed as
\begin{align}
\frac{T_\nu}{T_\gamma} &= \left(\frac{4}{11}\right)^{1/3} \left(\frac{1+2.046}{3}\right)^{1/4}=0.7164864 \, , \\
\Omega_\nu h^2 &= \frac{0.06~\mathrm{eV}}{94.07~\mathrm{eV}} \left(\frac{1+2.046}{3}\right)^{3/4} = 6.451439 \times 10^{-4}
~\Rightarrow~ 
\Omega_\nu = 1.437166 \times 10^{-3}. 
\end{align}

To mimic exactly the same settings, we must pass these last numbers to \class:

\class
\begin{lstlisting}
N_ncdm = 1
N_ur = 2.030666
T_ncdm = 0.7164864
Omega_ncdm = 0.001437166
\end{lstlisting}

\subsection{Linear matter power spectrum settings\label{app:linear}}

The following output settings allow \camb{} and \class{} to output the matter power spectrum with a fine enough sampling to avoid interpolation errors within the likelihood codes.

\camb
\begin{lstlisting}
transfer_high_precision = T
transfer_kmax = 50
transfer_k_per_logint = 50
\end{lstlisting}

\class
\begin{lstlisting}
P_k_max_h/Mpc = 50.
k_per_decade_for_pk = 50
k_per_decade_for_bao = 50
\end{lstlisting}

(The \cosmicfish{} internal mode expects a $k_\mathrm{max}$ in units of 1/Mpc: in this case we pass {\tt P\_k\_max\_1/Mpc = 50.})

\subsection{Nonlinear power spectrum settings \label{app:nonlinear}}

The following parameters are only relevant for the photometric likelihood, which requires a nonlinear power spectrum in input. We stick to the choice of \istfisher\ to use the \texttt{Halofit} version of \cite{Takahashi:2019hth}, including neutrino corrections found by \cite{Bird:2011rb} and implemented in both \camb{} and \class:

\camb
\begin{lstlisting}
do_nonlinear = 1
halofit_version = 4
\end{lstlisting}

Besides, it is crucial to change the value of a tolerance parameter in the version of the \texttt{Halofit} algorithm implemented in \camb. The file {\tt fortran/halofit.f90} contains a function {\tt THalofit\_GetNonLinRatios()} that computes a characteristic radius {\tt rmid} with a bisection method. The bisection accuracy is set by the line

{\tt     ~~~~if (abs(diff).le.0.001) then}

It is essential to substitute this tolerance by a new precision parameter or a hard-coded value equal to (or smaller than) $10^{-6}$. Otherwise, the error on {\tt rmid} is way too large given the \Euclid sensitivity, and leads to inaccurate derivatives in the Fisher matrix calculation. In our implementation, this new \camb{} precision parameter is called {\tt halofit\_tol\_sigma}. We discuss our settings for this parameter together with that of other important accuracy parameters in \cref{app:HP}.

The \texttt{Halofit} version implemented in \class{} coincides with the ``\texttt{Halofit} version 4'' of \camb. We need to ask \class{} to use a large enough $k_\mathrm{max}$ for the sake of convergence of the \texttt{Halofit} algorithm even at high redshift: 

\class
\begin{lstlisting}
non linear = halofit
nonlinear_min_k_max = 80.
\end{lstlisting}

Within \class, the tolerance parameter mentioned above is also called {\tt halofit\_tol\_sigma} and set by default to $10^{-6}$, which proves to be sufficient for computing \Euclid Fisher matrices with the \cosmicfish{} method. However, as discussed in \cref{app:stepsizes_MP}, the calculation of Fisher matrices with the \montepython{} method requires even higher precision. We discuss our settings for this parameter together with that of other important accuracy parameters in \cref{app:HP}.

\subsection{High-precision settings\label{app:HP}}

High-precision settings slow down the two codes substantially, but they guarantee excellent mutual agreement between them, as well as a low level of numerical noise that allows us to compute numerical derivatives w.r.t. cosmological parameters using smaller stepsizes. These settings relate mainly to the multipole $\ell_\mathrm{max}$ at which the various Boltzmann hierarchies are truncated, to the sampling of perturbations in wavenumber space, to integration time steps and to the use of various approximation schemes.

For \camb, we always use the following precision settings:

\camb
\begin{lstlisting}
do_late_rad_truncation = T
high_accuracy_default=T
transfer_interp_matterpower = T
accurate_reionization = F
\end{lstlisting}

There are three additional parameters playing a particularly important role, and for which we define three levels called P1, P2, P3:

\camb{} (P1 / P2 / P3)
\begin{lstlisting}
accuracy_boost = 2 / 2 / 3
l_accuracy_boost = 2 / 2 / 3
halofit_tol_sigma = 1.e-3 / 1.e-6 / 1.e-6
\end{lstlisting}

We compare P1, P2 and P3 in \cref{sec:acc}, but in the rest of this work we stick to P3.

For \class, we show in \cref{sec:acc} that default precision (DP) is usually sufficient when computing the \Euclid Fisher matrices with the  \cosmicfish{} method. However, the calculation of Fisher matrices with the \montepython{} method requires enhanced settings:

\class (HP)
\begin{lstlisting}
l_max_g=20
l_max_pol_g=15
radiation_streaming_approximation = 2
radiation_streaming_trigger_tau_over_tau_k = 240.
radiation_streaming_trigger_tau_c_over_tau = 100.
tol_ncdm_synchronous = 1.e-5
l_max_ncdm=22
ncdm_fluid_trigger_tau_over_tau_k = 41.

background_Nloga = 6000
thermo_Nz_log = 20000
thermo_Nz_lin = 40000
tol_perturbations_integration = 1.e-6

halofit_tol_sigma = 1.e-8
\end{lstlisting}

The first eight parameters increase the precision of the system of perturbation equations (by truncating Boltzmann hierarchies at higher multipoles, better sampling neutrino momenta and using approximations in a smaller region). The next four parameters reduce the integration stepsize in the ordinary differential equations describing respectively the background, thermodynamical and perturbation evolution. They also reduce interpolation errors when the perturbation equations require the evaluation of background and thermodynamical quantities at a given value of the scale factor $a$. The last parameter reduces random errors in \texttt{Halofit} caused by a bisection algorithm.

In DP mode, none of these parameters are passed to the code. We illustrate the impact of DP versus HP in Sec.~\ref{sec:acc}, and we stick to HP everywhere else when computing Fisher matrices. Instead, for MCMC runs, we always stick to DP settings.

\section{Fiducial parameter and choice of step sizes}\label{app:fiducialstep}

\subsection{Fiducial parameters\label{app:fiducial}}

In \cref{tab:pcosmo_input_WL} (resp. \cref{tab:pcosmo_input_GC}), we summarise our choice of fiducial values and step sizes for the cosmological and nuisance parameters of the photometric (resp. spectroscopic) probe. The fiducial values are the same as in \istfisher. In the last two columns of each table, we show the marginalised errors (inferred from MCMC runs), which play a role in the discussion of \cref{app:stepsizes_MP}.

\begin{table}[ht]
    \centering
    \caption{Fiducial values and MCMC marginalised errors for all free parameters in the photometric probe.}
    \begin{tabular}{|c||c|c|c|}
        \hline
        &fiducial&\multicolumn{2}{c|}{$\sigma_\alpha^\mathrm{marg.}$} \\
        parameter & value & pessimistic & optimistic \\
        \hline
        \hline
        $\Omega_{\mathrm{b},0}$ & 0.05 & 2.6 $\times 10^{-3}$ & 2.2 $\times 10^{-3}$ \\
        \hline
        $\Omega_{\mathrm{m},0}$ & 0.32 & 3.8 $\times 10^{-3}$ & 2.0 $\times 10^{-3}$ \\
        \hline
        $h$ & 0.67 & 2.1 $\times 10^{-2}$ & 1.3 $\times 10^{-2}$ \\
        \hline
        $n_{\rm s}$ & 0.96 & 1.1 $\times 10^{-2}$ & 3.8 $\times 10^{-3}$ \\
        \hline
        $\sigma_8$ & 0.815584 & 4.4 $\times 10^{-3}$ & 2.0 $\times 10^{-3}$ \\
        \hline
        $w_0$ & $-1$ & 4.5 $\times 10^{-2}$ & 2.8 $\times 10^{-2}$ \\
        \hline
        $w_a$ & 0 & 1.8 $\times 10^{-1}$ & 1.0 $\times 10^{-1}$ \\
        \hline
        \hline
        $\mathcal{A}_\mathrm{IA}$ & 1.72 & 1.8 $\times 10^{-1}$ & 1.2 $\times 10^{-1}$ \\
        \hline
        $\eta_\mathrm{IA}$ & $-0.41$ & 1.1 $\times 10^{-1}$ & 7.4 $\times 10^{-2}$ \\
        \hline
        \hline
        $b_1$ & 1.0998 & 7.8 $\times 10^{-3}$ & 2.4 $\times 10^{-3}$ \\
        \hline
        $b_2$ & 1.2202 & 9.0 $\times 10^{-3}$ & 3.3 $\times 10^{-3}$ \\
        \hline
        $b_3$ & 1.2724 & 9.8 $\times 10^{-3}$ & 3.8 $\times 10^{-3}$ \\
        \hline
        $b_4$ & 1.3166 & 1.0 $\times 10^{-2}$ & 4.2 $\times 10^{-3}$ \\
        \hline
        $b_5$ & 1.3581 & 1.1 $\times 10^{-2}$ & 4.7 $\times 10^{-3}$ \\
        \hline
        $b_6$ & 1.3998 & 1.2 $\times 10^{-2}$ & 5.1 $\times 10^{-3}$ \\
        \hline
        $b_7$ & 1.4446 & 1.3 $\times 10^{-2}$ & 5.6 $\times 10^{-3}$ \\
        \hline
        $b_8$ & 1.4965 & 1.3 $\times 10^{-2}$ & 5.8 $\times 10^{-3}$ \\
        \hline
        $b_9$ & 1.5652 & 1.4 $\times 10^{-2}$ & 6.3 $\times 10^{-3}$ \\
        \hline
        $b_{10}$ & 1.7430 & 1.6 $\times 10^{-2}$ & 7.1 $\times 10^{-3}$ \\
        \hline
    \end{tabular}
    \label{tab:pcosmo_input_WL}
\end{table}

\begin{table}[ht]
    \centering
    \caption{Fiducial values and MCMC marginalised errors for all free parameters in the spectroscopic probe.}
    \begin{tabular}{|c||c|c|c|}
        \hline
        &fiducial&\multicolumn{2}{c|}{$\sigma_\alpha^\mathrm{marg.}$} \\
        parameter & value & pessimistic & optimistic \\
        \hline
        \hline
        $\Omega_{\mathrm{b},0}$ & 0.05 & 2.0 $\times 10^{-3}$ & 1.9 $\times 10^{-3}$ \\
        \hline
        $\Omega_{\mathrm{m},0}$ & 0.32 & 1.1 $\times 10^{-2}$ & 9.5 $\times 10^{-3}$ \\
        \hline
        $h$ & 0.67 & 3.4 $\times 10^{-3}$ & 2.9 $\times 10^{-3}$ \\
        \hline
        $n_{\rm s}$ & 0.96 & 1.4 $\times 10^{-2}$ & 1.3 $\times 10^{-2}$ \\
        \hline
        $\sigma_8$ & 0.815584 & 1.2 $\times 10^{-2}$ & 1.1 $\times 10^{-2}$ \\
        \hline
        $w_0$ & $-1$ & 9.6 $\times 10^{-2}$ & 8.7 $\times 10^{-2}$ \\
        \hline
        $w_a$ & 0 & 3.3 $\times 10^{-1}$ & 2.8 $\times 10^{-1}$ \\
        \hline
        \hline
        $P_{S1}$ & 0 & 35 & 19 \\
        \hline
        $P_{S2}$ & 0 & 34 & 19 \\
        \hline
        $P_{S3}$ & 0 & 36 & 20 \\
        \hline
        $P_{S4}$ & 0 & 36 & 22 \\
        \hline
        $\ln[b_1 \sigma_8(z_1)]$ & $-0.3256$ & 1.4 $\times 10^{-2}$ & 1.3 $\times 10^{-2}$ \\
        \hline
        $\ln[b_2 \sigma_8(z_2)]$ & $-0.3160$ & 1.4 $\times 10^{-2}$ & 1.3 $\times 10^{-2}$ \\
        \hline
        $\ln[b_3 \sigma_8(z_3)]$ & $-0.3117$ & 1.4 $\times 10^{-2}$ & 1.3 $\times 10^{-2}$ \\
        \hline
        $\ln[b_4 \sigma_8(z_4)]$ & $-0.3203$ & 1.3 $\times 10^{-2}$ & 1.2 $\times 10^{-2}$ \\
        \hline
    \end{tabular}
    \label{tab:pcosmo_input_GC}
\end{table}

\subsection{Stepsizes in \cosmicfish\label{app:stepsizes_CF}}

In \istfisher, the different codes always computed the first-order derivatives $\partial_\alpha C_\ell$ or $\partial_\alpha P(k,\mu,z)$ with the SteM method or the $n$-point stencil method, which rely on the evaluation of the $C_\ell$'s or $P(k,z)$ at a few different values of the parameter $p_\alpha$ -- see Appendix B of \cite{Camera:2016owj} for further details. However, in this work, we checked explicitly that the \cosmicfish{} results stay the same when switching to a simpler two-sided derivative scheme, with steps roughly of the order of 1\% of the fiducial parameter values (or 0.01 in the case of $w_0$ and $w_a$). 

We attribute this feature mainly to our use of higher precision settings in \camb, and in particular, to the reduction of the tolerance parameter of \camb's \texttt{Halofit} version from $10^{-3}$ to $10^{-8}$, see \cref{app:nonlinear}. This leads to a significant reduction of the numerical noise in the \camb{} output, mainly for nonlinear spectra used by the photometric probe. Thus, it removes the need to average the noise over several output spectra, as done implicitly within the SteM scheme. The main \cosmicfish{} results of this work have been obtained with step sizes set to exactly 1\% of the fiducial values of \cref{tab:pcosmo_input_WL,tab:pcosmo_input_GC} (or to 0.01 for $w_0$ and $w_a$). We checked that changing the step sizes by a factor two does not impact the results. Note that the \cosmicfish{} does not use a finite difference method to evaluate derivatives with respect to the shot noise parameters because these are trivial (for $\alpha=p(z_i)$, one gets $\partial_\alpha P_\mathrm{obs}=1$).

\subsection{Stepsizes in \texttt{MontePython}/Fisher\label{app:stepsizes_MP}}

Since the \MPFisher{} and \cosmicfish{} pipelines need to compute some intrinsically different numerical derivatives (second derivatives of ${\cal L}$ versus first derivatives of $C_\ell$'s or $P(k)$'s), the step sizes also need to be optimised independently in the two cases.

As an overall guideline:
\begin{itemize}
    \item When the step sizes are too small, there is a risk that the difference $\Delta {\cal L}$ between likelihood values computed at $(p_\alpha\pm\Delta p_\alpha, p_\beta\pm\Delta p_\beta)$ is dominated by numerical errors rather than physical effects.
    \item When the step sizes are too big, we may exit from the region in which the likelihood is Gaussian, that is, $\chi^2(p_1, ..., p_N)$ is quadratic. The $\chi^2$ would then pick up higher order contributions, starting from cubic terms. As a result, the matrix $\partial_\alpha \partial_\beta (- \ln {\cal L})$ would strongly dependent on step sizes and would no longer stand for the Fisher matrix.
\end{itemize}
If the precision of the EBS and of the likelihood code are sufficient, there should exist a range of intermediate step sizes such numerical noise is under control while $F_{ij}$ is nearly constant as a function of step sizes. This can be checked explicitly: \begin{itemize}
    \item For diagonal coefficient of the Fisher matrix, one can plot $F_{\alpha \alpha} = \partial_\alpha^2 (-\ln {\cal L})$ as a function of the step $\Delta p_\alpha$. One should be able to see three regions: for too small $\Delta p_\alpha$, $F_{\alpha \alpha}$ should feature random fluctuations; For intermediate $\Delta p_\alpha$, it should remain flat; For too large $\Delta p_\alpha$, it should either raise or decrease, depending on the sign of higher order corrections. 
    \item For non-diagonal coefficient, one may plot $F_{\alpha \beta}$ as a two-dimensional function of ($\Delta p_\alpha$, $\Delta p_\beta$), and draw similar conclusions. 
\end{itemize}
After performing such tests, one can choose some arbitrary step sizes provided that they lay within the intermediate region. These tests also allow us to check the accuracy of the Einstein--Boltzmann and likelihood codes. If both codes are accurate enough for the purpose of computing the Fisher matrix, there will exist a wide range of intermediate step sizes for each parameter. If one of the codes is not sufficiently accurate, there will be no intermediate region, that is, derivative curves will show no transition between the unstable region and the non-Gaussian region.

Since doing all these tests can be very time-consuming, we do not want to perform them extensively before each new calculation of a Fisher matrix. In this work, we adopt a pragmatic approach:
\begin{enumerate}
\item We first compute the marginalised errors on each parameter $p_\alpha$, $\sigma_\alpha^\mathrm{marg.}$, using either \cosmicfish, or \MPMCMC{} runs, or preliminary runs of \MPFisher{} -- this makes no difference. We set these $\sigma_\alpha^\mathrm{marg.}$'s as reference values with respect to which we can calibrate our step sizes. Our values of $\sigma_\alpha^\mathrm{marg.}$ are listed in the last two columns of \cref{tab:pcosmo_input_WL,tab:pcosmo_input_GC}.
\item We compute a few diagonal and non-diagonal Fisher matrix elements $F_{\alpha \beta}$ with the step sizes ($\Delta p_\alpha = x \,\, \sigma_\alpha^\mathrm{marg.}$,$\Delta p_\beta = x \,\, \sigma_\beta^\mathrm{marg.}$) for various values of $x$ (typically in the range $0.01<x<1$), and we plot the curves $F_{\alpha \beta}(x)$.
\item We check that there exist an intermediate range of values of $x$ such that the curves are stable and flat. If not, we push the accuracy settings of the codes and we start the test again. If yes, we choose an arbitrary value of $x$ sitting comfortably within the intermediate region.
\item Once this test has been done once for each probe, we can make general recommendations to use a given accuracy in the codes and to set correctly the order of magnitude of the steps relatively to the marginalised errors. 
\item Then, there is no need to repeat this test again and again. The user can trust the fact that with such recommendations, the Fisher matrix is accurate and stable against small variations of the step sizes. At most, as a cross-check, the user can try once to increase or decrease all step sizes (for instance, by $\sim$30\%), recompute the Fisher matrix and verify that it remains stable.  
\end{enumerate}
We have performed such tests for each probe and for a few matrix elements $F_{\alpha \beta}$. We focused particularly on the non-diagonal element $F_{w_0 \, w_a}$, which is among the most difficult ones to compute accurately given the strong correlation between $w_0$ and $w_a$. 

\begin{figure}[h]
    \centering
    \includegraphics[width=0.65\textwidth]{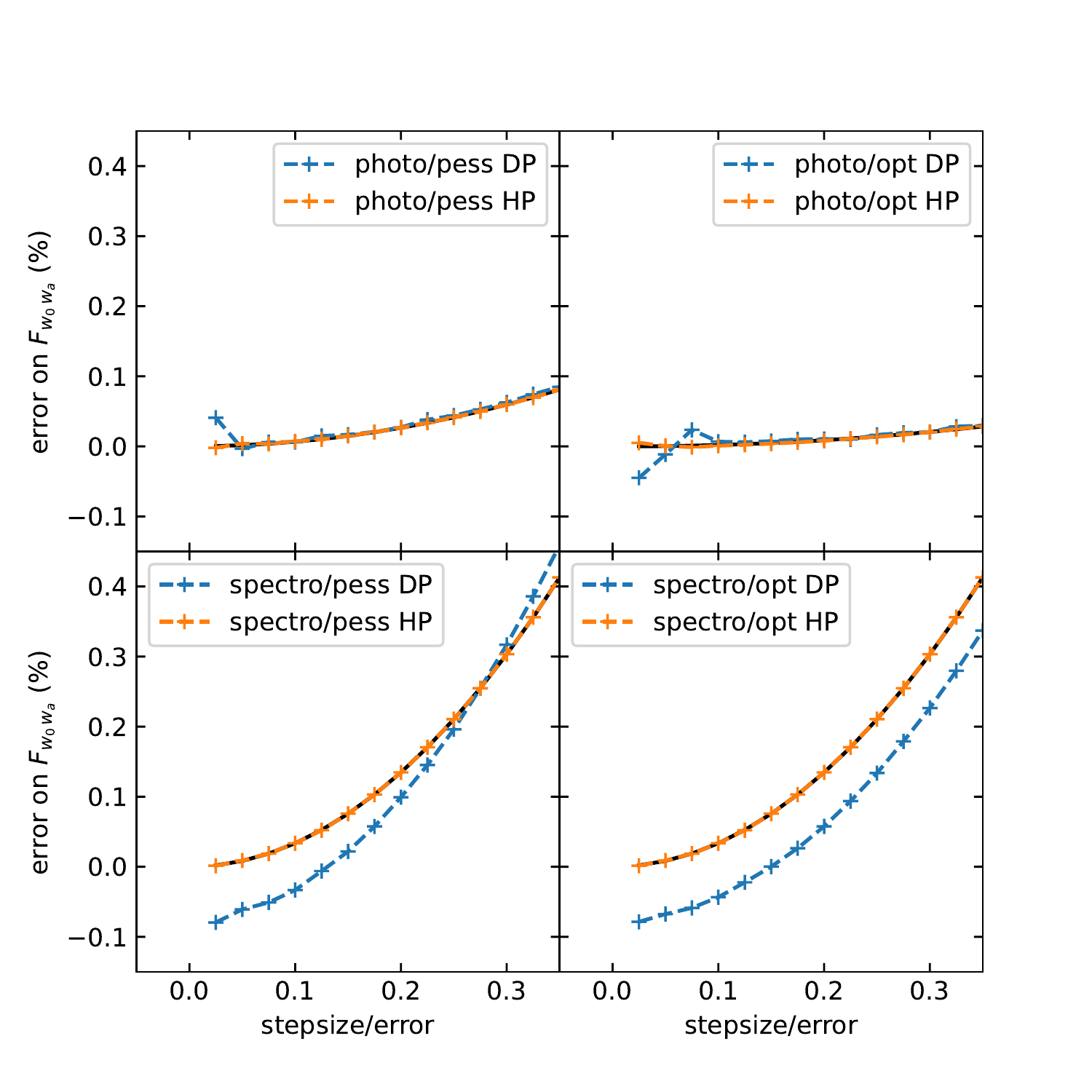}
    \caption{Percentage error on the estimate of the second derivative $F_{w_0 \, w_a}(x)$ as a function of the step-size-to-marginalised-error ratio $x$, for 15 values in the range $0.025\leq x \leq 0.4$ and each of the two probes (top: photometric /  bottom: spectroscopic) and case (left: pessimistic / right: optimistic). In each panel, we show the results from \class{} with default-precision (DP) or high-precision (HP) settings. The black line is a parabolic fit to the HP points.
    \label{fig:test_stepsizes}
    }
\end{figure}

For each probe, we have computed the function $F_{w_0 \, w_a}(x)$ for 40 values of $x$ ranging from 0.025 to 1 with a spacing $\Delta x=0.025$, while using \class{} with either the default-precision (DP) or high-precision (HP) settings defined in \cref{app:HP}. Once the value of $x$ has been set, we multiply it with the marginalised errors of \cref{tab:pcosmo_input_WL,tab:pcosmo_input_GC} to get our actual step sizes (these errors are derived from our MCMC runs, but we could have indifferently taken some marginalised errors from \cosmicfish). For sufficiently large values of $x$, the points lay on a parabola ($F_{w_0 \, w_a}(x) \propto x^2$) that corresponds to the leading-order deviation from a Gaussian likelihood (since an exact Gaussian likelihood would give $F_{w_0 \, w_a}(x) = \mathrm{constant}$). This happens typically for $x\geq0.1$ with HP, or above a larger threshold with DP. We then infer the asymptotic value $F^\mathrm{fit}_{w_0 \, w_a}(0)$ from a parabolic fit to the set of points computed with high-precision settings in the range $0.1 \leq x\leq 1$. Finally, in Fig.~\ref{fig:test_stepsizes}, we show the quantity
\begin{equation}
y \equiv 100 \left( \frac{F_{w_0 \, w_a}(x)}{F^\mathrm{fit}_{w_0 \, w_a}(0)}-1\right)
\end{equation}
as a function of the step-size-to-error ratio $x={\Delta p_\alpha} / {\sigma_\alpha^\mathrm{marg.}}$, for each accuracy setting. The quantity $y$ can be considered as the percentage error on the estimate of the second derivative $F_{w_0 \, w_a}(x)$ that comes from random numerical noise and/or the non-Gaussianity of the likelihood, that is, from the kind of errors that can be controlled by an adequate step size. (Of course, $y$ does not include contributions from some possible systematic errors in the EBS or in the likelihood code, but we can estimate those errors independently from the \camb-versus-\class{} and \cosmicfish-versus-\montepython{} comparison of \cref{sec:validation}.) We computed $y(x)$ in the range $0.025\leq x \leq 1$, but in Fig.~\ref{fig:test_stepsizes} we only show the range $0.025\leq x \leq 0.4$ which is the most interesting (for larger $x$, we obtain a nearly perfect parabola).

For the photometric survey, Fig.~\ref{fig:test_stepsizes} shows that, with DP, the points feature random fluctuations below $x<0.2$. Indeed, when computing the Fisher matrix, we get unstable results with DP and such small step sizes (e.g. the inverse Fisher matrix is not always positive definite and the marginalised errors fluctuate a lot with the step size). The solution consists in switching to HP: then, the points follow smoothly the parabola down to $x\sim0.1$, and even lower. Additionally, we see that non-Gaussian corrections are small: even at $x=0.4$ they affect $F_{w_0 \, w_a}$ by only 0.1\% (resp. 0.04\%) in the pessimistic (resp. optimistic) case. As expected, the optimistic case features a more Gaussian likelihood (thanks to its enhanced constraining power). Our overall recommendation for the photometric survey is to use HP settings with an $x$ roughly in the range $[0.1, 0.5]$, although it would also be possible to get good result while sticking to DP with 
$x$ roughly in the range $[0.3, 0.5]$. We did various tests to check explicitly that with such recommendations, we always get stable predictions for all unmarginalised and marginalised errors. However, in the result sections, we always used HP and $x=0.1$. The absolute step sizes $\Delta p_\alpha$ for the pessimistic and optimistic cases are thus given by the last two columns of \cref{tab:pcosmo_input_GC} divided by 10.

For the spectroscopic survey, Fig.~\ref{fig:test_stepsizes} shows that, with either DP or HP, the points are smoothly distributed and follow a parabola for any $x\geq0.05$. The non-Gaussianity is much stronger in this case: with $x=0.4$ the error on $F_{w_0 \, w_a}$ reaches 1\% (resp. 0.5\%) in the pessimistic (resp. optimistic) case. Thus one should restrict to $x < 0.1$ in order to keep the non-Gaussian contamination below approximately 0.1\%. For $x$ in the range $[0.05, 0.1]$, the DP and HP points are always following a smooth parabola, and indeed we checked that the Fisher matrix and errors are always stable in this range. However, the results obtained with HP are more correct because the physical effects of the parameters are captured with higher accuracy. In Fig.~\ref{fig:test_stepsizes}, this appears in the form of an overall shift of $F_{w_0 \, w_a}$ by approximately $-0.1$\% when sticking to DP. We observed similar shifts in other elements of the Fisher matrix. Instead, when further increasing precision beyond our HP settings, we checked that the results remain stable. Between DP and HP, the (un)marginalised Fisher errors only move by one or two percent, but for better accuracy we recommend using HP with $x$ in the range $[0.05, 0.1]$. In the result sections, we always used HP and $x=0.05$. The absolute step sizes $\Delta p_\alpha$ for the pessimistic and optimistic cases are then equal to the last two columns of \cref{tab:pcosmo_input_GC} divided by 20.

\newpage

\section{Contour plots for nuisance parameters \label{app:contours_nuisance}}

\subsection{Photometric survey with pessimistic settings}

\begin{figure}[h]
    \centering
    \includegraphics[width=0.95\textwidth]{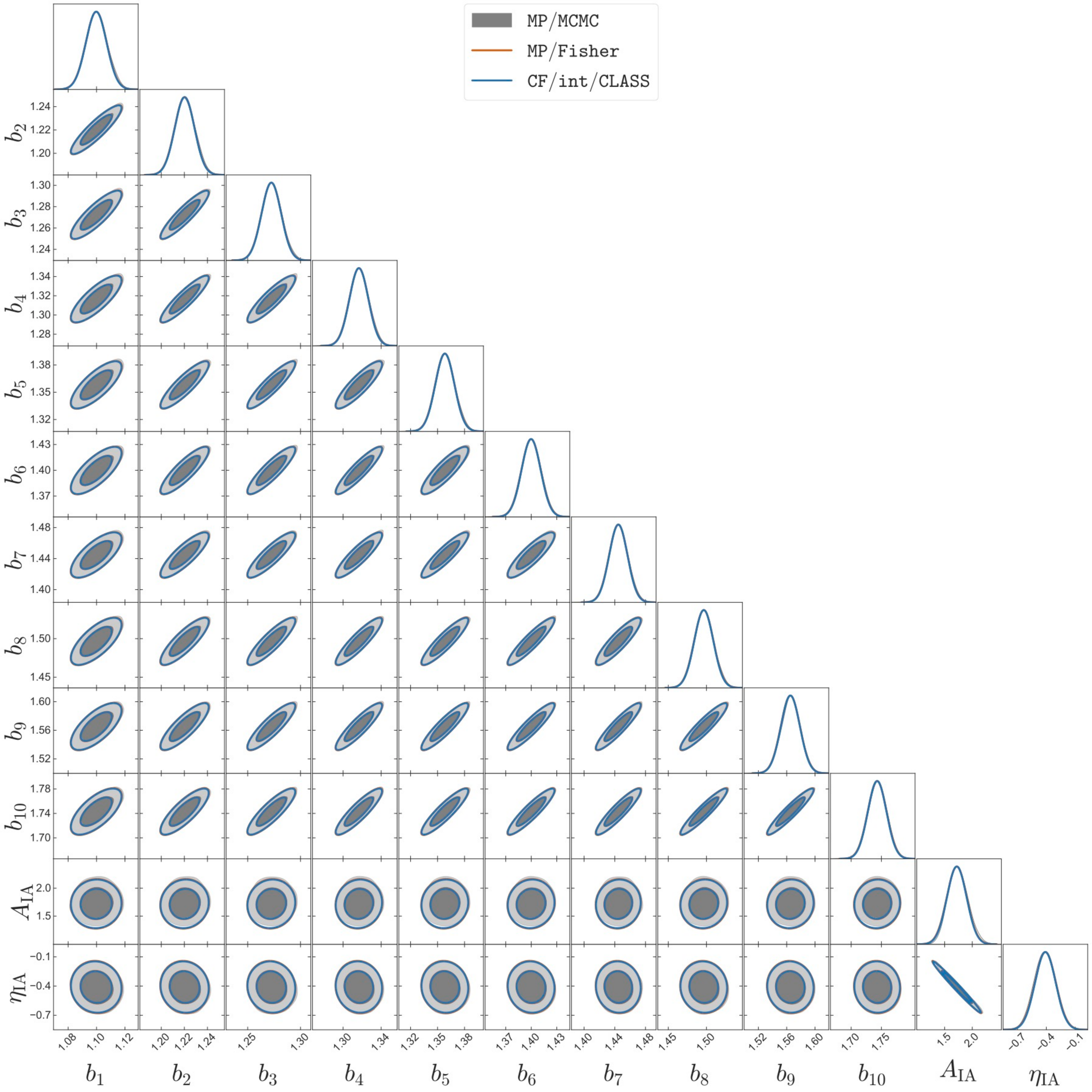}
    \caption{For the photometric survey with pessimistic settings, comparison of 1D posterior and 2D contours (for 68\% and 95\% confidence level) from different methods: \MPMCMC{} (grey lines/contours), \MPFisher{} (orange lines),  \CFintCLASS{} (blue lines). We only show here the nuisance parameters. 
    The triangle plot for cosmological parameters was shown in the main text. Plotted using \texttt{GetDist}.
    \label{fig:contours_nuisance_photo_pess}
    }
\end{figure} 

\begin{figure}[h]
    \centering
    \includegraphics[width=0.70\textwidth]{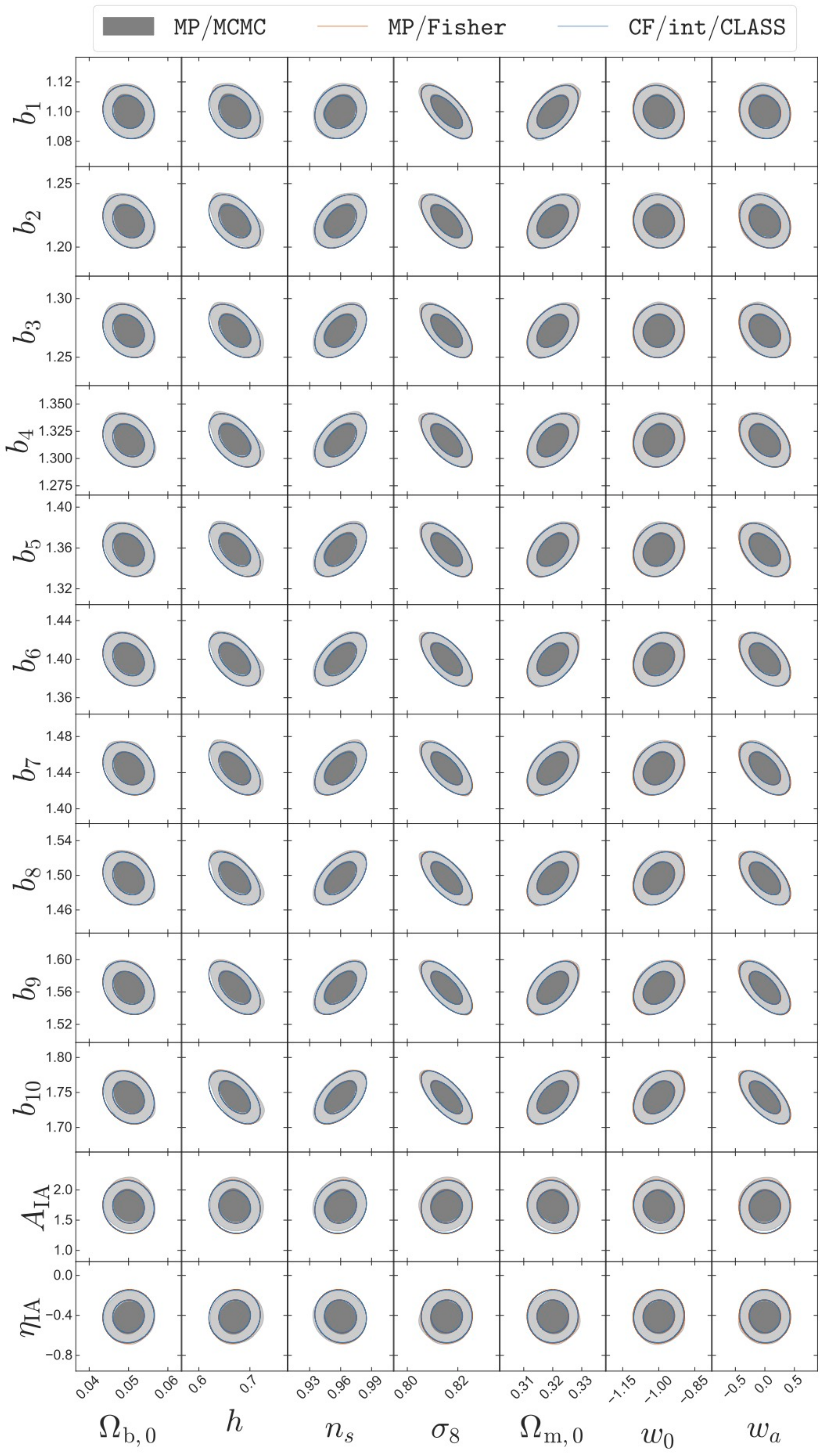}
    \caption{
    Same as Fig.~\ref{fig:contours_nuisance_photo_pess} for the correlation between cosmological and nuisance parameters. 
    \label{fig:contours_cross_photo_pess}
}
\end{figure}  

\newpage

\subsection{Spectroscopic survey with optimistic and pessimistic settings}

\begin{figure}[h]
    \centering
    \includegraphics[width=0.95\textwidth]{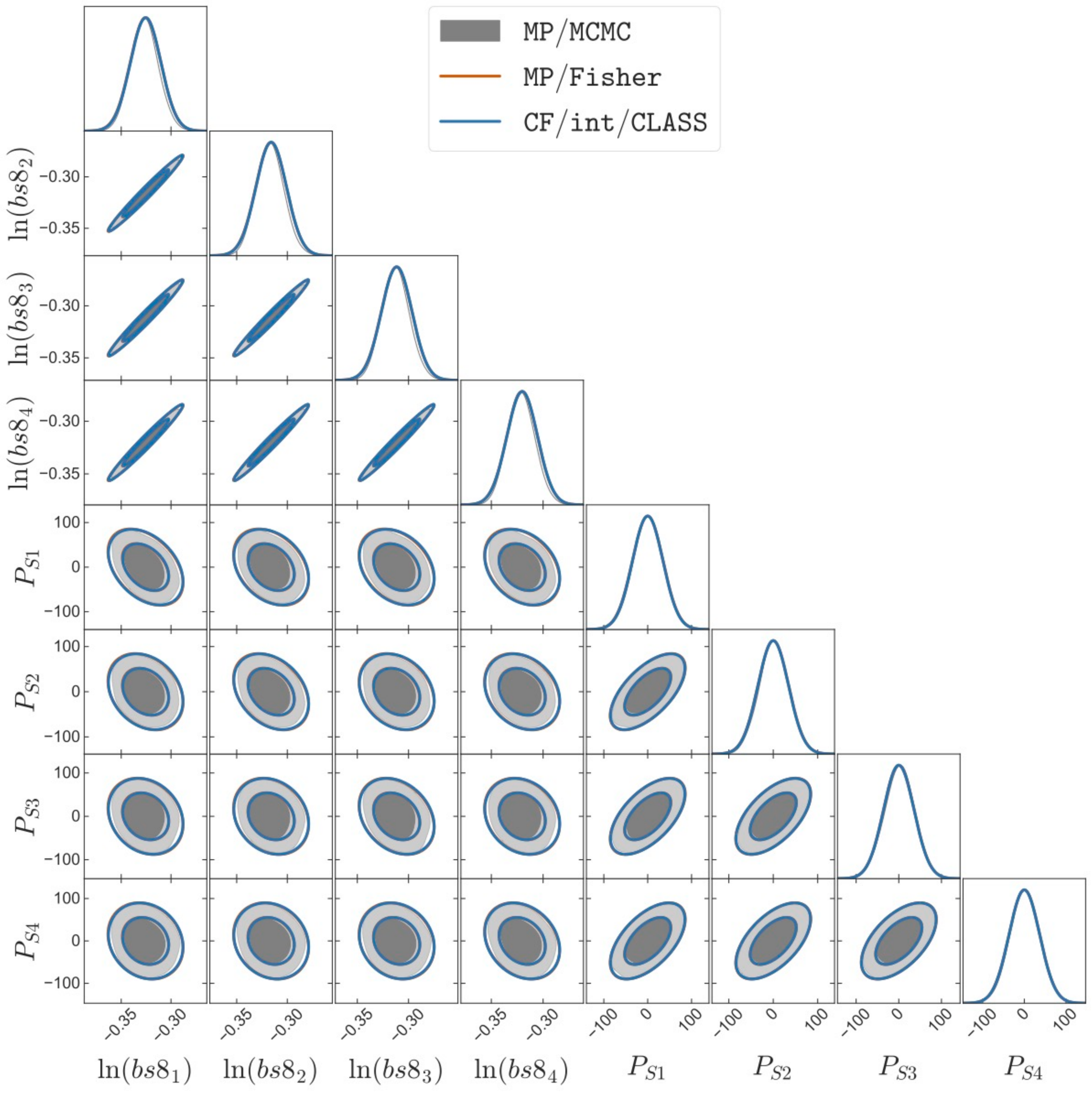}
    \caption{
    Same as Fig.~\ref{fig:contours_nuisance_photo_pess} for the nuisance parameters of the spectroscopic survey with pessimistic settings.
    For illustration purposes, we have labelled the parameters $\ln[b_i \sigma_8(z_1)]$ as $\ln(bs8_i)$.
    \label{fig:contours_nuisance_spectro_opt}
    }
\end{figure}

\begin{figure}[htbp]
    \centering
    \includegraphics[width=0.8\textwidth]{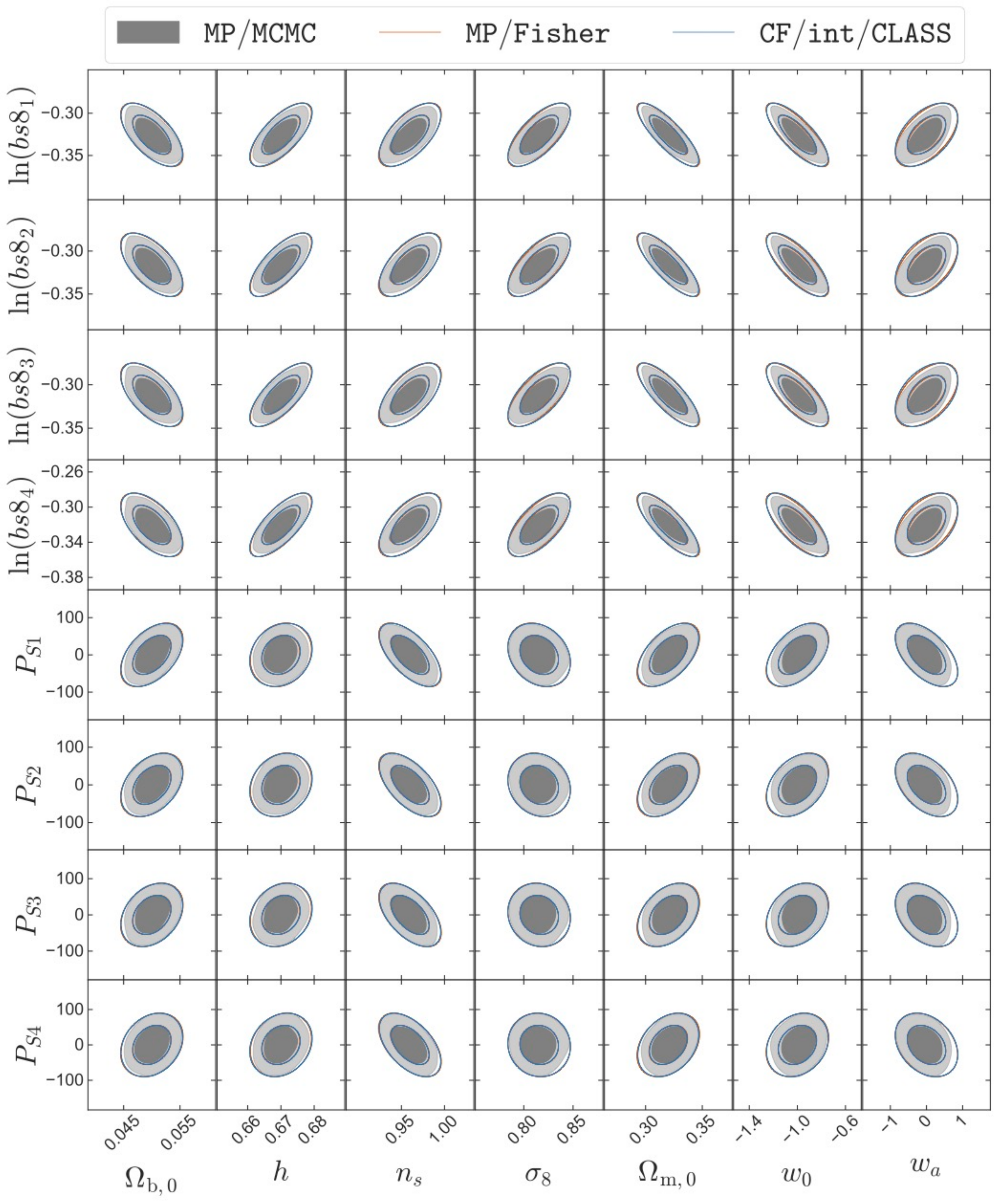}
    \caption{
    Same as Fig.~\ref{fig:contours_nuisance_spectro_opt} for the correlation between cosmological and nuisance parameters. 
    \label{fig:contours_cross_spectro_opt}
    }
\end{figure}

\section{Contour plots for combined probes with pessimistic settings\label{app:combined}}

\begin{figure}[h]
    \centering
    \includegraphics[width=0.95\textwidth]{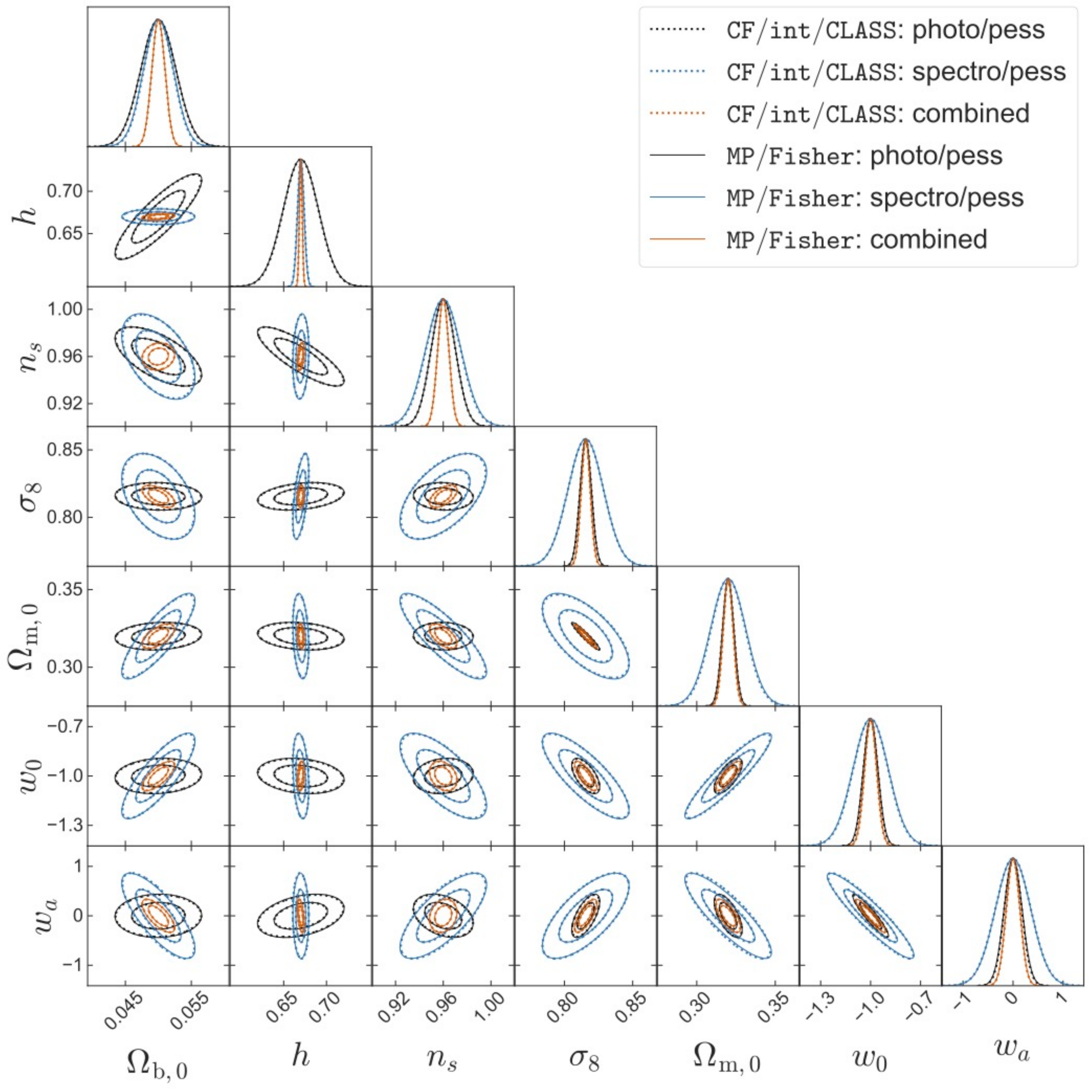}
    \caption{For the individual and combined photometric and spectroscopic surveys with pessimistic settings, 1D posterior and 2D contours on cosmological parameters from  \CFintCLASS{} (dotted lines) and \MPFisher{} (solid lines). Plotted using \texttt{GetDist.}}
    \label{fig:contours_cosmo_combined_pess_pess}
\end{figure}  

\end{document}